\documentclass[opre,nonblindrev]{informs3}

\definecolor{green}{rgb}{1,0.5,0}

\usepackage{endnotes}
\let\footnote=\endnote
\let\enotesize=\normalsize
\def\notesname{Endnotes}%
\def\makeenmark{$^{\theenmark}$}
\def\enoteformat{\rightskip0pt\leftskip0pt\parindent=1.275em
  \leavevmode\llap{\makeenmark}}

\usepackage{natbib}
 \bibpunct[, ]{(}{)}{,}{a}{}{,}%
 \def\bibfont{\small}%
\TheoremsNumberedThrough     % Preferred (Theorem 1, Lemma 1, Theorem 2)
\ECRepeatTheorems

\EquationsNumberedThrough    % Default: (1), (2), ...
\MANUSCRIPTNO{Appearing in \emph{Operations Research}: \url{https://doi.org/10.1287/opre.2022.2290}}

\usepackage[utf8]{inputenc}
\usepackage[T1]{fontenc}
\usepackage{enumerate}
\usepackage{mathrsfs}

\usepackage{amsmath,natbib,amssymb}

\usepackage{booktabs} % For formal tables

\usepackage{mathrsfs}
\usepackage{graphicx} 
\usepackage{bm}
\usepackage{dsfont}
\usepackage{fnpct}
\usepackage{mathtools}
\usepackage{comment}
\usepackage{mathrsfs}  
\mathtoolsset{showonlyrefs}
\usepackage{enumitem}
\usepackage{bbm}
\usepackage{makecell}

\usepackage{caption}
\usepackage{subcaption}

\usepackage[ruled,linesnumbered]{algorithm2e}
\LinesNumbered
\SetKw{KwRequire}{Require:}
\SetKw{KwFrom}{from}
\SetCommentSty{mycommfont}

\usepackage{tocloft} % influence appearance of ToC, LoF, and LoT

\RequirePackage{hypernat}

\usepackage[hidelinks]{hyperref}

\definecolor{green}{RGB}{0,0,0}%(0,95,15)

\hypersetup{linktocpage}

\usepackage{bookmark}

\newtheorem{fact}{Fact}
\newtheorem{Procedure}{Procedure}

\definecolor{DSgray}{cmyk}{0,0,0,0.7}
\definecolor{DSred}{cmyk}{0,0.7,0,0.7}
\definecolor{DSblue}{cmyk}{0.7, 0.7, 0, 0}

\newcommand{\revision}[1]{\textcolor{green}{#1}}

\newcommand{\alg}{\mbox{\sf Alg}}
\newcommand{\opt}{\mbox{\sf OPT}}
\newcommand{\ccomma}{\mathbin{\raisebox{0.5ex}{,}}}

\begin{document}

\RUNAUTHOR{Eckles et al.}
\RUNTITLE{Seeding with Costly Network Information}
\TITLE{Seeding with Costly Network Information}

% \ARTICLEAUTHORS{%
% \AUTHOR{Dean Eckles}
% \AFF{Sloan School of Management, Massachusetts Institute of Technology, \EMAIL{eckles@mit.edu}}
% \AUTHOR{Hossein Esfandiari}
% \AFF{Google Research, \EMAIL{esfandiari@google.com}}
% \AUTHOR{Elchanan Mossel}
% \AFF{Department of Mathematics, Massachusetts Institute of Technology, \EMAIL{elmos@mit.edu}}
% \AUTHOR{M. Amin Rahimian}
% \AFF{Department of Industrial Engineering, University of Pittsburgh, \EMAIL{rahimian@pitt.edu}}
% % Enter all authors
% } % end of the block

\ARTICLEAUTHORS{%
\AUTHOR{Dean Eckles$^{1}$, Hossein Esfandiari$^{2}$, Elchanan Mossel$^{3}$, M. Amin Rahimian$^{4}$}
\AFF{$^{1}$Sloan School of Management, MIT \quad $^{2}$Google Research \quad $^{3}$Department of Mathematics, MIT\\  $^{4}$Department of Industrial Engineering, University of Pittsburgh\\ \EMAIL{eckles@mit.edu, esfandiari@google.com, elmos@mit.edu, rahimian@pitt.edu}}
% Enter all authors
} % end of the block

\ABSTRACT{We study the task of selecting $k$ nodes, in a social network of size $n$, to seed a diffusion with maximum expected spread size, under the independent cascade model with cascade probability $p$. Most of the previous work on this problem (known as influence maximization) focuses on efficient algorithms to approximate the optimal seed set with provable guarantees given knowledge of the entire network; however, obtaining full knowledge of the network is often very costly in practice. Here we develop algorithms and guarantees for approximating the optimal seed set while bounding how much network information is collected. First, we study the achievable guarantees using a sublinear influence sample size. We provide an almost tight approximation algorithm with an additive $\epsilon n$ loss and show that the squared dependence of sample size on $k$ is asymptotically optimal when $\epsilon$ is small. We then propose a probing algorithm that queries edges from the graph and use them to find a seed set with the same almost tight approximation guarantee. We also provide a matching (up to logarithmic factors) lower-bound on the required number of edges. This algorithm is implementable in field surveys or in crawling online networks. Our probing takes $p$ as an input which may not be known in advance, and we show how to down-sample the probed edges to match the best estimate of $p$ if they are collected with a higher probability. Finally, we test our algorithms on an empirical network to quantify the tradeoff between the cost of obtaining more refined network information and the benefit of the added information for guiding improved seeding strategies.}

 %\HISTORY{Received: February 20, 2020. Revised: June 17, 2021; November 21, 2021. Accepted: December 30, 2021.}% 

\maketitle

\section{Introduction}\label{sec:intro}
Decision-makers in marketing, public health, development, and other fields often have a limited budget for interventions, such that they can only target a small number of people for an intervention. Thus, in the presence of social or biological contagion, they strategize about where in a network to intervene --- often where to seed a behavior (e.g., product adoption) by engaging in an intervention (e.g., giving a free product) \citep{domingos2001mining, kempe2003maximizing, hinz2011seeding, libai2013decomposing, banerjee2017using, godes2009firm}. The influence maximization problem is to choose a set of $k$ seeds with maximum expected spread size, given a known network and model of diffusion~\citep{domingos2001mining}. Following the seminal work of \cite{kempe2003maximizing} ---  who showed NP-hardness and efficient approximation through submodular influence maximization ---  a huge literature is devoted to developing fast algorithms that can be applied to massive scale social networks \citep[e.g.,][]{Chen:2009:EIM:1557019.1557047,wang2012scalable}. 

In this work, we address the problem of influence maximization when \revision{the} social network is unknown and so network information needs to be acquired through costly effort. This has applications in development economics --- e.g., adoption of microfinance \citep{banerjee2013diffusion} and insurance \citep{cai2015social}; public health --- e.g., adoption of water purification methods and multivitamins \citep{kim2015social}, spreading information about immunization camps \citep{banerjee2017using}, preventing misinformation about drug side effects \citep{chami2017social}, increasing HIV awareness among homeless youth \citep{yadav2017influence}, and adoption of contraception \citep{behrman2002social}; and education  --- e.g., reducing bullying and conflict among adolescents \citep{paluck2016changing}. In these settings, data about network connections is often acquired through costly surveys. In practice, collecting the entire network connection (edge) data can be difficult, costly, or even impossible. To reduce the cost of such surveys a few seeding strategies have been proposed to avoid collecting the entire network information by relying on stochastic ingredients, such as one-hop targeting, whereby one targets random network neighbors of random individuals \citep{kim2015social,chin2021evaluating,chami2017social}. Moreover, such methods have the advantage of scalability, since they can be implemented without mapping the entire network. This is also important in online social networks with billions of edges, where working with the entire contact lists might be impractical or limited by rate limits for third parties crawling these networks. Although the importance of influence maximization with partial network information has been noted and there are a few papers considering this problem \citep{mihara2015influence,mihara2017effectiveness,stein2017heuristic,wilder2018maximizing}, none of these previous works come with provable performance guarantees for general graphs.

To limit access of seeding algorithms to network information, we use an edge query model and provide tight guarantees of what is achievable with a bounded number of queries. We organize our edge queries by sequentially probing the graph nodes: we probe each node by revealing its incident edges with independent cascade probability $p$, proceed to probe its revealed neighbors, and repeat. Our approximation algorithm uses the revealed network information to seed $k$ nodes with guarantees that match hardness lower bounds (up to logarithms). 

We begin our analysis by a thought experiment (Section \ref{sec:spreadingQ}): assuming that network information is made available through ``influence samples'', i.e., by seeding random nodes and observing their spread outcomes, how many influence sample\revision{s} do we need to collect? We show that to seed $k$ nodes in a network of size $n$ with tight approximation guarantees, it is necessary (up to logarithms) and sufficient to collect $O(k^2\log n)$ influence samples. In Section \ref{sec:edgeQ}, we provide our main results by showing that the same approximation guarantees can be achieved using $O(p n^2 \log^4 n)$ edge queries (with a matching lower bound). Our probing mechanism for edge queries makes use of the independent cascade probability $p$ to sample edges; therefore, in subsection \ref{sec:discrepancy}, we study what happens when the probe and seed cascade probabilities (denoted by $p'$ and $p$, respectively) are different. We point out the hardness of giving general guarantees when $p'\neq p$ and propose a post-processing solution to correct for this discrepancy as long as $p'>p$, i.e., the edge data are collected with sufficiently high probability. In Section \ref{sec:valueofnetinfo}, we use our bounded-query framework to resolve a trade-off between the cost of acquiring network information and its benefit in increasing expected spread size. We provide discussion and concluding remarks in Section \ref{sec:conc}. Detailed comparisons with related works are provided in Appendix \ref{app:lit}. Detailed proofs are presented in Appendix \ref{app:proof}. In Appendix \ref{app:ext-inf-models}, we discuss the extension of our results to other influence models including independent cascade on directed graphs (Appendix \ref{app:ext}) and the linear threshold model\revision{,} for which we provide approximation guarantees using only ${O}(n k^2\log{n})$ edge queries (Appendix \ref{app:ext:linth}).

\subsection{Main contributions}\label{sec:intro:main-conrib}

We consider the independent cascade (IC) model of social contagion that is fairly well-studied since its use by  \cite{kempe2003maximizing}. In this model, network edges are ``active'' with probability $p$ independently of each other and all nodes with active connections to other active nodes become active. Motivated by applications to product and technology adoption, we refer to active nodes as adopters. Starting from a set of initial adopters, the adoption propagates through the network and the process terminates after a finite number of steps. Following the independent cascade model, every adopter has a single chance to activate each of its neighbors independently with probability $p$. The $k$-influence maximization problem, or $k$-IM in short, refers to the choice of $k$ initial adopters to maximize expected adoptions under this diffusion model. Let ${\opt}$ be the optimum value for this problem. A $\mu$-approximation algorithm outputs a set of $k$ initial adopters to guarantee that the expected number of adoptions is at least $\mu {\opt}$. In this work, we assume a query oracle access to the network graph and study the $k$-\revision{IM} problem with a limited number of queries.

We begin with a hypothetical scenario assuming that we can pay a cost to seed a random node and learn the outcome of the spreading process (e.g., imagine distributing traceable coupons to random individuals and asking them to pass the coupons to their friends; or a social network marketing firm that measures its audience by seeding ads and promotional goods randomly). We only learn the identity of the final adopters and do not use any information about the network edges through which the influence spreads. We collect several independent cascade outcomes by repeating this process and refer to them as ``influence samples''. We use these influence samples to seed $k$ nodes with optimality guarantees. We first show that an additive loss (e.g., $\epsilon n$) is necessary, given $o(n)$ influence samples (Theorem \ref{thm:additiveloss}, Subsection \ref{sec:spreadingQ}): 
\begin{repeattheorem}[Hardness of approximation with $o(n)$ influence samples.]
Let \revision{$\mu>0$ be any} constant. There is no $\mu$-approximation algorithm for influence maximization using $o(n)$ influence samples.
\end{repeattheorem}
Interestingly, we show that ${O}(k^2\log n)$ influence samples are enough to provide a $k$-IM solution with almost tight approximation guarantees. For example, \revision{if finding} a single seed on a star ($k=1$), with high probability all random samples are leaves of the star. However, based on the ${O}(\log n)$ spread outcomes our algorithm finds and seeds the center of the star. We also show that the quadratic order dependence on $k$ is the best possible. The following is a formal summary of our results from Theorems \ref{thm:main_spreading_queries} and \ref{thm:main_spreading_queries_lowerbound} in Subsection \ref{sec:spreadingQ}:
\begin{repeattheorem}[Approximation guarantees with bounded number of influence samples.]
    For any arbitrary $0< \epsilon \leq 1$, there \revision{exists} a polynomial-time algorithm for $k$-influence maximization that covers $(1-1/ e) {\opt}-\epsilon n$ nodes in expectation using no more than  ${O}_{\epsilon}(k^2\log n)$ influence samples. Moreover, there can be no approximation algorithms that provide $\mu\opt - \epsilon n$ guarantees for $k$-IM using $o(k^2)$ influence samples for a fixed $0<\mu<1$ and $0<\epsilon < \mu/k$. 
\end{repeattheorem} 

Notice that our bound on the number of influence samples depends logarithmically on $n$, therefore, when $k$ is poly-logarithmic we only use poly-logarithmic number of influence samples which is exponentially lower than the best known bound of $O({k}{n}\log n)$ for sample complexity of influence maximization on general graphs \cite[Section 2]{sadeh_et_al:LIPIcs:2020:11714}. We point out that our order $n$ improvement is only possible because we allow for an additive loss in our approximation guarantee. Detailed comparisons with this and other related works are presented in Appendix \ref{app:lit}.

Our main contribution is to show that similar approximation guarantees are possible as we bound the total number of edge queries, i.e., queries of the form $(v,i)$ that return the $i$-th neighbor of node $v$ with arbitrarily ordered neighborhoods. We propose a probing procedure to sequentially reveal random neighborhoods of the nodes, resulting in a snowball\revision{-like} sampling of the network edges. Notice that \emph{a single} simulation of \revision{the} independent cascade \revision{model} over the entire network (without using our subsampling and stopping constraints) requires $\Omega(p n^2)$ edge queries. In fact, we show that in the worst case one needs to query $\Omega(n^2)$ edges to guarantee that the expected number of covered nodes is at least a constant fraction of the optimum (Theorem \ref{thm:edge_query_complexity-additive-loss-is-unavoidable}, Section \ref{sec:edgeQ}):

\begin{repeattheorem}[Hardness of approximation with  $o({n^2})$ edge queries.]
Let $\mu$ be any constant. There is no $\mu$-approximation algorithm for influence maximization using $o({n^2})$ edge queries.
\end{repeattheorem}

We avoid the above impossibility by allowing for an $\epsilon n$ additive loss in our approximation guarantee. Subsequently, one natural question that arises is to study the relation between the required number of queries and the cascade probability. In particular, is it possible to find an \emph{approximately optimal} seed set using sub-quadratic number of queries when $p$ is desirably small? We resolve this question positively by showing that our probing scheme approximately preserves the greedy solution to the $k$-IM problem, achieving a $(1-1/e) {\opt}-\epsilon n$ guarantee using no more than ${O}_{\epsilon}(p n^2\log^4n + \sqrt{k p} n^{1.5}\log^{5.5}n + k n\log^{3.5}{n})$ edge queries. We also provide a matching lower bound (up to logarithms) to show that the linear order dependence on $p$ is tight. The following is a formal restatement of our results in Theorems \ref{thm:edge_query_complexity} and \ref{thm:main} of Subsection \ref{sec:sampling_alg}. 

\begin{repeattheorem}[Approximation guarantees with bounded number of edge queries.]
    For any arbitrary $\epsilon > 0$, there \revision{exists} a polynomial-time algorithm for influence maximization that covers $(1-1/ e) {\opt}-\epsilon n$ nodes in expectation, using ${O}_{\epsilon}(p n^2\log^4n + \sqrt{k p} n^{1.5}\log^{5.5}n + k n\log^{3.5}{n})$ queries, where ${\opt}$ is the expected number of nodes covered by the optimum solution to $k$-IM. Moreover, there can be no approximation algorithms that provide $\mu\opt - \epsilon n$ guarantees for $k$-IM using $o(pn^2)$ edge queries for a fixed $0<\mu\leq 1$ and $\epsilon < \mu^2/18$.
\end{repeattheorem}

To achieve this result, we apply some subsampling techniques with stopping constraints that enable us to \emph {approximately simulate} $O_{\epsilon}(k\log{n})$ independent cascades, starting from a random sample of $O_{\epsilon}(k\log{n})$ initial nodes and using only ${O}_{\epsilon}(p {n^2}\log^{4}{n} + \sqrt{k p} n^{1.5}\log^{5.5}{n} + k n\log^{3.5}{n})$ edge queries. We specify the dependence of the ${O}_{\epsilon}$ terms on $\epsilon$ when presenting our main results in Section \ref{sec:edgeQ}. \revision{Of note, our subsampling technique makes critical use of the independent cascade probability $p$ when deciding how many edges to query in the neighborhood of each node. In practice, the true value of $p$ is often subject to significant uncertainty. We address the dependency of our edge queries on $p$ with a hardness result in Section \ref{sec:discrepancy} and discuss how potential discrepancies may be corrected if $p$ is unknown at data collection but measurable afterwards.}

The most closely related result that provides approximation guarantees for $k$-IM with limited queries to an unknown graph is due to \cite{wilder2018maximizing}, who propose an algorithm for input graphs that are drawn from a particular family of stochastic block models. Their algorithm, which is tailored to that specific random graph model, consists of taking a random sample of $T$ nodes and exploring their extended neighborhoods in $R$ steps of a random walk. The outcome of the random walks is used to estimate the block sizes of each of the $T$ nodes, and this is achieved by revealing no more than $T  R \in O(\log^6 n)$ nodes. The $k$ nodes in the seed set are selected from the initial $T$ samples, such that the $k$ largest blocks are seeded uniformly at random. Unlike \cite{wilder2018maximizing}, we do not make any assumptions about inputs, so our results are applicable to general graphs. In the following subsection, we put our contributions in perspective by discussing related bodies of literature. Detailed discussions of  methodologically relevant work are provided in Appendix \ref{app:lit}. Our main results in Section \ref{sec:edgeQ} are presented for undirected graphs. In Appendix \ref{app:ext}, we provide the extension of the edge query model to directed graphs and show that the same approximation guarantees and query bounds hold true when nodes are queried for their influencers (their incoming edges).

\subsection{Related work}\label{sec:intro:related-lit}

Motivated by the difficulties of acquiring complete network data, we are interested in methods for targeting in networks without making explicit use of the full graph. Such methods have roots in multiple applied problems --- vaccination \citep{cohen2003efficient} and disease surveillance \citep{christakis2010social} --- in addition to seeding. One approach that has received substantial attention is a ``one-hop'' strategy [sometimes called ``nomination'' \citep{kim2015social} or ``acquaintance targeting'' \citep{cohen2003efficient,chami2017social}] that selects as seeds the neighbors of random nodes. This approach exploits a version of the friendship paradox that states: ``the friend of a random individual is expected to have more friends than a random individual,'' \citep{Lattanzi:2015:PRN:2684822.2685293,feld1991your}. For example, \cite{kim2015social} report on the results of field experiments that target individuals for delivery of public health interventions (spreading adoption of multivitamins and a water purification method). For one product, they argue that one-hop targeting (whereby a random individual nominates a friend to be targeted) leads to increased adoption rates, compared with random or in-degree targeting. Some other empirical work has been less encouraging (\citealp{chin2021evaluating}, cf. \citealp{kumar2019can}). While there are results about how these short random walks affect the degree distribution of selected nodes \citep{kumar2018network}, one-hop seeding currently lacks any theoretical guarantees under models of contagion. Furthermore, given the collection of data about the network neighborhoods of $k$ nodes, it is natural to ask whether this data can be more effectively used than just locally taking a random step, ignoring data collected from the other $k-1$ neighborhoods.

To address the challenges of seeding when obtaining network information is costly, we offer a framework for influence maximization using a bounded number of queries to the graph structure. In this framework, we investigate the expected spread size versus the increasing number of queries as we obtain more information about the network. In related work, \cite{akbarpour2020just} study the value of network information for seeding interventions.  We provide a detailed comparison with this work in Section \ref{sec:valueofnetinfo}, after clarifying our modeling assumptions and results.

In another related work, \cite{manshadi2020diffusion} study a model of spread where individuals contact their neighbors independently at random, and each contact leads to an adoption with some fixed probability. The contacts occur repeatedly; therefore, every cascade eventually spreads to the entire population. They characterize the time to reach a fraction of adopters as well as the contact cost (number of contacts made), in a random graph with a given degree distribution. They also propose optimal seeding strategies that only use the degree information. However, this model is not directly comparable to the influence maximization setup that we study. In our model, the realization of the influences is random and adoption spreads only through the realized edges. For us, the objective is to maximize the expected spread size and the incurred cost is in acquiring information about the influence structure (who influences whom).

Particularly relevant to our present study is recent work on influence maximization for unknown graphs \citep{mihara2015influence,mihara2017effectiveness,stein2017heuristic,wilder2017influence,wilder2018maximizing}. \cite{mihara2015influence} use a biased snowball sampling strategy to greedily probe and seed nodes with the highest degree; they later propose to improve their heuristic by including random jumps that avoid excessive local search in their snowball sampling strategy \citep{mihara2017effectiveness}. \cite{stein2017heuristic} explore applications of common heuristics and known algorithms in scenarios where parts of the network is completely unobservable. Although simulations of influence spread on synthetic and real social networks provide some evidence, none of these results come with provable performance guarantees in general graphs. To the best of our knowledge, the only available guarantee for influence maximization with unknown graphs is due to \cite{wilder2018maximizing}. However, as discussed above (section \ref{sec:intro:main-conrib}), this algorithm and analysis is tailored to graphs generated from a particular family of stochastic block models (roughly speaking, they use the outcome of the queries to estimate the size of each block and choose nodes to seed the largest blocks). Such an analysis does not apply to general graphs and the techniques that we use to provide performance guarantees for general graphs are significantly different. 

We rely on sketching techniques to summarize influence
functions using a bounded number of queries; in Subsection \ref{sec:bounding_probed_neoghborhoods}, we adopt high-level ideas from Bateni et al. \citeyearpar{Bateni:2017:AOS:3087556.3087585,Bateni:2018:ODS:3219819.3220081} for construction of our sketch (see Lemma \ref{lem:Limiting_neighborhood}). \cite{cohen2014sketch} also develop a sketch-based algorithm for influence maximization to bound the running time with approximation guarantees. In other related work, \cite{borgs2014maximizing} give a quasi-linear time algorithm for influence maximization based on reversed influence samples that use $O(k(n + m)\log{n})$ edge queries, where $m$ is the size of the input edge set. Although these algorithms achieve fast (nearly best possible) influence maximization, they may query all edges multiple times on some inputs because they are not directly concerned with limiting query access to unknown graphs. In Appendix \ref{app:lit}, we give a detailed comparison with these and other relevant works that are based on (reverse) influence sampling.

Our influence maximization guarantees also relate to the recent developments in stochastic submodular maximization \citep{karimi2017stochastic}, as well as optimization from samples \citep{CommunitiesLearningtoInfluence,balkanski2016power}. The key difference is that in our algorithms we make explicit use of the combinatorial structure of the collected data. This is in contrast to the optimization from samples framework, where only the sampled values of the submodular function are observed. Consequently, we are able \revision{to provide} guarantees for arbitrary inputs, and avoid some of the limitation\revision{s} of the optimization from samples \citep[cf.][]{balkanski2017limitations}. We provide more details about our relationship with this literature in Appendix \ref{app:lit}. 

Some prior work has addressed lack of perfect knowledge
about the spreading process by learning the influence model and potentially heterogeneous probabilities of spreading along \revision{each edge} \citep{goyal2010learning,gomez2012inferring}. However, rather than attempting to learn the model parameters from existing data, we are interested in data collection as an active, costly process that is performed to inform seeding interventions. To this end, we offer a methodology that coordinates data collection and influence maximization by limiting queries to the social network graph. In other online learning and bandit-based approaches, the learner can select different seed sets at each stage and receives feedback from seeding in previous stages \citep{SemibanditFeedback,wu2019factorization}. This is also related to adaptive seeding, where the initial choice of seeds influences what becomes available for seeding in a followup stage \citep{seeman2013adaptive,horel2015scalable,feng2020neighborhood}. The ability to seed nodes adaptively makes such setups incomparable to ours --- and inapplicable when practical considerations demand commencing seeding simultaneously.

\section{Problem setup and preliminary results}\label{sec:notation_set_up}

Consider \revision{a} graph $\mathcal{G} = (\mathcal{V},\mathcal{E})$ with the set of nodes $\mathcal{V}$, the set of edges $\mathcal{E}$ and a seed set $\mathcal{S} \subseteq  \mathcal{V}$. Starting from the seeded nodes in $\mathcal{S}$, adoption spreads along the edges of $\mathcal{E}$ with independent cascade probability $p$ according to the IC model in Section \ref{sec:intro:main-conrib}. Given $\mathcal{S}$, for $v \in \mathcal{V}$, let $\phi(v,\mathcal{S})$ be the probability that $v$ adopts when the nodes in $\mathcal{S}$ are seeded. The influence function, $\Gamma$, maps each seed set, $\mathcal{S}$, to its value, $\Gamma(\mathcal{S}) = \sum_{v\in\mathcal{V}}\phi(v,\mathcal{S})$, which is the expected number of nodes that adopt if the nodes in set $\mathcal{S}$ are seeded.

\begin{definition}[$k$-IM]\label{def:kIM} Given graph $\mathcal{G}$, the $k$-influence maximization ($k$-IM) problem is to choose a seed set $\mathcal{S}\subset \mathcal{V}$ with $\mbox{\emph{card}}(\mathcal{S}) =k$ to maximize $\Gamma(\mathcal{S})$. We use $\Lambda = \argmax_{\mathcal{S},\mbox{\emph{card}}(\mathcal{S}) =k}\Gamma(\mathcal{S})$ to denote any such solution and use ${\opt} = \Gamma(\Lambda)$ to denote the optimal value.
\end{definition}

\begin{definition}[Approximations]\label{def:alphaAPPROX} Given graph $\mathcal{G}$, any $\Lambda^{\alpha}\subset\mathcal{V}$, $\mbox{\emph{card}}(\Lambda^{\alpha}) = k$, satisfying $\Gamma(\Lambda^{\alpha}) \geq \alpha {\opt}$ is an $\alpha$-approximate solution to $k$-IM.
\end{definition}

An important result in influence maximization is that $\Gamma$ is a non-negative, monotone, submodular set function \citep{kempe2003maximizing,kempe2005influential,kempe2015maximizing,doi:10.1137/080714452}. Subsequently, it can be approximately maximized by sequentially selecting $k$ seeds with the largest marginal gains, i.e., the greedy algorithm, which makes $O(nk)$ oracle calls to  $\Gamma$ and achieves a $1-(1-1/k)^k \geq (1-1/e)$ approximation guarantee \citep{nemhauser1978analysis}. The greedy algorithm sets a gold standard for influence maximization that is NP-hard to improve upon --- indeed, $k$-IM generalizes the maximum coverage problem and suffers its hardness of approximation beyond a $1-(1-1/k)^k$ factor. Here, we achieve \emph{roughly} the same guarantee without oracle access to $\Gamma$ and using only a limited number of queries to the graph $\mathcal{G}$. In our approach, rather than optimizing the influence function on the original graph, we do so on a subgraph that is properly sampled from the original graph. As its main property, we show that for the appropriate choice of $\alpha$ and $\epsilon$, an $\alpha$-approximate solution to $k$-IM on this subgraph has an influence on the original graph that is lower-bounded by  $\alpha {\opt} - \epsilon n$. We can thus achieve similar worst-case guarantees using only partial information about the network.

Accessing the input graph by performing edge queries is a common technique in sublinear time algorithms that inspect only a small portion of their input before providing an output \citep{alon2009combinatorial,alon2000efficient,chazelle2005approximating,esfandiari2018metric,DBLP:conf/stoc/Indyk99}. Formally, we assume that the input graphs are represented by an adjacency list defined as a collection of lists, $\{(\mathcal{N}_{\nu},\mbox{card}(\mathcal{N}_{\nu})),\nu\in\mathcal{V}\}$, where each list, $\mathcal{N}_{\nu}$, consists of all the neighbors of node $\nu$ in some arbitrary (but fixed) order, and is accompanied by its length. Our query oracle model is defined such that given a vertex $\nu$ and an index $1\leq i \leq \mbox{card}(\mathcal{N}_{\nu})$, the algorithm can query who is the $i$-th neighbor of $\nu$ \citep{gonen2011counting}:

\begin{definition}[Edge Query]\label{def:edgequery} Given a vertex $\nu\in\mathcal{V}$ and an index $i \in \{1,\dots, \mbox{card}(\mathcal{N}_{\nu})\}$, an edge query with parameters $(\nu,i)$ reveals the $i$-th neighbor of $\nu$.
\end{definition}

We use edge queries as part of a probing mechanism, whereby a node is asked to reveal her neighbors each with probability $p$. \revision{Formally, probing node $\nu$ is defined as follows: }
\revision{\begin{definition}[Probing]\label{def:probing} Given a vertex $\nu\in\mathcal{V}$, a probing with parameters $(\nu,p)$ performs a sequence of $(\nu,i)$ edge queries for every index $i$ that remains after eliminating the elements of the index set $\{1,\ldots,\mbox{card}(\mathcal{N}_{\nu})\}$, independently at random, with probability $1-p$. 
\end{definition}}

\revision{Of note, the total number of edge queries that are performed as a result of a $(\nu,p)$ probing is a binomial random variable with size parameter $\mbox{card}(\mathcal{N}_{\nu})$ and success probability $p$.} In addition to probing nodes and running edge queries, our algorithm also needs a subsample of initial nodes that are chosen at random without replacement from the node set. Given this initial sample, we repeatedly probe the extended neighborhoods of the initial sample using edge queries. In our analysis, we bound the total number of edge queries that our algorithm makes in order to achieve the desired approximation guarantees. We also provide a query complexity lower-bound to show that the probing algorithm is order optimal for achieving the desired approximation guarantees, with as few queries as possible. 

Although accessing the input graph locally by revealing the ordered neighborhoods of its nodes is a common query oracle model, our probing setup is also motivated by practical methods of network sampling such as snowball sampling, link-tracing, and respondent-driven sampling (RDS) that are popular in public health surveillance, social policy research, sociology, and survey design applications \citep{heckathorn2017network}. The principle utility of such methods is in constructing samples of hidden (hard to reach) populations (e.g., when estimating prevalence of HIV among drug injectors). In these situations, research begins with a convenience sample of initial subjects which is then expanded by tracing their network links in waves, until the target sample size is attained (\citealp{heckathorn2017network,salganik2004sampling}, cf. \citealp{goel2010assessing}). Following our probing procedure, researchers can decide which links to trace in the neighborhood of a probed node, randomly by simulating independent biased coin flips with head probability $p$. More broadly, our PROBE algorithm (Algorithm~\ref{alg:probe}, Section~\ref{sec:edgeQ}) can be integrated with social network data collection software --- e.g., the Trellis mobile platform \citep{LUNGEANU2021171} used in \citet{kim2015social} and other studies --- to generate survey sampling plans for researchers in the field.

\subsection{Approximation guarantee with a bounded number of influence samples}\label{sec:spreadingQ}

To demonstrate the challenges of seeding with partial network information, we present a thought experiment whereby one can pay a cost to learn the outcome of a spreading process when a single node is seeded. Imagine giving out coupons (or lottery tickets) to random individuals and observing their usage spread as a way of collecting network information. Formally, we define an ``influence sample'' as the outcome of seeding a random node:

\begin{definition}[Influence Sample]\label{def:influencesample} Each influence sample consists of all nodes that become active, after a single node is chosen uniformly at random (with replacement) and seeded.
\end{definition}

\begin{figure}[t]
\centering
\includegraphics[scale=0.5]{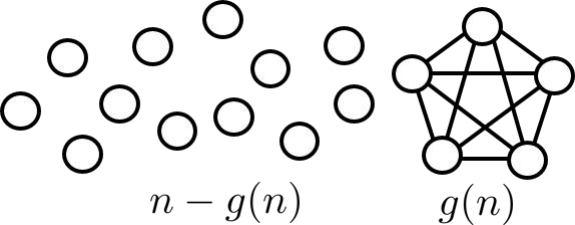}
\caption{Using $f(n)\in o(n)$ influence samples on a graph comprised of a clique of size $g(n)=\sqrt{{n}/{f(n)}}$ and $n - g(n)$ isolated nodes, one cannot achieve an approximation factor that is better than $o(1)$.}
\label{fig:f_n_g_n}
\end{figure}

As a theoretical exercise, we ask how many influence samples we need to collect to be able to provide a $k$-IM approximate solution. We first show that one cannot hope to provide a constant factor approximation guarantee, $\mu \opt$, for any $\mu>0$, using $o(n)$ influence samples. Our hard example consists of a graph with a small clique and many isolated nodes (Figure \ref{fig:f_n_g_n}). In such a structure, using $o(n)$ influence samples one is unlikely to observe the small clique and cannot achieve better than an $o(1)$ approximation factor. This example is similar to one in \citet[Theorem 1]{wilder2018maximizing}, but we improve their $O(n^{1-\epsilon})$ lower bound to $o(n)$. The proof details are in Appendix \ref{app:proof:thm:additiveloss}. 

\begin{theorem}\label{thm:additiveloss}
Let $0< \mu < 1$ be any constant. There is no $\mu$-approximation algorithm for influence maximization using $o(n)$ influence samples.
\end{theorem} 

Knowing that a multiplicative approximation guarantee is impossible with $o(n)$ influence samples, we next ask how many influence samples we need for providing a  $(1-1/e){\opt}-\epsilon n$ guarantee with fixed $\epsilon>0$. Algorithm~\ref{alg:sample} provides such a guarantee using $k\rho \in {O}_{\epsilon}(k^2\log(nk))$ influence samples, where $\rho = \rho_{\epsilon}^{n,k} = \color{green} \lceil \normalcolor  (81 k/\epsilon^3)\log(6nk/\epsilon) \color{green} \rceil \normalcolor $ is the number of influence samples we collect to choose one seed. The total number of influence samples that we use in Algorithm~\ref{alg:sample} is 
\begin{align}
T_{\epsilon}^{n,k} = k \rho_{\epsilon}^{n,k} = k \color{green} \left\lceil \normalcolor  \frac{81 k \log ({6nk}/{\epsilon})}{\epsilon^3} \color{green} \right\rceil \normalcolor \in O \color{green}\left(\normalcolor\frac{k^2 \log{n}}{\epsilon^3} + \frac{k^2 \log (1/\epsilon)}{\epsilon^3}  \color{green}\right)\normalcolor \cdot \label{eq:influencesamplesbound}
\end{align}

\begin{algorithm}[thb]
    \SetAlgoLined
    \DontPrintSemicolon
    \vspace{5pt}
    \KwIn{Influence sampling access to graph $\mathcal{G}$, sample size $\rho$, and seed set size $k$}
    \KwOut{$\Lambda^{\star}$, approximate seed set of size $k$ with value at least $(1-1/e)\opt-\epsilon n$}
    Initialize $\Lambda^{\star} \leftarrow \varnothing$.\;
    \For{$i$ \KwFrom $1$ \KwTo $k$}{
        Collect $\rho$ influence samples and call them $A^i_1$, $\ldots$, $A^i_{\rho}$. \;
        \tcp{Discard the influence samples that intersect with already chosen seeds ($\Lambda^{\star}$):}
        \For{$j$ \KwFrom $1$ \KwTo $\rho$}{
            \If{$A^i_{j} \cap \Lambda^{\star}\neq \varnothing$}{
                    $A^i_{j} \leftarrow \varnothing$.\;
                    }
        }
        \tcp{Choose the $i$-th seed based on the remaining influence samples:}
        \For{$u\in \mathcal{V}\setminus\Lambda^{\star}$}{
            \For{$j$ \KwFrom $1$ \KwTo $\rho$}{
                $X^{i}_{u,j} \leftarrow \mathds{1}\{u \in A^i_j\}$.\;
            }
            $X^i_u \leftarrow \sum_{j=1}^{\rho}X^{i}_{u,j}$.\;
        }
        $v^{\star} \leftarrow \argmax_{u \in \mathcal{V}\setminus{\Lambda^{\star}}} X^i_u$.\;
        ${\Lambda^{\star}} \leftarrow {\Lambda^{\star}} \cup \{v^{\star}\}$.\;
        }
    \Return{$\Lambda^{\star}\cdot$} \;
\caption{INF-SAMPLE$(\rho,k)$}
\label{alg:sample}
\end{algorithm}  

\begin{figure}[thb]
\centering
\begin{subfigure}[b]{0.32\textwidth}
\includegraphics[width=\textwidth]{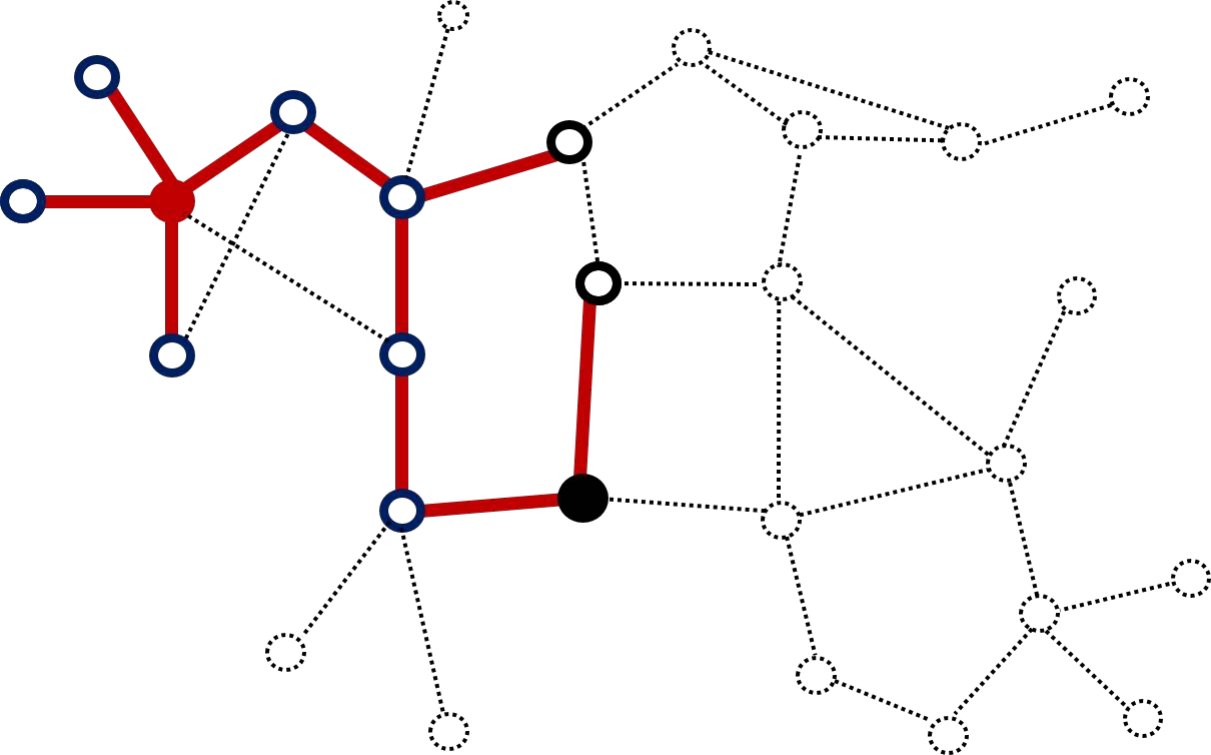}
\caption{ }
\label{fig:spread_queries_red}
\end{subfigure}~\begin{subfigure}[b]{0.32\textwidth}
\includegraphics[width=\textwidth]{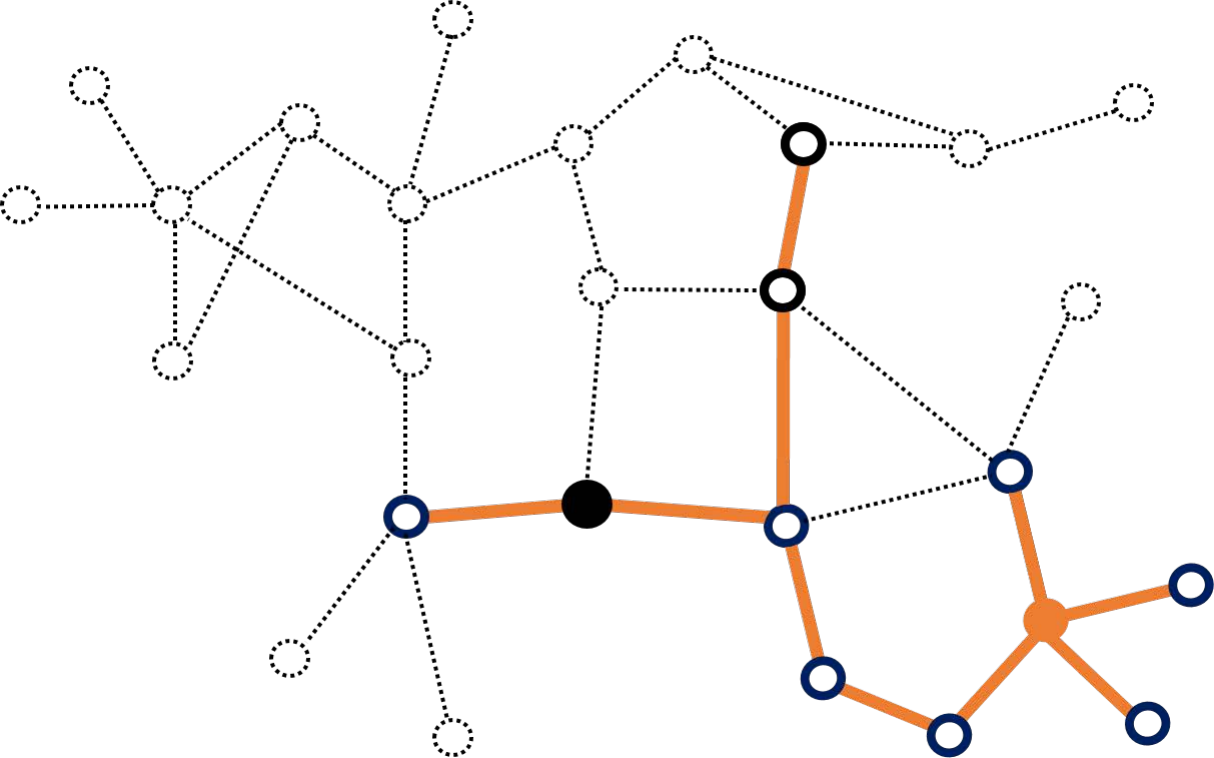}
\caption{ }
\label{fig:spread_queries_orange}
\end{subfigure}~\begin{subfigure}[b]{0.32\textwidth}
\includegraphics[width=\textwidth]{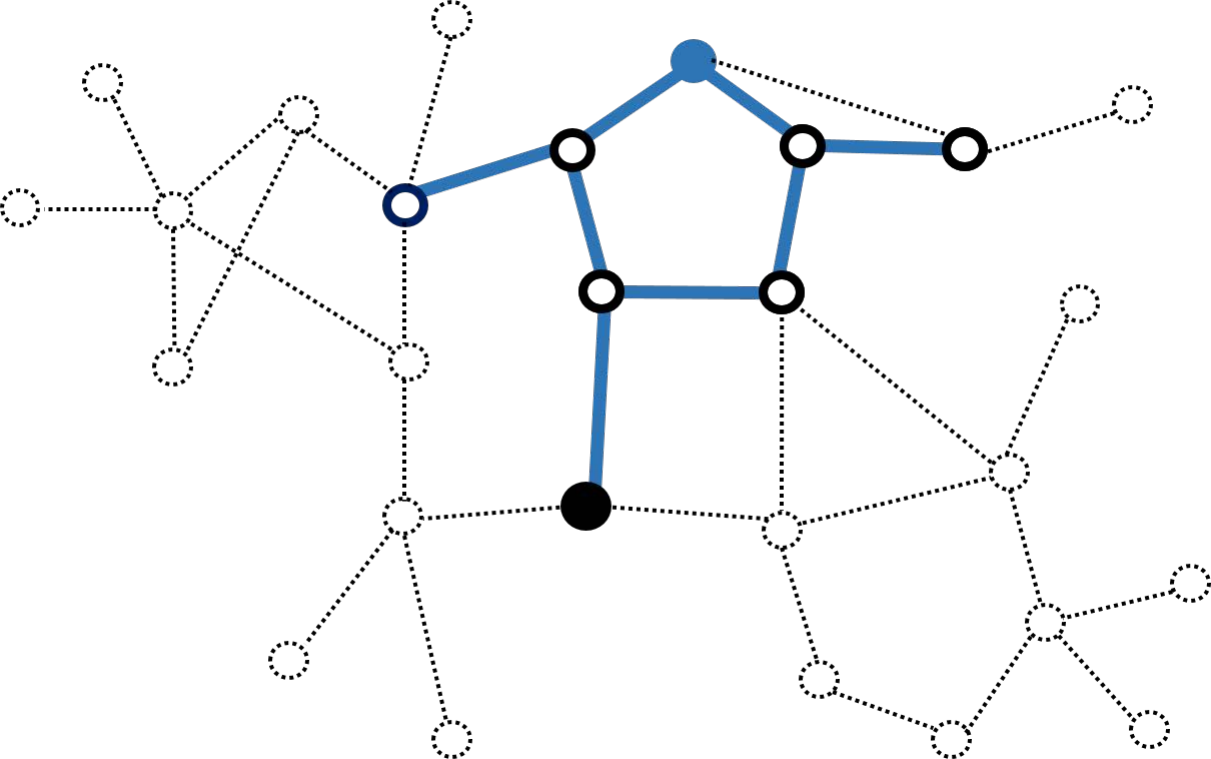}
\caption{ }
\label{fig:spread_queries_blue}
\end{subfigure}
\caption{Three influence samples are depicted in (\subref{fig:spread_queries_red}) red, (\subref{fig:spread_queries_orange}) orange, and (\subref{fig:spread_queries_blue}) blue. In each influence sample, the random initial node is marked in the same color as the cascade. The node that appears most across different samples is marked in black. The dotted segments are not observed in the samples.}
\label{fig:spread_queries}
\end{figure}

The following theorem formalizes our guarantees for Algorithm~\ref{alg:sample}. Its proof is in Appendix \ref{app:proof:them:main_spreading_queries}. The main idea is that nodes that appear in many influence samples are good candidates for seeding since they are reached by many random nodes. For example, in Figure~\ref{fig:spread_queries} the black node is the only node that appears in all influence samples and is the best candidate for seeding. To prevent overlap with the previously chosen seeds, at each step we discard those influence samples that contain any of the already chosen seeds (belonging to $\Lambda^{\star}$). The crux of the argument is in realizing that $(n/\rho) X^i_{u}$ is an unbiased estimator of the expected marginal gain from adding $u$ to the seed set. By controlling the deviation of $X^i_{u}$ from $\mathbb{E}[X^i_{u}]$, we can approximate every step of the greedy algorithm by choosing $u$ from $\mathcal{V}\setminus{\Lambda^{\star}}$ to maximize $X^i_{u}$.

\begin{theorem}\label{thm:main_spreading_queries}
    For any arbitrary $0< \epsilon \leq 1$, there \revision{exists} a polynomial-time algorithm for influence maximization that covers $(1-1/ e) \opt-\epsilon n$ nodes in expectation in ${O}(k^2\log({n}/{\epsilon})/{\epsilon^3})$ time, using no more than  $ k \color{green} \left\lceil \normalcolor 81 k \log( {6nk}/{\epsilon})/{\epsilon^3} \color{green} \right\rceil \normalcolor \in {O}(k^2\log({n}/{\epsilon})/{\epsilon^3})$ influence samples. 
\end{theorem}

\begin{figure}[t]
\centering
\includegraphics[scale=0.5]{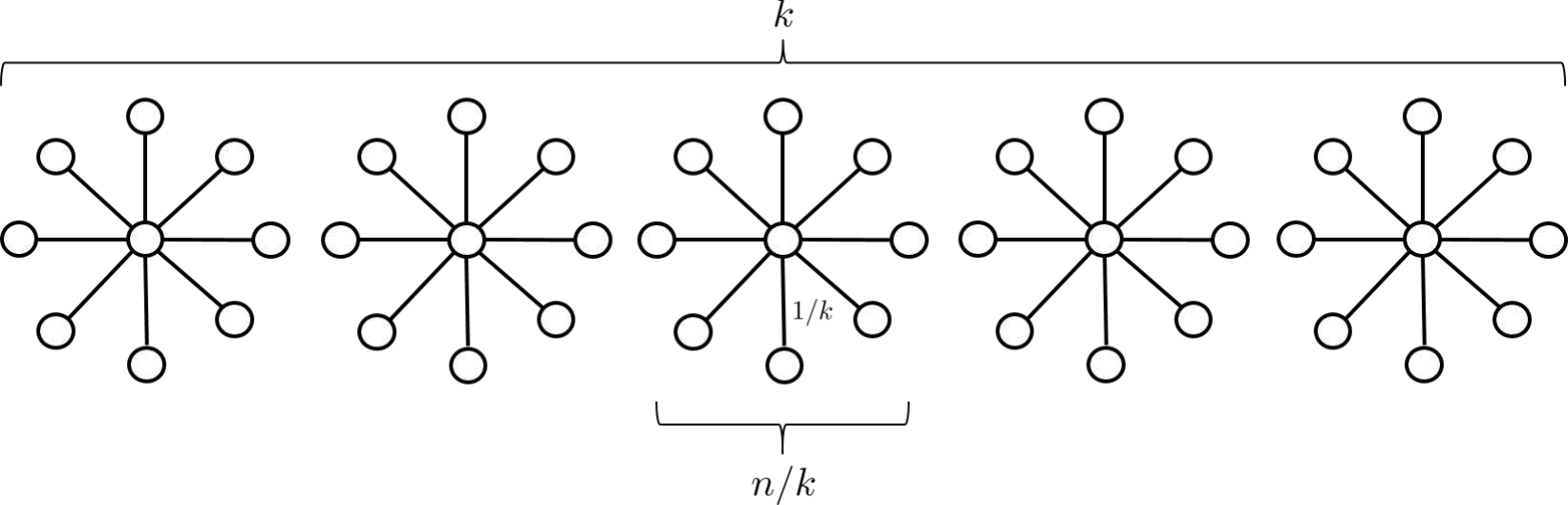}
\caption{Using $o(k^2)$ influence samples on a graph comprised of $k$ stars of size $n/k$ each, with cascade probability $p=1/k$, one cannot achieve a $\mu\opt - \epsilon n$  approximation guarantee for $\epsilon < \mu/k$.}
\label{fig:k_stars_hard_example}
\end{figure}

We end this section by a lower bound on the required number of influence samples for achieving the $(1-1/e) \opt-\epsilon n$ approximation guarantee. In particular, we show that the $k^2$ asymptotic rate for $T_{\epsilon}^{n,k}$ in \eqref{eq:influencesamplesbound} is optimal  for $\epsilon < \mu/k$. Our hard example consists of a collection of $k$ stars of size $n/k$ each, with independent cascade probability $p  = 1/k$; see Figure \ref{fig:k_stars_hard_example}. The optimum achieves $k (n/k^2) = n/k$ expected spread size by seeding the centers of each of the $k$  stars. In Appendix \ref{app:proof:thm:main_spreading_queries_lowerbound}, we show that any algorithm that uses $o(k^2)$ influence samples can at most discover a small expected fraction of the center nodes, and therefore, fails to guarantee a $\mu\opt - \epsilon n$ expected spread size when $\epsilon < \mu/k$.

\begin{theorem}\label{thm:main_spreading_queries_lowerbound}
     Fix $0<\mu<1$ and $0<\epsilon < \mu/k$. There can be no approximation algorithms that provide $\mu\opt - \epsilon n$ guarantees for $k$-IM using $o(k^2)$ influence samples. 
\end{theorem}

While these results characterize the challenges of seeding with limited network information, influence samples are often not a practical query model in a number of settings of interest and do not readily extend to directed graphs (Appendix Appendix \ref{app:ext}). Thus, we turn to another, more widely-applicable query method.

\section{Approximation guarantees with bounded edge queries}\label{sec:edgeQ}

\begin{figure}[t]
\centering
\begin{subfigure}[b]{0.32\textwidth}
\includegraphics[width=\textwidth]{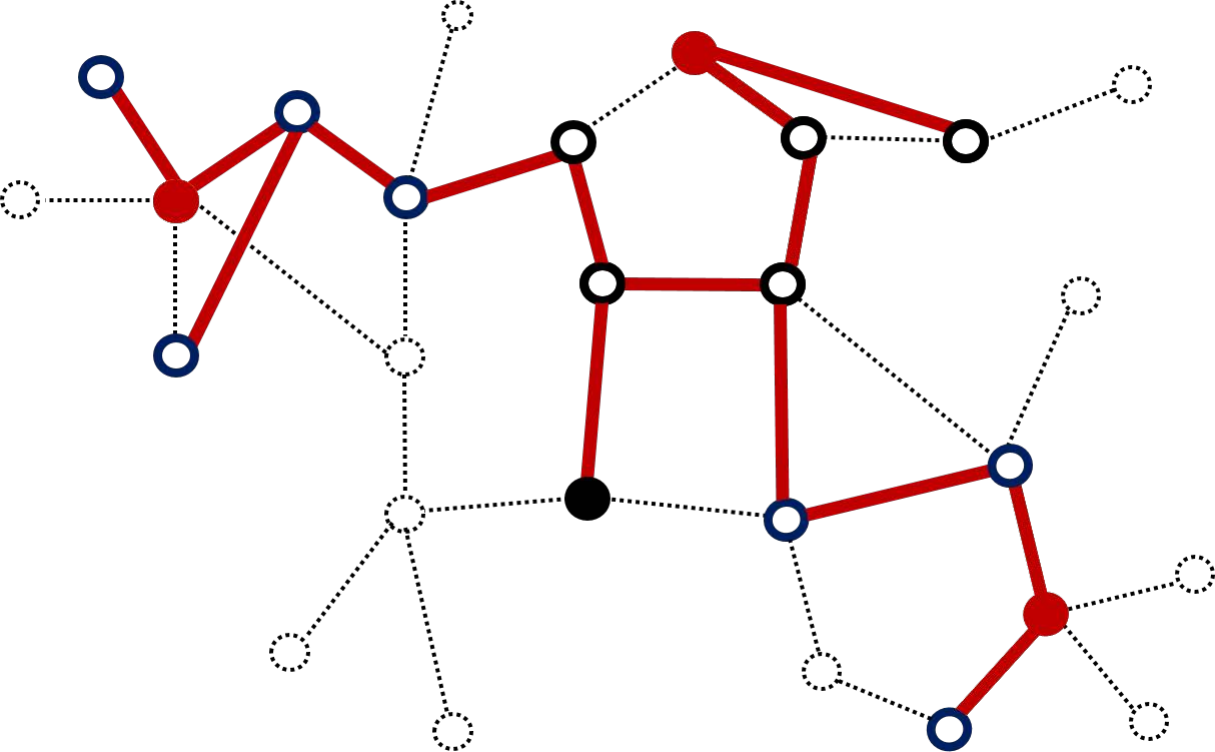}
\caption{ }
\label{fig:edge_query_red}
\end{subfigure}~\begin{subfigure}[b]{0.32\textwidth}
\includegraphics[width=\textwidth]{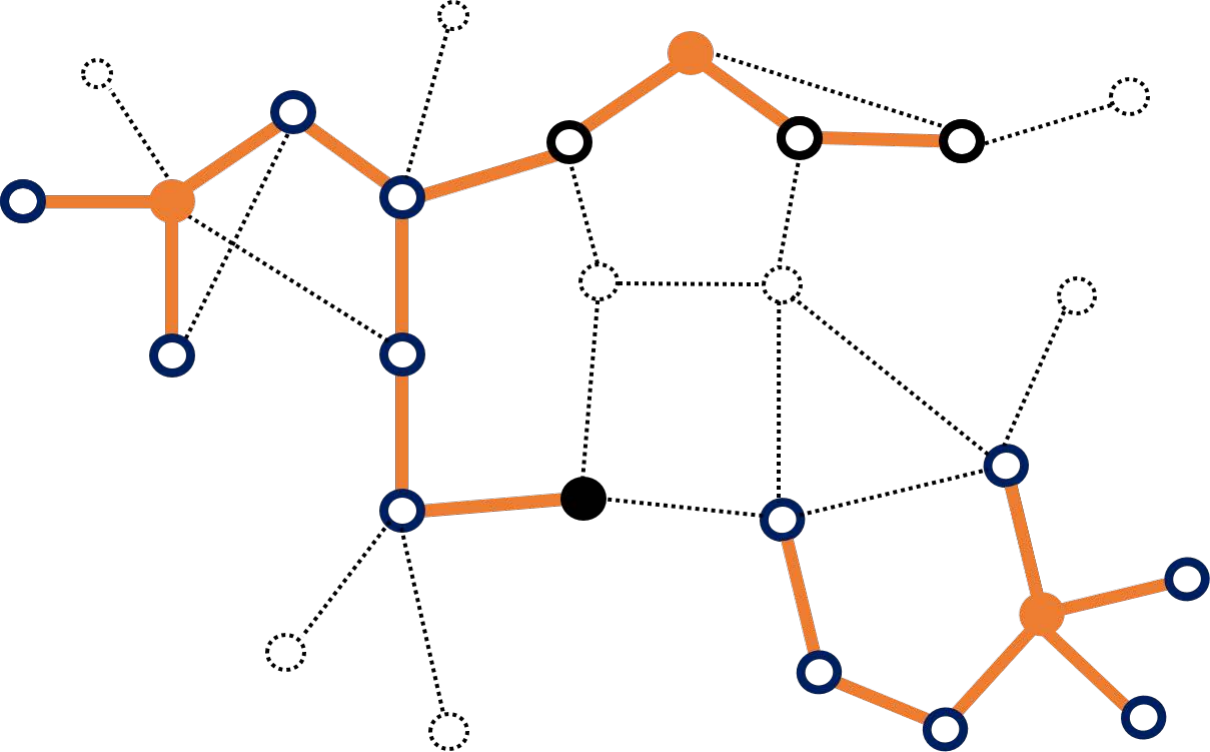}
\caption{ }
\label{fig:edge_query_orange}
\end{subfigure}~\begin{subfigure}[b]{0.32\textwidth}
\includegraphics[width=\textwidth]{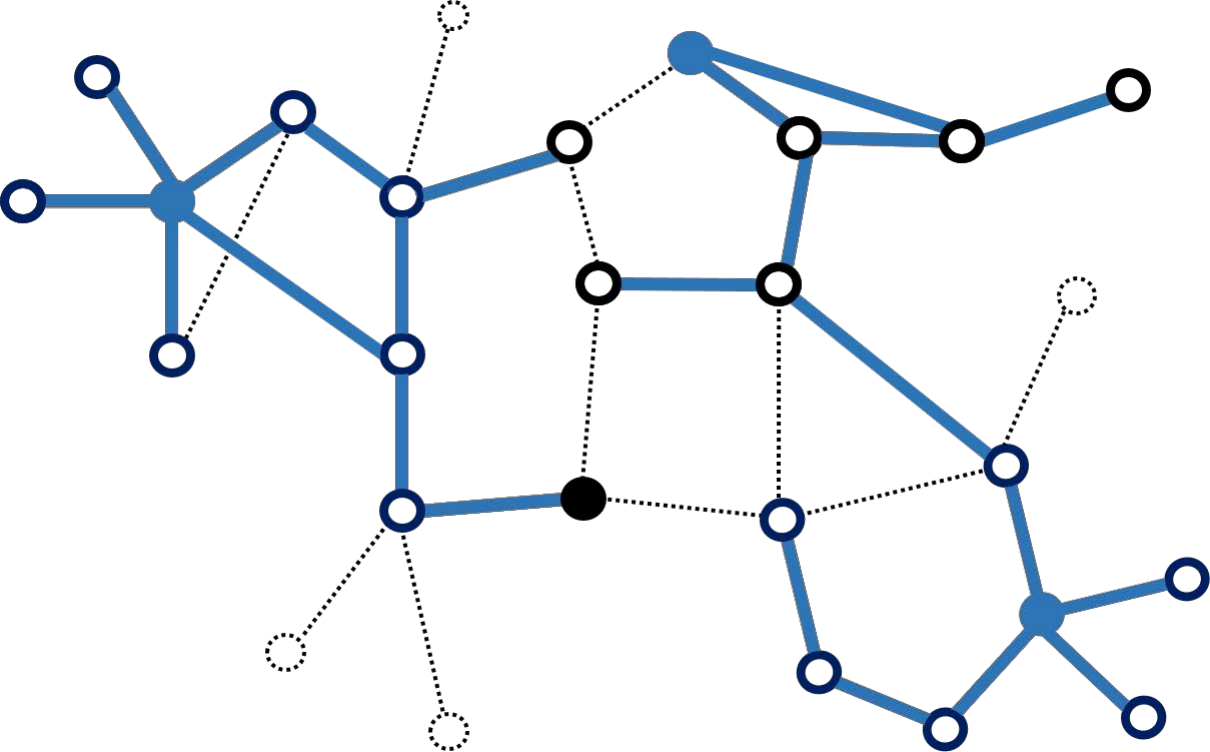}
\caption{ }
\label{fig:edge_query_blue}
\end{subfigure}
\caption{Three example cascades obtained through edge queries, depicted in (\subref{fig:edge_query_red}) red, (\subref{fig:edge_query_orange}) orange, and (\subref{fig:edge_query_blue}) blue. All cascades start from the same random initial nodes which are marked in the same color as the cascades. The node that is marked in black scores as high as or higher than other nodes across the three cascades. The dotted sections consist of unsampled edges and nodes.}
\label{fig:edge_queries}
\end{figure}

In this section, we present an algorithm to perform edge queries by probing the extended neighborhood of a random subsample of the network nodes (Algorithm~\ref{alg:probe}: PROBE), as well as an algorithm to output an approximate seed set, given the outcome of the edge queries (Algorithm~\ref{alg:seed}: SEED). The main idea of the PROBE algorithm is to simulate multiple independent cascades starting from a set of random initial nodes; by querying each node about its neighbors and repeating the same for the revealed neighbors, neighbors of neighbors, etc. The SEED algorithm takes in the output of the PROBE algorithm and chooses seeds that are connected to \revision{the most initial nodes} along the queried edges. In Figure~\ref{fig:edge_queries}, we depict an example of three cascades that are obtained through edge queries. Consider the node that is marked in black. Its queried connected component has three initial nodes in \ref{fig:edge_queries}(\subref{fig:edge_query_red}), two initial nodes in \ref{fig:edge_queries}(\subref{fig:edge_query_orange}), and one initial node in \ref{fig:edge_queries}(\subref{fig:edge_query_blue}). The value of each connected component is the number of initial nodes in that component and adding them gives the total value of the marked node: $1+2+3=6$. To prevent overlap with already chosen seeds, we set the value of a connected component to zero once one of its nodes is seeded. Using these valuations, the SEED algorithm approximates the greedy algorithm by sequentially adding the most valuable candidates to the seed set and updating the value of the connected components.

\begin{figure}[t]
\centering
\includegraphics[scale=0.5]{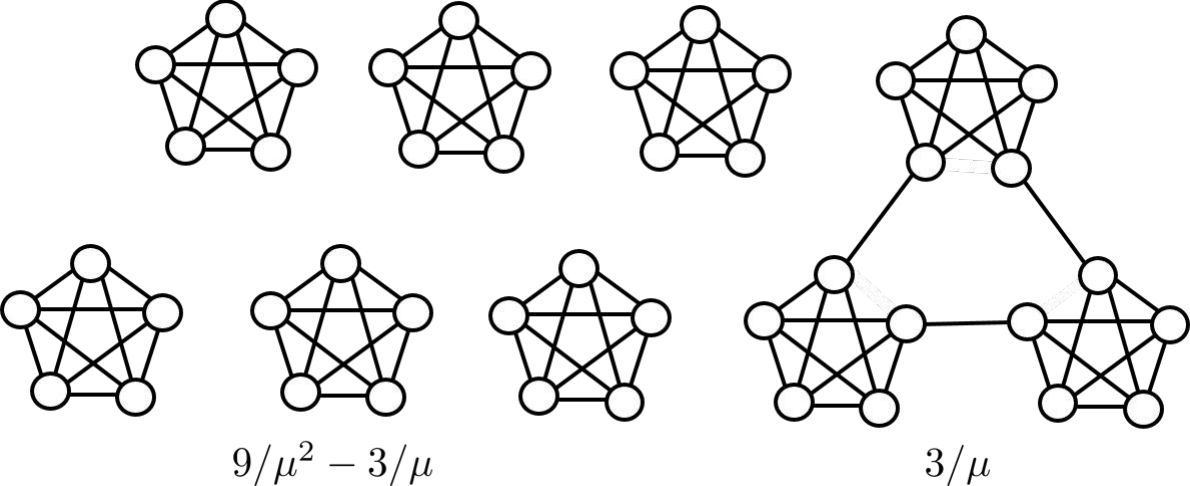}
\caption{To lower-bound the required number of edge queries, our hard example consists of $9/\mu^2$ cliques, $3/\mu$ of which are connected in a circle. An algorithm that makes $o(n^2)$ edge queries may detect the connected cliques with probability at most $\mu/3$. The expected performance of any such algorithm is worse than a factor $\mu$ of the optimum.}
\label{fig:mu_n}
\end{figure}

Our analysis consists of a $(1-1/e)\opt - \epsilon n$ lower bound on the expected spread size of the chosen seed set, as well as upper bounds on the total number of edges that are queried by the PROBE algorithm and the subsequent run-time of the SEED algorithm. Before digging any further into the analytical details, let us provide a companion hardness result that parallels Theorem \ref{thm:additiveloss} for influence samples, showing that a multiplicative approximation guarantee is, in general, not achievable using a nontrivial number of edge queries (thus the additive loss in our lower bound). We provide a hard example containing $(9/\mu^2)\binom{n \mu^2 /9}{2}$ edges for which one cannot provide a $\mu$-approximation while querying $o(n^2)$ edges. To this end, we consider an arbitrary algorithm that makes less than $C_{\mu} n^2$ edge queries, for some constant $C_{\mu}$ that is specified in Appendix \ref{app:proof:thm:edge_query_complexity-additive-loss-is-unavoidable}. Our hard example consists of a collection of $9/\mu^2$ cliques of size $n \mu^2 /9$ each. We choose $3/\mu$ of these cliques at random and connect them as in Figure \ref{fig:mu_n}. With $k = p = 1$, an optimal algorithm will seed one of the nodes in the connected cliques and achieves $(3/\mu)(n \mu^2 /9) = n \mu /3$ spread size. However, an algorithm that makes less than $C_{\mu} n^2$ queries cannot detect the connected clique with probability more than $\mu/3$. In Appendix \ref{app:proof:thm:edge_query_complexity-additive-loss-is-unavoidable}, we show that the expected spread size from seeding the output of any such algorithm is less than $n\mu^2/3$, i.e., less than a factor $\mu$ of the optimum. 

\begin{theorem}\label{thm:edge_query_complexity-additive-loss-is-unavoidable}
Let $0< \mu \leq 1$ be any fixed constant. There can be no $\mu$-approximation algorithm for influence maximization using $o(n^2)$ edge queries.
\end{theorem}

The hallmark of our analysis is in identifying an auxiliary submodular function, ${\Gamma}_{\delta} :2^{\mathcal{V}} \to \mathbb{R}$, to approximate our submodular function of interest  ${\Gamma} :2^{\mathcal{V}} \to \mathbb{R}$. The approximation is such that $|{\Gamma}_{\delta}(\mathcal{S}) - {\Gamma}(\mathcal{S})| \leq \epsilon n$ for all seed sets $\mathcal{S}$ of size $k$, with high probability. Here $\epsilon$ is the quality of approximation and it depends on $\delta$, which parameterizes the approximator $({\Gamma}_{\delta})$. Following the notation introduced in Definitions \ref{def:kIM} and \ref{def:alphaAPPROX}, we use ${\Lambda}_{\delta}$ and ${\Lambda}$ to denote the maximizers of ${\Gamma}_{\delta} $ and ${\Gamma} $ with constrained size $k$. The following Lemma (proved in Appendix \ref{app:proof:lem:approx_tech}) is true for any set function $\Gamma$ and its approximator $\Gamma_{\delta}$. It allows us to bound the loss that is incurred from optimizing ${\Gamma}_{\delta}$ in place of ${\Gamma}$.

\begin{lemma}\label{lem:approx_tech} Consider set functions ${\Gamma}_{\delta} $ and ${\Gamma} $ that map subsets of ${\mathcal{V}}$ to  $\mathbb{R}$ with their respective maximum values, ${\opt}_{\delta}$ and ${\opt}$, on subsets of size $k$. Assume that for all seed sets of size $k$, $\mathcal{S}$, we have $|{\Gamma}_{\delta}(\mathcal{S}) - {\Gamma}(\mathcal{S})| \leq \epsilon n$. Let ${\Lambda}'_{\delta}$ be any approximate maximizer of size $k$ for ${\Gamma}_{\delta}$, satisfying ${\Gamma}_{\delta}({\Lambda}'_{\delta}) \geq \alpha {\opt}_{\delta} - \beta n$. Then ${\Lambda}'_{\delta}$ also satisfies ${\Gamma}({\Lambda}'_{\delta}) \geq \alpha {\opt} - (\beta+(\alpha+1)\epsilon)n$.  
\end{lemma}

We start with a random set of $n\rho = O(k\log n)$ initial nodes and fix them for the subsequent steps (Subsection \ref{sec:sampling_nodes}). We then proceed to perform edge queries by probing their extended neighborhoods repeatedly (Subsection \ref{sec:probing_neighborhoods}). \revision{In this way}, we obtain $T = O(k\log n)$ cascades all starting from the same set of initial nodes (see Figure \ref{fig:edge_queries}). In Subsection \ref{sec:bounding_probed_neoghborhoods}, we argue that one does not need to continue probing the extended neighborhood of an initial node if the size of its revealed connected component is large enough (Lemma \ref{lem:Limiting_neighborhood}). This is a key observation that allows us to upper-bound the total number of edge queries in Theorem \ref{thm:total_cost_bound}. In Subsection \ref{sec:sampling_alg}, we propose the SEED algorithm to choose $k$ seeds (approximately optimally), based on the outcome of the edge queries (the output of PROBE), and prove a $(1-1/e)\opt - \epsilon n$ approximation guarantee with bounds on the number of edge queries (Subsection \ref{sec:space_bound}; Theorem \ref{thm:total_cost_bound}) and the run time (Subsection \ref{sec:fast_implement}; Theorem \ref{thm:main}).

Recall from Section \ref{sec:notation_set_up} and Definition \ref{def:edgequery} (edge queries) that each probing consists of independent random draws from the probed node's ordered neighborhood. Even if a node is probed more than once across different cascades, field survey researchers who devise their sampling plans based on the PROBE algorithm would only need to trace each revealed link once (having identified the unique links beforehand by recording $T$ random draws from an ordered set). After the field survey is concluded, the data from all traced links can be collected to reconstruct the output of the PROBE algorithm based on the outcome of the random draws. In our analysis, we upper bound the total number of queries that the PROBE algorithm makes in all of the $T$ cascades; therefore, the practical cost of tracing links during a field survey would be lower when the same edges appear in multiple cascades (e.g., the incident edge to the black node in Figure \ref{fig:edge_query_orange} is queried again in Figure \ref{fig:edge_query_blue}). In other applications --- e.g., for a web crawler that follows the PROBE algorithm to mine data from online social networks \citep{catanese2011crawling}, bounding the total number of queried edges is a direct concern not only for scalability, but also to control the data collection costs and time.  

\subsection{Sampling the initial nodes}\label{sec:sampling_nodes}

Recall our goal is to choose a seed set that (approximately) maximizes the influence function $\Gamma$. In this subsection, we show that we can estimate the value of $\Gamma$  by choosing a large enough set of nodes uniformly at random. To begin, fix $0 < \rho < 1$ and choose $\color{green}\lceil\normalcolor n\rho \color{green}\rceil\normalcolor$ nodes uniformly at random. We call these the \emph{initial nodes} and denote them by ${\mathcal{V}}_{\rho}$. Given ${\mathcal{V}}_{\rho}$, for any set $\mathcal{S} \subset \mathcal{V}$ we estimate the value of $\Gamma(\mathcal{S}) = \sum_{v\in\mathcal{V}}\phi(v,\mathcal{S})$ by:
\begin{align}
\Gamma_{\rho}\left(\mathcal{S}\right) := \frac{1}{\rho}  \sum_{v \in {\mathcal{V}}_{\rho}} \phi(v,\mathcal{S})\cdot \label{eq:SamplingNodes}
\end{align}
That is, we approximate the expected size of the cascade using the adoption probabilities of these initial nodes. To proceed, also define 
\begin{align}
\rho^{n,k}_{\epsilon,\delta} := \frac{(2+\epsilon)(k\delta\log n + \log2 )}{2\epsilon^2 n}\cdot
\end{align}

In the next Lemma, we bound the difference between $\Gamma$ and $\Gamma_{\rho}$ for $\rho \geq \rho^{n,k}_{\epsilon,\delta}$. The proof is in Appendix \ref{app:proof:lem:sample_loss_bound}. In the proof, we use a standard concentration argument to control the deviation of $\Gamma_{\rho}(\mathcal{S})$ from $\Gamma(\mathcal{S})$ for a fixed $\mathcal{S}$, and then a union bound to make the inequality true for any $\mathcal{S}$. 

\begin{lemma}[Bounding the sampling loss]
\label{lem:sample_loss_bound}   Let $\rho^{n,k}_{\epsilon,\delta} \leq \rho \leq 1$. With probability at least $1-e^{-\delta}$, for all seed sets $\mathcal{S}$ of size $k$ we have $|\Gamma_{\rho}(\mathcal{S}) - \Gamma(\mathcal{S})| \leq  \epsilon n$.
\end{lemma}

\subsection{Probing the extended neighborhoods of the initial nodes}\label{sec:probing_neighborhoods}

Note that our definition of $\Gamma_{\rho}$ in \eqref{eq:SamplingNodes} is in terms of $\phi(v,\mathcal{S})$, which can only be computed given the knowledge of the entire graph. However, when access to graph information is restricted (network information is made available only through edge queries)  we need to replace $\phi(v,\mathcal{S})$ by a proper estimate. To this end, we sample the graph edges through the probing procedure introduced in Section \ref{sec:notation_set_up}. Consider the $\color{green}\lceil\normalcolor n\rho \color{green}\rceil\normalcolor$  initial nodes in ${\mathcal{V}}_{\rho}$. For each initial node, we probe its neighborhood, keeping the edges with probability $p$. We then proceed to probe the neighborhoods of the revealed nodes, etc. We never probe a node more than once, and each edge receives at most one chance of being sampled. The probing stops after a finite number of steps (bounded by $n$).  We repeat this probing procedure $T$ times and obtain $T$ subsampled graphs that we denote by $\mathcal{G}^{(1)}_{\rho}, \ldots, \mathcal{G}^{(T)}_{\rho}$. 

We can now estimate $\phi(v,\mathcal{S})$ for $v$ belonging to ${\mathcal{V}}_{\rho}$ using the $T$ subsampled graphs $\mathcal{G}^{(1)}_{\rho}, \ldots, \mathcal{G}^{(T)}_{\rho}$ as follows. For $i=1,\ldots,T,$ and $v \in {\mathcal{V}}_{\rho}$, set $Y^{(i)}(v,\mathcal{S}) = 1$ if $v$ has a path to $\mathcal{S}$ in $\mathcal{G}^{(i)}_{\rho}$, otherwise set $Y^{(i)}(v,\mathcal{S}) = 0$. Our estimate of $\phi(v,\mathcal{S})$ for $v \in {\mathcal{V}}_{\rho}$ and $\mathcal{S} \subset \mathcal{V}$ is 
\begin{align}
    {\phi}^{(T)}(v,\mathcal{S}) := \frac{1}{T}\sum_{i=1}^{T}Y^{(i)}(v,\mathcal{S}).
    \label{eq:phi_T}
\end{align} We can similarly construct an estimate for the influence function that we want to optimize: 
\begin{align}
  {\Gamma}^{(T)}_{\rho}(\mathcal{S}) := \frac{1}{\rho} \sum_{v\in\mathcal{V}_{\rho}}{\phi}^{(T)}(v,\mathcal{S})\cdot  
  \label{eq:Gamma_rho}
\end{align} To proceed, define 
\begin{align}
T_{\epsilon,\delta}^{n,k} := \color{green} \left\lceil \normalcolor \frac{3(\delta+\log 2) (k+1)\log n }{\epsilon^2}  \color{green} \right\rceil \normalcolor \cdot
\end{align}Our next result bounds the difference between ${\Gamma}^{(T)}_{\rho} $ and ${\Gamma}_{\rho} $ for $T \geq T_{\epsilon,\delta}^{n,k}$. In the proof, we use concentration and union bound to ensure that ${\phi}^{(T)}(v,\mathcal{S})$ remains close to ${\phi}(v,\mathcal{S})$ for all $v \in {\mathcal{V}}_{\rho}$ and $\mathcal{S} \subset \mathcal{V}$. The proof details are in Appendix \ref{app:proof:lem:probing_chernoff_bound}. 

\begin{lemma}[Bounding the probing loss]\label{lem:probing_chernoff_bound} Let $T \geq T_{\epsilon,\delta}^{n,k}$. With probability at least $1 -e^{-\delta}$, for all sets $\mathcal{S}$ of size $k$ we have $|\Gamma^{(T)}_{\rho}(\mathcal{S}) - \Gamma_{\rho}(\mathcal{S})| \leq \epsilon n$.
\end{lemma}

\subsection{Limiting the probed neighborhoods}\label{sec:bounding_probed_neoghborhoods}

Here we consider a variation of the probing procedure described in the previous subsection whereby we stop probing when we hit a threshold $\tau$ of nodes in a connected component. Note that the probing may stop even before hitting $\tau$ nodes if no new edges are activated. Limiting the probed neighborhoods in this manner helps us bound the total number of edges that are used in our sketch (see Subsection \ref{sec:space_bound} and Theorem \ref{thm:total_cost_bound}). In fact, we show that it is safe to stop probing as soon as there are
$\tau =  \tau_{\epsilon}^{n,k} $ nodes in a connected component where $\tau_{\epsilon}^{n,k} := \color{green} \lceil \normalcolor -{(n\log\epsilon)}/{(\epsilon k)} \color{green} \rceil \normalcolor$.

Let us denote the $T$ subsampled graphs obtained through limited probing by $\mathcal{G}^{(1)}_{\rho,\tau}, \ldots, \mathcal{G}^{(T)}_{\rho,\tau}$. Moreover, let $\Gamma^{(T)}_{\rho,\tau}$ be our estimate of the influence function that is constructed based on $\mathcal{G}^{(1)}_{\rho,\tau}, \ldots, \mathcal{G}^{(T)}_{\rho,\tau}$ in the exact same way as in \eqref{eq:phi_T} and \eqref{eq:Gamma_rho}. This new estimator is, itself, a submodular function since it can be expressed as a sum of coverage functions. Our following result ensures that by optimizing ${\Gamma}^{(T)}_{\rho,\tau}$ instead of ${\Gamma}^{(T)}_{\rho}$, we do not loose more than $(1-\epsilon)$ in our approximation factor. The proof follows a probabilistic argument similar to \citet[Lemma 2.4]{Bateni:2017:AOS:3087556.3087585}. The crux of the argument is in constructing a random set whose expected value on $\Gamma^{(T)}_{\rho,\tau}$ is no less than $1-\epsilon$ of the optimum on $\Gamma^{(T)}_{\rho}$. We do so by starting from the optimum set on $\Gamma^{(T)}_{\rho}$ and replacing $\epsilon k$ of its nodes at random. Taking $\tau$ large enough allows us to argue that any node whose connections are affected by limiting the probed neighborhoods should belong to a large component, of size $\tau =  \tau_{\epsilon}^{n,k} $, and such large components are likely to be covered by one of the $\epsilon k$ random nodes. The complete proof is in Appendix \ref{app:proof:lem:Limiting_neighborhood}.

\begin{lemma}[Bounding the loss from limited probing]\label{lem:Limiting_neighborhood}
For $0 < \rho < 1$ and $0 < \epsilon \leq 1$, consider the limited probing procedure with the probing threshold set at $\tau =  \tau_{\epsilon}^{n,k} $. Then any $\alpha$-approximate solution to $k$-IM for $\Gamma^{(T)}_{\rho,\tau}$ is an $\alpha(1-\epsilon)$-approximate solution to $k$-IM for $\Gamma^{(T)}_{\rho}$.
\end{lemma}

Algorithm~\ref{alg:probe} summarizes the limited probing procedure for performing edge queries on the input graph ($\mathcal{G}$). The output is a sketch comprised of the $T$ independent subsampled graphs ($\mathcal{G}^{(1)}_{\rho,\tau}, \ldots, \mathcal{G}^{(T)}_{\rho,\tau}$) that fully determine the estimator $\Gamma^{(T)}_{\rho,\tau}$.

\begin{algorithm}[htb]
    \SetAlgoLined
    \DontPrintSemicolon
    \vspace{5pt}
    \KwIn{Edge query access to graph $\mathcal{G}$, cascade probability $p$ and probing parameters $\rho$, $T$ and $\tau$}
    \KwOut{Subsampled graphs $\mathcal{G}^{(1)}_{\rho,\tau}, \ldots, \mathcal{G}^{(T)}_{\rho,\tau}$ and initial node set $\mathcal{V}_{\rho}$}
    Choose $\color{green} \lceil \normalcolor n\rho \color{green} \rceil \normalcolor $ nodes uniformly at random without replacement and call them $\mathcal{V}_{\rho}$.\;
    \For{$i$ \KwFrom $1$ \KwTo $T$}{
        Initialize $\mathcal{X} \leftarrow \varnothing$, $\mathcal{V}^{(i)}_{\rho,\tau} \leftarrow \mathcal{V}_{\rho}$, $\mathcal{E}^{(i)}_{\rho,\tau} \leftarrow \varnothing$ and $\mathcal{G}^{(i)}_{\rho,\tau} \leftarrow (\mathcal{V}^{(i)}_{\rho,\tau},\mathcal{E}^{(i)}_{\rho,\tau})$. \;
        \tcp{Construct $\mathcal{G}^{(i)}_{\rho,\tau}$ by probing the unexplored nodes ($\mathcal{V}^{(i)}_{\rho,\tau}\setminus\mathcal{X}$):}
        \While{$\mathcal{V}^{(i)}_{\rho,\tau}\setminus\mathcal{X} \neq \varnothing$}{
            Draw a node, $\nu$, randomly from $\mathcal{V}^{(i)}_{\rho,\tau}\setminus\mathcal{X}$ add it to the explored nodes: $\mathcal{X} \leftarrow \mathcal{X} \cup \{\nu\}$. \;
            Draw a random integer according to $\mbox{Binomial}\left(\mbox{card}(\mathcal{N}_{\nu}),p\right)$ distribution and call it $I$. \;
            Draw a random subset of size $I$ from $\{1,\dots, \mbox{card}(\mathcal{N}_{\nu})\}$ and call it $\mathcal{I}$. \;
            \While{size of the connected component of $\nu$ in $\mathcal{G}^{(i)}_{\rho,\tau}$ is less than $\tau$}{
                Draw an index, $\iota$, randomly from $\mathcal{I}$ and remove it: $\mathcal{I} \leftarrow \mathcal{I} \setminus \{\iota\}$.\;
                Perform a $(\nu,\iota)$ edge query to graph $\mathcal{G}$ to reveal the $\iota$-th neighbor of $\nu$ and call it $\nu_\iota$.\;
                \If{$\nu_\iota \not\in \mathcal{X}$}{
                    \tcp{Add the newly discovered node and edge to $\mathcal{G}^{(i)}_{\rho,\tau}$:}
                    $\mathcal{V}^{(i)}_{\rho,\tau} \leftarrow \mathcal{V}^{(i)}_{\rho,\tau}\cup \{\nu_\iota\}$ \;
                    $\mathcal{E}^{(i)}_{\rho,\tau} \leftarrow \mathcal{E}^{(i)}_{\rho,\tau}\cup \{\nu, \nu_\iota\}$\;
                    }
            }
        
        }
        $\mathcal{G}^{(i)}_{\rho,\tau} \leftarrow (\mathcal{V}^{(i)}_{\rho,\tau},\mathcal{E}^{(i)}_{\rho,\tau})$.\;
    }
    \Return{$\mathcal{G}^{(1)}_{\rho,\tau}, \ldots, \mathcal{G}^{(T)}_{\rho,\tau}$ and $\mathcal{V}_{\rho}$.}
\caption{PROBE$(\rho,T,\tau,p)$}
\label{alg:probe}
\end{algorithm} 

\subsection{Influence maximization on the sampled graph}\label{sec:sampling_alg}
Lemmas \ref{lem:sample_loss_bound}, \ref{lem:probing_chernoff_bound}, and \ref{lem:Limiting_neighborhood} provide the following appropriate choices of the PROBE algorithm parameters $\rho$, $T$ and $\tau$:
\begin{align}
&\rho = \rho^{n,k}_{\epsilon,\delta} = \frac{(2+\epsilon)(\delta{k}\log n + \log2 )}{2\epsilon^2 n} \in O\color{green}\left(\normalcolor\frac{\delta{k}\log n}{\epsilon^2 n}\color{green}\right)\normalcolor\ccomma
\\& T =   T_{\epsilon,\delta}^{n,k}  =  \color{green} \left\lceil \normalcolor  \frac{3(\delta+\log 2) (k+1)\log n }{\epsilon^2}  \color{green} \right\rceil \normalcolor \in O \color{green}\left(\normalcolor\frac{\delta k\log n}{\epsilon^2}\color{green}\right)\normalcolor\ccomma
\\&\tau =  \tau_{\epsilon}^{n,k}  = \color{green} \left\lceil \normalcolor \frac{-n \log{\epsilon}}{\epsilon k} \color{green} \right\rceil \normalcolor \cdot   \label{eq:edgequeriesparametersetup}
\end{align} The following lemma (proved in Appendix \ref{app:proof:lem:bounding_approximation_loss}) combines our results so far (Lemmas \ref{lem:approx_tech} to \ref{lem:Limiting_neighborhood}) to show that with $\rho$, $T$ and $\tau$ set according to \eqref{eq:edgequeriesparametersetup} any $\alpha$-approximate solution, $\Lambda^{\star}$, to $k$-IM on $\Gamma^{(T)}_{\rho,\tau}$ satisfies $\Gamma(\Lambda^{\star}) \geq \alpha' \opt  - \epsilon'n$ for appropriate choices of $\alpha'$ and $\epsilon'$; thus providing an approximate solution to the original $k$-IM problem on $\Gamma$.

\begin{lemma}[Bounding the total approximation loss]\label{lem:bounding_approximation_loss}
Consider any  $0 < \epsilon, \alpha < 1$, and fix $\rho = \rho_{\epsilon,\delta}^{n,k}$, $T = T_{\epsilon,\delta}^{n,k}$ and $\tau = \tau_{\epsilon}^{n,k}$ according to \eqref{eq:edgequeriesparametersetup}. Moreover, let $\alpha'  = \alpha(1-\epsilon)$ and $\epsilon' = 2(\alpha(1-\epsilon)+1)\epsilon$. With probability at least $1 - 2e^{-\delta}$, any $\alpha$-approximate solution to \revision{the} $k$-IM problem on $\Gamma^{(T)}_{\rho,\tau}$ has value at least $\alpha' \opt - \epsilon' n$ on the original problem.
\end{lemma}

In Subsection \ref{sec:space_bound}, we bound the total number of edges that are queried by PROBE$(\rho,T,\tau,p)$. The output of the PROBE algorithm is the set of $T$ subsampled graphs ($\mathcal{G}^{(1)}_{\rho,\tau}, \ldots, \mathcal{G}^{(T)}_{\rho,\tau}$). From these $T$ subsampled graphs, we construct the estimator, $\Gamma^{(T)}_{\rho,\tau}$, and then use a submodular maximization algorithm to find a $(1-1/e-\epsilon)$-approximate solution to $k$-IM on $\Gamma^{(T)}_{\rho,\tau}$ for any $\epsilon>0$.  In Subsection \ref{sec:fast_implement}, we describe a fast implementation of submodular maximization on the sketch (the output of PROBE) that runs in ${O}_{\epsilon}(p {n^2}\log^{4}{n} + \sqrt{k p} n^{1.5}\log^{5.5}n + k n\log^{3.5}{n})$ time.  

\subsubsection{Bounding the total number of edge queries}\label{sec:space_bound}
Our edge query upper bound includes the following terms: 
\begin{align}
{E}^{n,k}_{\epsilon,p} & : =  p{\tau_{\epsilon}^{n,k}(\tau_{\epsilon}^{n,k} - 1)}/2 \in O\color{green}\left(\normalcolor\frac{p n^2 \log^2 (1/\epsilon)}{\epsilon^2 k^2}\color{green}\right)\normalcolor \ccomma \label{eq:Enkep}\\
C^{n,k}_{\epsilon,\delta} & : =  n \rho_{\epsilon,\delta}^{n,k} T_{\epsilon,\delta}^{n,k}\left( 1 + {E}^{n,k}_{\epsilon,p} + \sqrt{\delta(\tau_{\epsilon}^{n,k} \log{n}+\log T_{\epsilon,\delta}^{n,k}){E}^{n,k}_{\epsilon,p}}\right) \\ & \in O\color{green}\left(\normalcolor\frac{\delta^2 \log^2 (1/\epsilon)}{\epsilon^6}p n^2\log^2n + \frac{\delta^{3} \log^{1.5}(1/\epsilon)}{\epsilon^{5.5}}\sqrt{k p} n^{1.5}\log^{2.5}n + \frac{\delta^{2}}{\epsilon^{4}}{k^2}\log^{2}n\color{green}\right)\normalcolor.
\end{align} In Theorem \ref{thm:total_cost_bound}, we bound the total number of edge queries, denoted by $q$, in terms of ${E}^{n,k}_{\epsilon,p}$ and $C^{n,k}_{\epsilon,\delta}$. Our proof in Appendix \ref{app:proof:thm:total_cost_bound} relies critically on how we limit the probed neighborhoods (Subsection \ref{sec:bounding_probed_neoghborhoods}).  Roughly speaking, the output of PROBE$(\rho,T,\tau,p)$ consists of at most $n \rho T$ components of size no more than $\tau$ (barring the less than $n{\rho}T$ edges that may connect them). Moreover, since each edge is revealed with probability $p$, the expected number of edges in each of these components is at most $p{\tau(\tau - 1)}/2$. Subsequently, concentration allows us to give a high probability upper bound on the total number of edges that appear in the output of PROBE$(\rho,T,\tau,p)$. We can, similarly, also bound the total number of edges that are queried but discarded since they have been pointing to already probed nodes (see steps $10$ and $11$ of the PROBE algorithm).

\begin{theorem}[Bounding the edge queries]\label{thm:total_cost_bound} Consider any  $0 < \epsilon, \alpha < 1$, fix $\rho = \rho_{\epsilon,\delta}^{n,k}$, $T = T_{\epsilon,\delta}^{n,k}$ and $\tau = \tau_{\epsilon}^{n,k}$ according to \eqref{eq:edgequeriesparametersetup}, and denote the total number of edge queries during a single run of \emph{PROBE}$(\rho,T,\tau,p)$ by $q$. For $n \geq \sqrt{\delta + \log T}$ with probability at least $1 - 3e^{-\delta}$, $q$ can be bounded as follows:
\begin{align}
    q \leq Q^{n,k}_{\epsilon,\delta} & :=  2C^{n,k}_{\epsilon,\delta} + \left(2+\sqrt{2}\right)T_{\epsilon,\delta}^{n,k} n\sqrt{\delta + \log T_{\epsilon,\delta}^{n,k}}  
    \\ &  \in O\color{green}\left(\normalcolor\frac{\delta^2 \log^2 (1/\epsilon)}{\epsilon^6}p n^2\log^2n + \frac{\delta^{3} \log^{1.5}(1/\epsilon)}{\epsilon^{5.5}}\sqrt{k p} n^{1.5}\log^{2.5}n + \frac{\delta^{2}}{\epsilon^{4}}{k^2}\log^{2}n \right.\\ 
      & \left. + \frac{\delta^{1.5} \log^{0.5}(1/\epsilon)}{\epsilon^{2}}k\sqrt{\log k} n\log{n}\sqrt{\log\log{n}} \color{green}\right)  
    \\ & \subset {O}_{\epsilon,\delta}\color{green}\left(\normalcolor p n^2\log^{2}{n} + \sqrt{k p} n^{1.5}\log^{2.5}n + k n\log^2{n}\color{green}\right)\normalcolor. 
    \label{eq:edgequeryupperbound}
\end{align}
\end{theorem}

 It is worth highlighting that to provide our main approximation guarantee in Theorem \ref{thm:main}, we set $\delta = 2\log n$; see Appendix \ref{app:proof:thm:main}. In Appendix \ref{app:proof:thm:edge_query_complexity}, we prove a matching (up to logarithmic factor) lower bound by giving a hard example where it is impossible to provide a $\mu\opt-\epsilon n$ approximate guarantee using $o(pn^2)$ edge queries. We allow $p$ to vary with $n$ and prove the hard case for $k=1$ and $p \in \Omega(\log n/n)$, whereby the first term in \eqref{eq:edgequeryupperbound} is dominant and we have $Q^{n,k}_{\epsilon,\delta} \in {O}_{\epsilon,\delta}(p n^2\log^2n)$; of note, replacing $p\in O(\log n/n)$ in \eqref{eq:edgequeryupperbound} yields $Q^{n,k}_{\epsilon,\delta} \in O_{\epsilon,\delta}(k n \log^3{n})$. Our hard example builds on our previous construction in Figure \ref{fig:mu_n}, which is a variant of the so-called ``caveman graph'' where edges are rewired to link different cliques. In Appendix \ref{app:proof:thm:edge_query_complexity}, we provide a construction consisting of $9/\mu^2$ cliques, $3/\mu$ of which are connected around a circle, by choosing $-p^{-1}\log(\gamma\mu/6)$ edges randomly from each clique and rewiring them to connect to the next clique on the circle, while preserving the node degrees. Here $0 < \gamma < \mu/6$ is a constant that is chosen arbitrarily and then fixed. With $k = 1$, an optimal algorithm seeds one of the nodes in the $3/\mu$ connected cliques and achieves an expected spread size of at least $n(1-\gamma)\mu/3$. To show that no approximation algorithm can provide a $\mu\opt-\epsilon n$ guarantee using $o(pn^2)$ edge queries, we consider an arbitrary algorithm that makes less than $p C^{\epsilon}_{\mu,\gamma} n^2$ edge queries, where $C^{\epsilon}_{\mu,\gamma}$ is a constant that is specified in Appendix \ref{app:proof:thm:edge_query_complexity}. The probability that such an algorithm detects the $3/\mu$ connected cliques is at most $\mu/3 - 3\epsilon/\mu$. Hence, the expected spread size from seeding the output of such an algorithm cannot exceed $n(1-\gamma)\mu^2/3 - \epsilon n$ which is strictly less than $\mu\opt - \epsilon n$.

\begin{theorem}\label{thm:edge_query_complexity}
Let $0< \mu \leq 1$ be any constant. There can be no approximation algorithms that provide $\mu\opt - \epsilon n$ guarantees for $k$-IM using $o(pn^2)$ edge queries when $\epsilon < \mu^2/18$.
\end{theorem} 

\revision{Of note, the upper bound $(\mu^2/18)$  on $\epsilon$ in Theorem \ref{thm:edge_query_complexity} is necessary for limiting the probability of querying a rewired edge from the connected cliques. An upper bound of $\mu/3 - 3\epsilon/\mu>0$ on this probability implies an overall $\mu\opt - \epsilon n$ upper bound on the performance of any algorithm on this hard input. It seems unavoidable that this hardness should hold only for small $\epsilon$ relative to $\mu$, however, the exact quadratic dependence on $\mu$ may be improvable with a significantly different construction.}
 
\subsubsection{Bounding the total running time} \label{sec:fast_implement} In this subsection, we provide a fast implementation of our algorithm for influence maximization on the sampled graph. In fact, we can achieve a running time that is linear in the number of queried edges. First note that ${\Gamma}_{\rho,\tau}^{(T)}$ is, by definition, a coverage function, ergo a submodular function. Hence, we can use the randomized greedy algorithm of \cite{mirzasoleiman2015lazier} to provide a $(1-1/e-\epsilon)$ approximation guarantee. We start with $\Lambda^{\star}\leftarrow\varnothing$ and as in any greedy algorithm, we only use two types of operations: 
\begin{itemize}
    \item We query the marginal increase of a node $v$ on the current set $\Lambda^{\star}$, denoted by:\\ $\Delta(v|\Lambda^{\star}):= {\Gamma}_{\rho,\tau}^{(T)}(\Lambda^{\star} \cup \{v\}) - {\Gamma}_{\rho,\tau}^{(T)}(\Lambda^{\star})$.
    \item We choose a node $v^{\star}$  with maximal marginal increase and add it to the seed set: \\ $\Lambda^{\star} \leftarrow \Lambda^{\star} \cup \{v^{\star}\}$.
\end{itemize} 
The only difference is that the search for the node $v^{\star}$ is restricted to a subset $\mathcal{R}$ of size $(n/k)\log(1/\epsilon)$ that is drawn uniformly at random from $\mathcal{V}\setminus\Lambda^{\star}$. 

\begin{algorithm}[htb]
    \SetAlgoLined
    \DontPrintSemicolon
    \vspace{5pt}
    \KwIn{Subsampled graphs $\mathcal{G}^{(1)}_{\rho,\tau}, \ldots, \mathcal{G}^{(T)}_{\rho,\tau}$ and initial node set $\mathcal{V}_{\rho}$}
    \KwOut{$\Lambda^{\star}$, $(1-1/e-\epsilon)$-approximate solution to $k$-IM for $\Gamma_{\rho,\tau}^{T}$}
    Initialize $\Lambda^{\star} \leftarrow \varnothing$.\;
    Find the connected components of $\mathcal{G}^{(1)}_{\rho,\tau}, \ldots, \mathcal{G}^{(T)}_{\rho,\tau}$.\;
    Initialize the value of every connected component to be equal to the number of its nodes in $\mathcal{V}_{\rho}$.\;
    \For{$i$ \KwFrom $1$ \KwTo $k$}{
        \tcp{Selecting the $i$-th seed following a randomized greedy step:}
        Choose a subset $\mathcal{R}$ from $\mathcal{V}\setminus\Lambda^{\star}$ randomly with $\mbox{card}(\mathcal{R}) = (n/k)\log(1/\epsilon)$. \;
        \For{$v \in \mathcal{R}$}{
            Set $\Delta(v|\Lambda^{\star})$ equal to the sum of the values of the connected components containing $v$.\;
        }
        $v^{\star} \leftarrow \argmax_{v\in\mathcal{R}} \Delta(v|\Lambda^{\star})$.\;
        ${\Lambda^{\star}} \leftarrow {\Lambda^{\star}} \cup \{v^{\star}\}$.\;
        Update the values of the connected components containing $v^{\star}$ to zero.
        }
    \Return{$\Lambda^{\star}$.}    
\caption{SEED$(\epsilon,k)$}
\label{alg:seed}
\end{algorithm}  

In  Algorithm~\ref{alg:seed}: SEED$(\epsilon,k)$, we provide efficient implementations for the above operations. Our implementations are based on the structure of ${\Gamma}_{\rho,\tau}^{(T)}$, as determined by the $T$ subsampled graphs ($\mathcal{G}^{(1)}_{\rho,\tau}, \ldots, \mathcal{G}^{(T)}_{\rho,\tau}$). First using a graph search (e.g., DFS) we find the connected components of each of the $T$ subsampled graphs and count the number of initial nodes (belonging to $\mathcal{V}_{\rho}$) in each connected component. We refer to this count for each connected component as the ``value'' of that component. The main idea is that maximizing ${\Gamma}_{\rho,\tau}^{(T)}$ is equivalent to finding a seed set, $\Lambda^{\star}$, such that the total value of all connected components containing at least one seed in $\Lambda^{\star}$ is maximized. If a connected component already contains (i.e., is covered by) some nodes in $\Lambda^{\star}$, then the marginal increase due to that component should be zero. This is achieved by setting the value of a component to zero after adding a seed from that component to $\Lambda^{\star}$ --- see step $11$ of Algorithm~\ref{alg:seed}.

In Algorithm \ref{alg:seeding-queries} we run PROBE and SEED under the right parameter setting to achieve a $(1-1/e)\opt -\epsilon' n$ approximation guarantee for influence maximization. Theorem \ref{thm:main} formalizes our guarantees by combining our conclusions from Lemma \ref{lem:bounding_approximation_loss} and Theorem \ref{thm:total_cost_bound}, as well as the analysis of the performance of fast submodular maximization (randomized greedy) in \cite{mirzasoleiman2015lazier}. The proof is in Appendix \ref{app:proof:thm:main}.

\begin{algorithm}[htb]
    \SetAlgoLined
    \DontPrintSemicolon
    \vspace{5pt}
    \KwIn{Edge query access to graph $\mathcal{G}$ on $n$ nodes and an approximation loss $\epsilon'> 0$}
    \KwOut{An approximate $k$-IM solution, $\Lambda^{\star}$, satisfying $\Gamma(\Lambda^{\star}) \geq (1-1/e)\opt - \epsilon' n$}
    \tcp{Setting the parameters $(\epsilon, \delta, \rho, T, \tau)$:} 
    $\epsilon \leftarrow \epsilon'/7 \cdot$\; 
    $\delta \leftarrow 2\log{n} \cdot$\;
    $\rho \leftarrow  {(2+\epsilon)(\delta{k}\log n + \log2 )}/{(2\epsilon^2 n)} \cdot$\;
    $T \leftarrow \color{green} \lceil \normalcolor {3(\delta+\log 2) (k+1)\log n }/{\epsilon^2} \color{green} \rceil \normalcolor \cdot$\;
    $\tau \leftarrow \color{green} \lceil \normalcolor {n \log (1/\epsilon)}/{(\epsilon k)} \color{green} \rceil \normalcolor \cdot$\;
    \tcp{Running PROBE followed by SEED:}
    Run PROBE$(\rho,T,\tau,p)$ with edge query access to graph $\mathcal{G}$ to obtain $\mathcal{G}^{(1)}_{\rho,\tau}, \ldots, \mathcal{G}^{(T)}_{\rho,\tau}$ and $\mathcal{V}_{\rho}$.\;
    Run SEED$(\epsilon,k)$ with $\mathcal{G}^{(1)}_{\rho,\tau}, \ldots, \mathcal{G}^{(T)}_{\rho,\tau}$ and $\mathcal{V}_{\rho}$ as inputs to obtain $\Lambda^{\star}$.\;
    \Return{$\Lambda^{\star}$.}
\caption{$k$-IM with a bounded number of edge queries}
\label{alg:seeding-queries}
\end{algorithm}  

\begin{theorem}\label{thm:main}
    For any $0 < \epsilon' \leq 1$ and $n \geq (30/\epsilon')^2$, there \revision{exists} an algorithm for influence maximization that covers $(1-1/ e) \opt-\epsilon' n$ nodes in expectation with ${O}_{\epsilon'}(p n^2\log^4 n + \sqrt{k p} n^{1.5}\log^{5.5} n + k n\log^{3.5}{n})$ expected run time and using no more than ${O}_{\epsilon'}(p n^2\log^4 n + \sqrt{k p} n^{1.5}\log^{5.5} n + k n\log^{3.5}{n})$ edge queries in expectation; the dependence of the constant factors on $\epsilon'$ is the same as \eqref{eq:edgequeryupperbound} with $\epsilon  = \epsilon'/7$.
\end{theorem}

\subsection{Discrepancy between the query and cascade probabilities}\label{sec:discrepancy}

So far we have assumed that the cascade probability ($p$) is known or otherwise available to perform PROBE$(\rho,T,\tau,p)$ in step 6 of Algorithm~\ref{alg:seeding-queries}. While $p$ can be measured beforehand, in practice, such measurements are subject to significant uncertainty. Even when the network is fully observed, there can be uncertainty about $p$ \citep{chen2016robust,HeKempeKDD2016}; however, here $p$ plays a role in data collection, so it is important to consider whether other values of $p$ can be entertained after data collection.
\revision{
Motivated by the possibility that $p$ may be unknown and our best estimate of $p$ can change after the fact (i.e., after data collection), let us assume that the edge data is collected according to PROBE$(\rho,T,\tau,p')$ where the ``probe probability'' ($p'$) is different from the target cascade probability ($p$). Note that the setting of PROBE parameters ($\rho$, $T$ and $\tau$) in steps 1 to 5 of Algorithm~\ref{alg:seeding-queries} is not dependent on $p'$. Let $\mathcal{G'}^{(1)}_{\rho,\tau}$, $\ldots$, $\mathcal{G'}^{(T)}_{\rho,\tau}$ be the subsampled graphs output by PROBE$(\rho,T,\tau,p')$. The following hardness result shows that when $p$ is unknown, controlling the discrepancy between the query and seed cascade probabilities by a multiplicative ratio (i.e., requiring $|\max\{p,p'\}/\min\{p,p'\}| < 1+\delta$) is not enough for providing general $\mu\opt - \epsilon n$ approximation guarantees:
}\revision{
\begin{theorem}\label{thm:hardness_discrepency} Fix $0<\mu<1$ and $0<\delta < 1$. Let $p'$ and $p$ be the independent cascade probabilities for probing and seeding. There can be no approximation algorithms that provide $\mu\opt - \epsilon n$ guarantees for $k$-IM when the querying and seeding cascade probability are different, $p\neq p'$, with $\max\{p,p'\}/\min\{p,p'\} = 1+\delta$, $c_{\delta} = 1+(\delta - \delta^2)/2$ and 
\begin{align}
\epsilon < \frac{8\mu (\delta-\delta^2)c_{\delta}}{41(2-\mu) + 16(\delta-\delta^2)c_{\delta}}.    
\end{align}
\end{theorem}}
 
\revision{Stability and robustness of influence maximization are well-studied in light of uncertain cascade probabilities \citep{chen2016robust,wilder2017uncharted,HeKempeKDD2016,He:2018:SRI:3271478.3233227}. \citet[Theorem 3]{chen2016robust} use phase transition behavior in Erd\H{o}s-R\'enyi graphs to show that an $O(1/n)$ additive perturbation is enough to reduce the robust ratio (defined as the maximum over all seed sets of the minimum of approximation ratio of a seed set over the parameter range) to $O(\log n/n)$. In Appendix \ref{app:proof:thm:hardness_discrepency}, we use a similar construction to prove Theorem \ref{thm:hardness_discrepency}. Let us denote $\hat{p} = \max\{p,p'\}$ and $\check{p} = \min\{p,p'\}$. Our hard example} consists of a clique, $\mathcal{G}_1$, of size $2\epsilon{n}/\mu$, and a second subgraph, $\mathcal{G}_2$, that is obtained from the realization of a random graph with edge probability $(1-\delta/2)/(n_{\epsilon,\mu}\check{p})$ on the remaining $n_{\epsilon,\mu}:=(1-2\epsilon/\mu)n$ nodes. We choose $\check{p} =\Omega(\log n /n)$ such that active edges on $\mathcal{G}_1$ constitute a connected component of size of $\epsilon n$, with high probability as $n \to \infty$. However, when the cascade probability is $\check{p}$, the active edges on $\mathcal{G}_2$ constitute a random graph with edge probability $(1-\delta/2)/n_{\epsilon,\mu}$ so that the the largest connected component on $\mathcal{G}_2$ is with high probability $O(\log n)$. For $k = 1$, it is optimal to choose one of the $2\epsilon{n}/\mu$ nodes in $\mathcal{G}_1$ as the seed when the cascade probability is $\check{p}$. On the other hand, if the cascade probability is $\hat{p} = (1+\delta)\check{p}$, then the active edges induce a random graph with edge probability $c_\delta/n_{\epsilon,\mu}$ on $\mathcal{G}_2$ where $c_\delta = 1+(\delta-\delta^2)/2 > 1$. With high probability as $n \to\infty$, this random graph contains a giant connected component of size $f_\delta n$, satisfying $f_{\delta} = 1 - e^{-f_{\delta}c_{\delta}} > 1 - e^{-c_{\delta}} > 2\epsilon{n}/\mu$. In Appendix \ref{app:proof:thm:hardness_discrepency}, we use common techniques from connectivity analysis of random graphs and the emergence of the giant connected component therein to bound the expected spread size from seeding a node in either of the two subgraphs ($\mathcal{G}_1$ and $\mathcal{G}_2$). Subsequently, we show that for $\epsilon$ sufficiently small, any $k$-IM approximation algorithm that provides a $\mu\opt - \epsilon n$ guarantee, should necessarily seed $\mathcal{G}_1$ when the cascade probability is $\check{p}$ and $\mathcal{G}_2$ when the cascade probability is $\hat{p}$. Therefore, no such approximation algorithm exists when when the probe and cascade probabilities are different, i.e., with $\hat{p} \neq \check{p} = 1+\delta > 1$.

\revision{
If $p$ is known and the PROBE$(\rho,T,\tau,p')$ data is collected such that $p' > p$, then one can use a simple pruning procedure to correct for the difference between $p'$ and $p$ by removing the edges of $\mathcal{G'}^{(1)}_{\rho,\tau}$, $\ldots$, $\mathcal{G'}^{(T)}_{\rho,\tau}$ independently with probability $1 - p/p'$. Given the output of PROBE$(\rho,T,\tau,p')$, the algorithm in Appendix \ref{app:alg:discrepency} performs such a pruning and returns a corrected set ${\mathcal{G}}^{(1)}_{\rho,\tau}, \ldots, {\mathcal{G}}^{({T})}_{\rho,\tau}$ that exactly simulates the output of PROBE$(\rho,T,\tau,p)$. On the other hand, the dependent sampling process by which $\mathcal{G'}^{(1)}_{\rho,\tau}$, $\ldots$, $\mathcal{G'}^{(T)}_{\rho,\tau}$ are generated prevents us from using a similar post-processing (e.g., by combining $\mathcal{G'}^{(1)}_{\rho,\tau}$, $\ldots$, $\mathcal{G'}^{(T)}_{\rho,\tau}$ into union graphs) when $p' < p$. The reason is that different edges have different probability of appearing in a subsampled graph ($\mathcal{G'}^{(i)}_{\rho,\tau}$), depending on their network location and realization of other edges in $\mathcal{G'}^{(i)}_{\rho,\tau}$ --- edges that are connected to more influential nodes or other queried edges are more likely to be queried. In practical applications of edge queries (e.g., when running surveys to help diffuse health interventions), one can incentivize survey participants to reveal more edges and trace enough of them to ensure $p'>p$, albeit at an increased cost.
}

\section{Costs and benefits of network information}\label{sec:valueofnetinfo}

We can study the value of network information by examining how the expected spread size changes as more queries are used to select the seeds; that is, we can vary $T$ and $\rho$ in our algorithms. In this section, simulations of spread sizes with increased queries on an empirical network indicate the existence of an inflection point, whereby the first few queries improve the performance significantly before hitting a notably diminished returns. When this is the case, we can extract the benefits of the network information using just a few queries.

In particular, we conduct simulations with the Pennsylvania State University (Penn State) Facebook social network, with $41,536$ nodes, average degree $65.59$, and a total of $1,362,220$ edges. It is the largest network in a collection of Facebook social networks in 100 U.S. colleges and universities described by \cite{traud2012social}. Note that although the social network is known by the platform in advance, the friends lists are not available to, e.g., the electronic commerce companies that operate on the Facebook platform and can be collected through costly effort. 

\begin{figure}[t]
\centering
\begin{subfigure}[b]{0.49\textwidth}
\includegraphics[width = \textwidth]{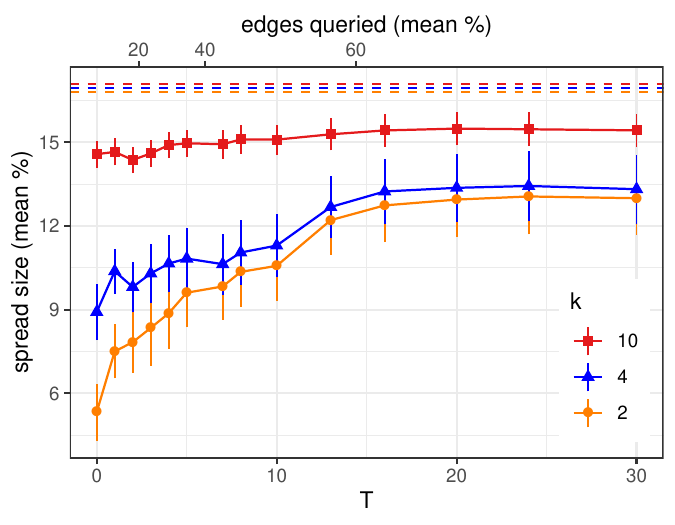}
\caption{$n\rho = 10$}
\label{fig:spread_size_vs_T_nrho_10}
\end{subfigure}~\begin{subfigure}[b]{0.49\textwidth}
\includegraphics[width = \textwidth]{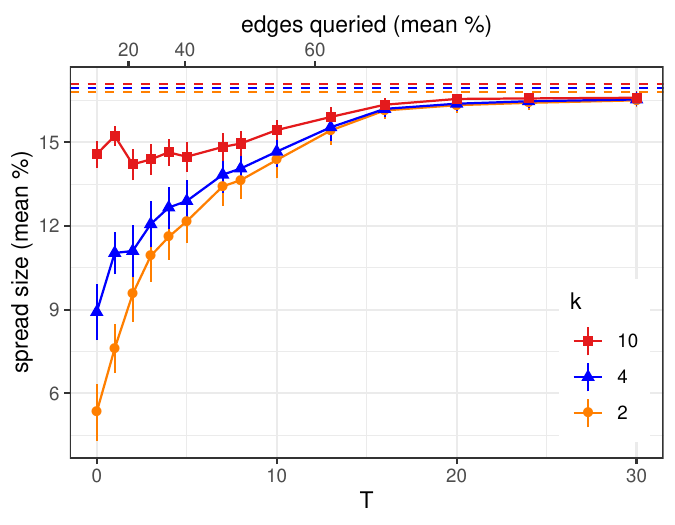}
\caption{$n\rho = 100$}
\label{fig:spread_size_vs_T_nrho_100}
\end{subfigure}
\caption{The mean spread sizes from seeding the output of Algorithm~\ref{alg:seed} applied to the Penn State Facebook social network as $T$ is increased for \revision{(\subref{fig:spread_size_vs_T_nrho_10})} $10$ and (\subref{fig:spread_size_vs_T_nrho_100}) $100$ initial nodes. To estimate the influence of each output seed set, we average the spread sizes over $500$ independent cascades with $p = 0.01$. To generate the $T$ subsampled graphs ($\mathcal{G}^{(1)}_{\rho,\tau}, \ldots, \mathcal{G}^{(T)}_{\rho,\tau}$) that are input to the SEED algorithm, we run the PROBE algorithm starting from $n\rho$ initial nodes, randomly chosen, and vary $T$ over a logarithmic scale: \revision{$T \in \{0, 1, 2, 3, 4, 5, 7, 8, 10, 13, 16, 20, 24, 30\}$}. Note that the $T = 0$ case corresponds to random seeding (using no network information at all). For each $T$, we run the PROBE algorithm $50$ times to generate $50$ random inputs for the SEED algorithm. The vertical axes in (\subref{fig:spread_size_vs_T_nrho_10}) and (\subref{fig:spread_size_vs_T_nrho_100}) show the mean spread sizes and $95\%$ confidence intervals that are computed over the $50$ outputs of the SEED algorithm for each $T$. Their top axes show the average number of revealed edges that is computed over the $50$ random inputs for each $T$. The complete-information, greedy baseline at each $k$ is marked by a dashed line.}
\label{fig:performance_vs_edge_queries}
\end{figure}

Figure \ref{fig:performance_vs_edge_queries} shows the performance of Algorithms~\ref{alg:probe} and \ref{alg:seed} (PROBE \& SEED) on the Penn State Facebook social network. Running the PROBE algorithm with higher values of $T$ leads to discovery of more nodes and edges from the social network. The vertical axes show the mean spread sizes from seeding the output of Algorithm~\ref{alg:seed} for each $T$ using $50$ random inputs. Recall that each input is a set of $T$ probed samples $\mathcal{G}^{(1)}_{\rho,\tau}, \ldots, \mathcal{G}^{(T)}_{\rho,\tau}$ that is obtained through Algorithm~\ref{alg:probe}.  The output performance improves with increasing $T$, \revision{since with more nodes and edges revealed,} the output seed set can be better optimized; nevertheless, there are diminishing returns to the increasing network information. Figure \ref{fig:spread_size_vs_T_nrho_100} shows that we can extract the benefits of complete network information using just $T = 30$ iterations: with enough information, the mean spread size from seeding the output of the algorithm saturates at the complete-information (deterministic) greedy algorithm output, and acquiring more network information does not improve the performance beyond that. 

It is worth noting that the random variations in the algorithm output --- hence, the width of the confidence intervals --- also decreases with the increasing network information in the input. There are two sources of randomness in the SEED algorithm's performance: $T<\infty$ and $\rho<1$. The output variance for large $T$ remains non-vanishing in Figure \ref{fig:spread_size_vs_T_nrho_10}; however, increasing the number of initial nodes, i.e., the size of the sample set $(\mbox{card}(\mathcal{V}_{\rho})=n\rho)$, from $10$ nodes in Figure \ref{fig:spread_size_vs_T_nrho_10} to $100$ nodes in Figure \ref{fig:spread_size_vs_T_nrho_100} allows us to remove the remnant randomness from the algorithm output at large $T$. 

\begin{figure}[t]
\centering
\begin{subfigure}[b]{0.49\textwidth}
\includegraphics[width = \textwidth]{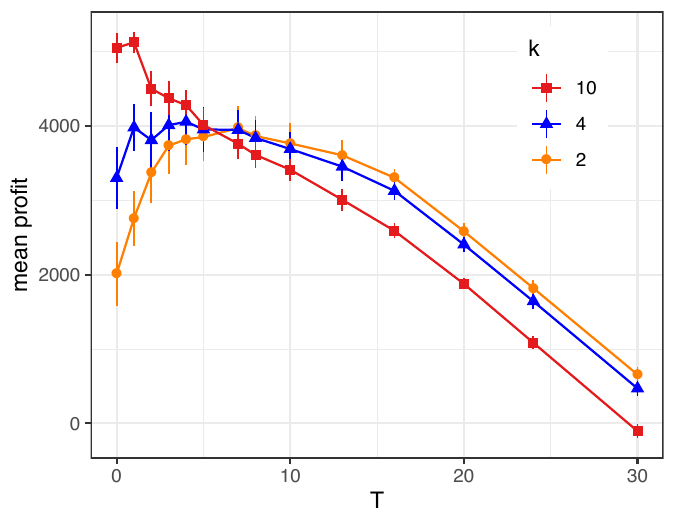}
\caption{$C_S = 100$, $C_T = 200$}
\label{fig:profit_vs_query_seedcost_100_iterationcost_200}
\end{subfigure}~\begin{subfigure}[b]{0.49\textwidth}
\includegraphics[width = \textwidth]{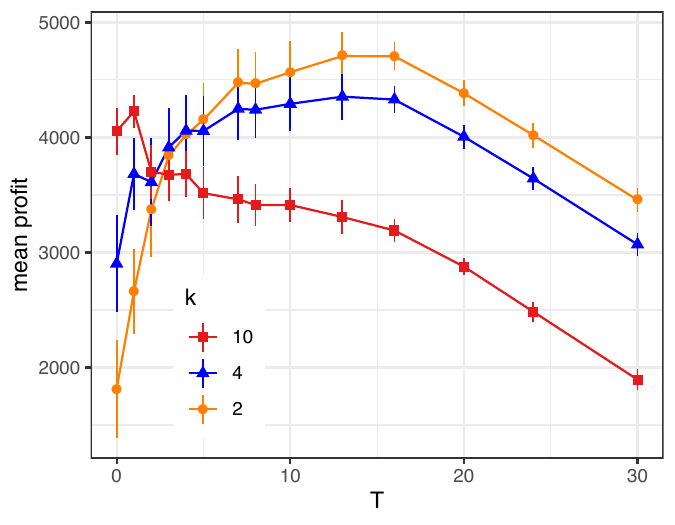}
\caption{$C_S = 200$, $C_T = 100$}
\label{fig:profit_vs_query_seedcost_200_iterationcost_100}
\end{subfigure}
\caption{The mean profit from seeding the output of Algorithm~\ref{alg:seed}, given the cost of seeds ($C_S$ per seed), iterations cost ($C_T$ per iteration), and unit revenue per adopter. The vertical axis shows the mean profits and confidence intervals that are computed from $50$ executions of the algorithms with increasing $T$ and $100$ initial nodes. (\ref{fig:profit_vs_query_seedcost_100_iterationcost_200}) When $C_S = 100$ and $C_T = 200$, the maximum expected profit is achieved at $k=10$ with no queries. (\ref{fig:profit_vs_query_seedcost_200_iterationcost_100}) Increasing the cost of seeds to $C_S = 200$ and decreasing the cost of iterations to $C_T = 100$ changes the optimal operating point to $k=2$ seeds with $T=13$ iterations.}
\label{fig:profit_vs_query}
\end{figure}

Our algorithms and these simulation results add important nuance to recent discussions of the value of network information. \cite{akbarpour2020just} show that, for special classes of random graph inputs, seeding $k+x$ nodes at random (using no information about the network) for some $x \in \omega(1)$ is enough to outperform the optimum spread size with $k$ nodes, as the network size increases ($n\to\infty$). They conclude that the benefits of acquiring network information to identify the optimal $k$ seeds can be offset by seeding a few more nodes at random (without using any network information).
We complement the results of \cite{akbarpour2020just} \revision{in two ways. First, we  measure how many queries are needed to yield the same expected spread sizes achievable using full knowledge of the network. Second, in our framework we can make the trade-off between acquiring network information and using more seeds explicit} by seeding more nodes and reducing the number of queries to keep the performance fixed. If we assume a cost, $C_S$, for each seeded node and another cost, $C_T$, for running each PROBE iteration, and a unit revenue per each adoption, then there is a number of iterations that is expected profit maximizing for a given number of seeds. In Figure \ref{fig:profit_vs_query}, the maximum expected profit for $C_S = 100$ and $C_T = 200$ is achieved with $k=10$ and $T = 0$ iterations, i.e., randomly seeding with the largest seed set size considered (Figure \ref{fig:profit_vs_query_seedcost_100_iterationcost_200}). However, increasing the cost of the seeds to $C_S = 200$ and decreasing the cost of the iterations to $C_T = 100$ reverses this result (Figure \ref{fig:profit_vs_query_seedcost_200_iterationcost_100}), and the optimum operating point shifts to $k=2$ and $T= 13$.

\section{Discussion and future directions}\label{sec:conc}

We addressed the problem of choosing the $k$ most influential nodes when the social network is unknown and accessing it is costly. We analyzed two ways of acquiring network information by observing influence spread of random nodes (influencing sampling) and by revealing the identity of neighboring nodes in an adjacency list (edge queries). We provided polynomial-time algorithms with almost tight approximation guarantees using a bounded number of queries to the graph structure (Theorems \ref{thm:main_spreading_queries} and \ref{thm:main}). We also provided hardness results to show that multiplicative approximation guarantees are generally impossible (Theorems \ref{thm:additiveloss} and \ref{thm:edge_query_complexity-additive-loss-is-unavoidable}) and to lower-bound the query complexity and show tightness of our query bounds for providing approximation guarantees with an additive loss (Theorems \ref{thm:main_spreading_queries_lowerbound} and \ref{thm:edge_query_complexity}). Finally, we showed the utility of our bounded-query framework for studying the trade-off between the cost of acquiring more network information and the benefit of increasing the spread size. 

The preceding results were for the independent cascade model over undirected graphs with a homogeneous cascade probability. In Appendix \ref{app:ext}, we discuss the extension of our results to directed graphs. In Appendix \ref{app:ext:linth}, we explain how our techniques can be applied to other commonly posited models of diffusion, and in particular, we prove the following extension for the linear threshold model \citep{kempe2015maximizing}:

\begin{theorem}\label{theo:main_lin_threshold_queries}
    For any arbitrary $0< \epsilon \leq 1$, there \revision{exists} a polynomial-time algorithm for influence maximization under the linear threshold model that covers $(1-1/ e) {\opt}-\epsilon n$ nodes in expectation in ${O}(n k^2\log({n}/{\epsilon})/{\epsilon^3})$ time, using no more than $nk\color{green} \lceil \normalcolor 81 k  \log( {6nk}/{\epsilon})/{\epsilon^3} \color{green} \rceil \normalcolor \in {O}(n k^2\log({n}/{\epsilon})/{\epsilon^3})$ edge queries. 
\end{theorem}

Results that address problem of seeding with partial network information are nascent and we foresee many directions for future research in this area. It is possible to provide tighter approximation guarantees or better query bounds if the input graph follows a known distribution, e.g., the stochastic block model \citep{wilder2018maximizing} or if individuals can directly report on who is influential \revision{by, e.g.,} recalling frequent origins of past cascades \citep{banerjee2017using, flodgren2011local}. Thus, an important venue for future work is to explore other ways of acquiring information about the graph structure. For example, one can draw inspiration from the graph sampling literature to devise new query methods \citep{leskovec2006sampling} to obtain subsampled graphs that preserve enough network information to perform influence maximization satisfactorily. We are particularly interested in queries that measure the spread of influence subject to time constraints. This is especially relevant in practice when spending time on data collection is costly and decision-makers have preference for earlier rather than later adoptions \citep{libai2013decomposing} \revision{or prefer diffusions among certain subgroups more than others. We speculate that operational considerations such as unequal, time-critical adoptions and privacy concerns in data collection open new venues for future methodological and applied works that build on the same foundation as ours.}

\revision{Broadly, our results highlight the importance of thinking through data collection in conjunction with planned interventions. Natural sampling methods can be re-designed to optimize for intervention outcomes (Algorithms \ref{alg:sample} and \ref{alg:probe}). Beyond the  co-design of sampling and targeting algorithms presented here, it is important to plan the data collection efforts with attention to their intervention contexts. For example, in the absence of reliable information about the spread, collecting network data without attention to diffusion parameters can lead to unsatisfactory outcomes (Theorem \ref{thm:hardness_discrepency}). Our lower bounds (Theorems \ref{thm:main_spreading_queries_lowerbound} and \ref{thm:edge_query_complexity}) point to the implied cost of data collection, which in practice can pose a major bottleneck. Notwithstanding, explicit understanding of these costs helps practitioners in value of information analysis and in deciding on how to allocate their limited resources between data collection and increased intervention (Section \ref{sec:valueofnetinfo}). Future work can bring out other trade-offs that are inherent in this co-design framework, e.g., by focusing on the privacy costs of data collection and its benefits to various subgroups (measured in terms of the intervention outcomes). With the increasing prevalence of data-driven intervention designs and policy practices, understanding such trade-offs has important implications for social welfare.}

\ACKNOWLEDGMENT{The authors gratefully acknowledge the research assistantship of Md Sanzeed Anwar through the support from the Institute for Data, Systems, and Society at MIT. \revision{The authors would like to thank the two anonymous reviewers and the associate editor whose close reading of the paper and thoughtful comments were instrumental to the development of the paper through its revisions. Eckles acknowledges a grant from Amazon, which partially supported Rahimian while at MIT.} Mossel is supported by ONR grant N00014-16-1-2227, NSF grant CCF-1665252 and ARO MURI grant W911NF-19-0217. Rahimian acknowledges computing hardware, software, and research consulting provided through the Pitt Center for Research Computing (Pitt CRC).
The authors are listed in alphabetical order.}

%\bibliographystyle{informs2014} 
%\bibliography{ref}

\clearpage

\begin{APPENDICES}

\phantomsection

\section{Additional related work}\label{app:lit}

Our INF-SAMPLE algorithm uses $O(k^2\log{n})$ influence samples to seed $k$ nodes approximately optimally (Theorem \ref{thm:main_spreading_queries}). Our $(1-1/e)\opt-\epsilon n$ approximation guarantee for INF-SAMPLE matches the $(1-1/e)\opt$ multiplicative factor that is tight for $k$-IM but suffers an $\epsilon{n}$ additive loss. Theorem \ref{thm:additiveloss} shows that the additive loss is impossible to avoid with a sub-linear sample size. Theorem \ref{thm:main_spreading_queries_lowerbound} shows that even an $\epsilon n $ additive guarantee is impossible if the number of samples is sub-quadratic in $k$, therefore, the quadratic order dependence of sample size on $k$ is also hard to improve. In the PROBE algorithm, we organize our edge queries to approximately simulate $T\in O(k\log{n})$ independent cascades each starting from the same set of $n\rho\in O(k\log n)$ randomly sampled nodes. The nuanced implementation of PROBE allows us to construct these simulations using only a bounded number of edge queries, while still preserving enough information in its output to select $k$ seeds approximately optimally by applying the SEED algorithm (Theorem \ref{thm:main}).

Since an earlier version of the present work \citep{costlyseedingEC19}, sample complexity of influence maximization is treated thoroughly by \cite{sadeh_et_al:LIPIcs:2020:11714}. Through a careful analysis, \cite{sadeh_et_al:LIPIcs:2020:11714} are able to upper-bound the variance of the number of adopters at a fixed diffusion step in term of the expected number of adopters and the diffusion step. This variance upper bound allows them to efficiently control the estimation error of the spread sizes by relaxing the relative error requirements when seed sets produce small spreads. Subsequently, \cite{sadeh_et_al:LIPIcs:2020:11714} give a sample complexity bound of $O({k}{t}\log{n})$ where $t$ is the number of diffusion steps. This is the best known upper bound on the number of simulations required for achieving tight $k$-IM approximation guarantees but is still super-linear in the worst-case because the number of diffusion steps  $t$ can be of the order of the network size $n$. It is worth noting that the $O(k{t}\log{n})$ bound applies to i.i.d. simulations of the spread and is achieved using an adaptive sample size --- their implementation of the approximate greedy algorithm uses $O({k^3}{t}\log n)$ simulations, and they are not directly concerned with limiting access to the input graph. On a broader scope, reverse influence sampling, i.e., collecting influence samples on the transposed graph with edge directions reversed --- proposed by \cite{borgs2014maximizing} ---  has been successfully applied in the design of $k$-IM algorithms with practical efficiency and near-optimal run time \citep{tang2014influence,tang2015influence,nguyen2016stop}. For undirected graphs we can collect our influence samples on the original graph (as opposed to its transpose), and by allowing for an $\epsilon n$ additive loss, we can achieve our $(1-1/e)\opt-\epsilon n$ guarantee on general graphs using ${O}(k^2\log n)$ influence samples; thus avoiding the linear dependence on $n$ which is prohibitive when network information is limited and costly.

More broadly, the influence sampling framework relates to recent advances in application of stochastic oracles for submodular maximization with imperfect information. \citet{karimi2017stochastic} consider a class of discrete optimization problems where the objective function is expressed as an expectation over submodular functions and can be estimated by sample averaging but is not explicitly available  and cannot be used as a black box oracle for the greedy algorithm. Their approach is by lifting the objective function into the continuous domain using a concave upper-bound on its multilinear extension. The concave upper-bound is guaranteed to be no more than $e/(e-1)$-times the optimization objective and can be maximized efficiently through projected stochastic gradient ascent. The finally transfer the solution back to the discrete domain using a randomized rounding technique that preservers the quality of approximation in expectation. The fast convergence result of \citet{karimi2017stochastic} applies to the total number  of gradient steps required for maximizing the concave relaxation of the objective function and is given by $T= \color{green} \lceil \normalcolor B^2\rho^2/\epsilon^2 \color{green} \rceil \normalcolor $ where $\epsilon$ is an additive loss in the approximation guarantee, $\rho$ is a bound on the gradient norms, and $B$ is a bound on the norm of the continuous optimization variable \citep[Theorem 2]{karimi2017stochastic}. In the case of $k$-IM on an $n$ node network while subgradients can be estimated by BFS on influence samples, with $\rho = \sqrt{n}$ and $B=\sqrt{k}$, we need a total of $T=O(nk)$ stochastic gradient steps, thus suffering the same prohibitive linear dependence on $n$ \citep[Lemma 4]{karimi2017stochastic}.         

Another related body of literature studies optimizing submodular functions based on input-output data pairs that are sampled from a distribution over feasible inputs (e.g., uniformly over all input sets of size $k$). In the $k$-IM setup, this learning-theoretic framework implies that the observed data consist of pairs of initial nodes and expected number of adopters for cascades initiated from those nodes. \citet{balkanski2017limitations} show hardness of approximation for maximizing coverage functions  under cardinality constraints (including $k$-IM) using polynomially many samples from any distribution, whereas for monotone submodular functions with bounded curvature more positive results can be achieved with polynomial sample size \citep{balkanski2016power}. Our influence sampling framework is principally different from the learning-theoretic framework of optimization from samples because we cannot observe the exact value of the influence function but instead see a random realization of the adopters for each input. Moreover, each influence sample starts from a random initial node and our inputs are not constrained to be subsets of size $k$. \citet{CommunitiesLearningtoInfluence} consider an adaptation of the optimization from samples framework to $k$-IM where each sample consists of the number of adopters in a random cascade. They offer an algorithm to list nodes in the order of their decreasing expected marginal contributions to a random set, and then iteratively remove those whose marginal contribution significantly overlaps with an earlier node on the list. Estimating marginal contributions in this framework requires polynomially many samples \citep[Lemma 15; the exact dependence on $n$ is not clarified but $O(n)$ appears to suffice]{CommunitiesLearningtoInfluence}, while the performance guarantees are applicable only to stochastic block model random input graphs, \citep[Theorems 6 and 12]{CommunitiesLearningtoInfluence}.

\section{Proofs \& other mathematical details}\label{app:proof}

\subsection{Proof of Theorem \ref{thm:additiveloss}: Hardness of approximation with influence samples}\label{app:proof:thm:additiveloss}

Pick an arbitrary function $f(n)\in o(n)$, and let $g(n)=\sqrt{{n}/{f(n)}}$. Note that $g(n)\in \omega(1)$. Consider an algorithm $\alg$ that uses $f(n)$ influence samples. Whe show that $\alg$ cannot be a $\mu$-approximation. Our hard example consists of a clique of size $g(n)$, chosen uniformly at random, and $n-g(n)$ isolated nodes and we aim to seed one node (Figure \ref{fig:f_n_g_n}). One can bound the probability that $\alg$ queries a node from the clique by
\begin{align*}
    1-\left(1-\frac{g(n)}{n}\right)^{f(n)}= 1-\left(1-\frac{g(n)}{n}\right)^{\frac{n}{g(n)^2}} \leq     1-\left(1-\frac 1 e\right)^{\frac{1}{g(n)}}.
\end{align*}
Moreover, since $g(n)\in \omega(1)$ we have $1-\big(1-\frac 1 e\big)^{\frac{1}{g(n)}}\in o(1)$. If $\alg$ does not see a node via influence samples, it seeds one of the nodes of the clique with probability at most $\frac{g(n)}{n-f(n)} \in o(1)$. Therefore, the expected number of nodes covered by $\alg$ is at most $o(1) g(n) + 1$, which means that the approximation factor of $\alg$ is $\frac{o(1) g(n) + 1}{g(n)} \in o(1)$ as claimed.

\subsection{Proof of Theorem \ref{thm:main_spreading_queries}: Approximation guarantees with bounded influence samples}\label{app:proof:them:main_spreading_queries}

Recall our notation in the INF-SAMPLE algorithm. The output of the algorithm, $\Lambda^{\star}$, is a set of $k$ nodes that are chosen, one by one, in $k$ iterations. Let us use ${\Lambda^{\star}}^{i}$ to denote the first selected $i$ seeds. In the $i$-th iteration, we choose $\rho$ initial nodes at random (with replacement) and collect influence samples. We use $A^i_j$ to denote the $j$-th influence sample collected during the $i$-th iteration. We reset $A^{i}_j$ to $\varnothing$ if it contains any of the $i-1$ nodes selected in the previous iterations. We consider the pool of candidates, $u\in \mathcal{V}\setminus{\Lambda^{\star}}^{i-1}$, and choose the $i$-th seed to be the one that appears in the most $A^{i}_j$'s. To put this in mathematical notation, let $X^i_{u,j} = \mathds{1}\{u \in A^i_j\}$ be the indicator that $u$ belongs to $A^i_j$, and set $X^i_u = \sum_{j=1}^{\rho}X^{i}_{u,j}$ to count the number of times that $u$ appears in any of the subsets $A^i_{1}$, $\ldots$, $A^i_{\rho}$. Subsequently, in step $i$, we choose  $v^{\star} = \argmax_{u \in \mathcal{V}\setminus{\Lambda^{\star}}^{i-1}} X^i_u$ and add it to ${\Lambda^{\star}}$.

We analyze the iterations of Algorithm~\ref{alg:sample} and show that for $\epsilon' = \epsilon/3 $ and $\rho = \rho_{\epsilon}^{n,k} =\color{green} \lceil \normalcolor {3 k \log ( {2nk}/{\epsilon'})}/{\epsilon'^3} \color{green} \rceil \normalcolor $, the output of INF-SAMPLE$(\rho,k)$ satisfies the desired approximation guarantee. Let us define random variable $N^i_u$ to be the expected number of nodes that are covered by ${\Lambda^{\star}}^{i-1} \cup \{ u\}$ but not  by ${\Lambda^{\star}}^{i-1}$. Note the probability that $X^i_{u,j}=1$ is equal to ${N^i_u}/{n}$. Therefore, we have $\mathbb{E}[(n/\rho)X^i_{u}]= \mathbb{E}[N^i_{u}]$. Moreover, notice that choosing $\nu^{\star} \in \mathcal{V}\setminus{\Lambda^{\star}}^{i-1}$ to maximize $\mathbb{E}[N^i_{u}]$ is equivalent to one step of the greedy algorithm. This is equivalent to choosing $\nu^{\star} \in \mathcal{V}\setminus{\Lambda^{\star}}^{i-1}$ to maximize $\mathbb{E}[X^i_{u}]$, since $\mathbb{E}[(n/\rho)X^i_{u}]= \mathbb{E}[N^i_{u}]$.

Next note that due to submodularity, the marginal values only decrease as we add more elements. Hence, if we stop at the $i$-th iteration satisfying $\mathbb{E}[N^i_{u}] < \epsilon' n/k$ for all candidates $u \in \mathcal{V}\setminus{\Lambda^{\star}}^{i-1}$, then in total we do not loose more than $k(\epsilon n/k)=\epsilon n$. For the sake of analysis, let us assume that the algorithm stops when $\mathbb{E}[N^i_{u}] < \epsilon' n/k$ for all $u$, i.e., the algorithm stops if it selects $k$ seeds or $\mathbb{E}[N^i_{u}] < \epsilon' n/k$ for all $u$, whichever comes first. This allows us to lower-bound the expected spread size from seeding the output of Algorithm~\ref{alg:sample}. In reality, any additional node that the algorithm selects will only improve the expected spread size of its output. Henceforth, without loss of generality, we assume that $\max_u\mathbb{E}[N^i_{u}] \geq \epsilon' n/k$ which means that we have $\mathbb{E}[X^i_{u}] \geq \epsilon' \rho/k$.
 
Recall that $X^i_u$ is the sum of i.i.d. binary random variables $X^{i}_{u,j}$. Hence, by the Chernoff bound we have 
\begin{align}
    \mathbb{P}\left[|X^{i}_{u} - \mathbb{E}[X^{i}_{u}]| \geq  \epsilon' \mathbb{E}[X^{i}_{u}]\right] &\leq 2\exp\left(-\frac{\epsilon'^2\mathbb{E}[X^{i}_{u}]}{3}\right)& \mathbb{E}[X^{i}_{u}] \geq \epsilon'\rho /k \\
    &\leq 2\exp\left(-\frac{\epsilon'^3 \rho}{3 k}\right) &  \rho = \color{green} \left\lceil \normalcolor  \frac{81 k \log ({6nk}/{\epsilon})}{\epsilon^3} \color{green} \right\rceil \normalcolor  =  \color{green} \left\lceil \normalcolor  \frac{3 k \log( {2nk}/{\epsilon'})}{\epsilon'^3} \color{green} \right\rceil \normalcolor \\
    &=  \frac{\epsilon'}{n k}\cdot &  
\end{align} Union bound over all $u \in \mathcal{V}$ and $1 \leq i\leq k$ provides that with probability at least $1-\epsilon'$, $(1-\epsilon')\mathbb{E}[X^{i}_{u}] \leq X^{i}_{u} \leq  (1+\epsilon')\mathbb{E}[X^{i}_{u}]$ for all $u$ and $i$.
This implies that the seed that our algorithm selects has marginal increase at least $\frac{1-\epsilon'}{1+\epsilon'} \geq 1-2\epsilon'$ times that of the greedy algorithm. Such an algorithm is called $(1-2\epsilon')$-approximate greedy in~\cite{golovin2011adaptive} and it is proven to return a $(1-1/e - 2\epsilon')$-approximate solution~\cite{badanidiyuru2014fast,golovin2011adaptive,kumar2015fast}. Therefore, we can bound the expected value of the output solution of INF-SAMPLE$(\rho,k)$ as follows:
\begin{align}
    \mathbb{E}[\Lambda^{\star}] &\geq (1-\epsilon')[(1-1/e-\epsilon'){\opt} - \epsilon' n] \\&\geq (1-\epsilon')[(1-1/e){\opt} - 2\epsilon' n] \\ & \geq (1-1/e){\opt} - 3\epsilon' n = (1-1/e){\opt} - \epsilon n.
\end{align}

\subsection{Proof of Theorem \ref{thm:main_spreading_queries_lowerbound}: Lower-bounding the required number of influence samples} \label{app:proof:thm:main_spreading_queries_lowerbound}

Fix any $0 < \mu < 1$, $0 < \epsilon < \mu/k$, and set $\beta = 0.5(\mu- k\epsilon - 1 - 2/k) $. Consider our hard example in Figure \ref{fig:k_stars_hard_example} with $n > k^2$. Note that given the complete network information in this example, the optimum strategy is to seed the $k$ centers of each of the $k$ stars, which achieves $\opt  = k(n/k^2) = n/k$ expected spread size. Let $\alg$ be any algorithm that seeds optimally using less than $\beta k^2$ influence samples. We will show that the expected spread size from seeding the output of $\alg$ is strictly less than $\mu\opt - \epsilon n$.  To upper-bound the expected spread size from seeding the output of $\alg$, consider a reduction from the following problem, which we call SEED-STAR:
\begin{problem}[SEED-STAR] The input graph is restricted to be a collection of $k$ stars of size $n/k$ each (as in Figure \ref{fig:k_stars_hard_example}), and the algorithm observes a fixed number of influence sample from the input graph. Furthermore, if an influence sample contains the center of a star subgraph, then the entire star subgraph is revealed to the algorithm. Given the influence samples and revealed star subgraphs, the SEED-STAR problem is to choose $k$ seeds such that their expected spread size is maximal.
\end{problem} Let $\alg'$ be any optimal algorithm using less than $\beta k^2$ influence samples for the SEED-STAR problem. Consider the outputs of $\alg$ and $\alg'$ on the hard example in Figure \ref{fig:k_stars_hard_example}. Because both $\alg$ and $\alg'$ seed optimally given $\beta k^2$ influence samples, and $\alg'$ has strictly more information than $\alg$, the expected spread size from seeding the output of $\alg'$ is at least as large as $\alg$; hence, it suffices to upper-bound the expected spread size from seeding the output of $\alg'$. Note that an optimal seed set of size $k$ for $\alg'$ should include the centers of all the revealed stars. If less than $k$ stars are revealed, then it is optimal for $\alg'$ to choose the remaining seeds randomly from among the sampled leaf nodes that do not belong to any of the revealed stars. Any such leaf node will contribute 
\begin{align}
    1 + \frac{1}{k} + \frac{1}{k^2} \color{green} \left( \normalcolor \frac{n}{k}-2 \color{green} \right) \normalcolor < 1 + \frac{1}{k} + \frac{n}{k^3}.
\end{align} Because there at most $k$ such leaf nodes the total contribution of the leaf nodes to the expected spread size of the out of $\alg'$ is at most $k + 1 + {n}/{k^2}$. Next we show that the probability of a new star being revealed to $\alg'$ as a result of an influence sample is always strictly less than $2/k$. To see why, note that if the initial node of the influence sample belongs to a previously seen star then this probability is zero. Conditioned on the initial node belonging to a new star, then influence sample will reveal the star if it is its center node, which happens with probability $k/n$, or if it is a leaf node  and its incident edge is activated, which happens with probability at most $1/k$. Hence, the probability of a new start being revealed at any influence sample is upper-bounded by 
\begin{align}
    \frac{k}{n} + \frac{1}{k} <  \frac{2}{k}, \label{eq:new_star_prob_upperbound}
\end{align} where the last inequality is because of the assumption $n > k^2$. Therefore, the expected number of stars that are revealed by $\beta k^2$ influence samples is at most $2\beta k$. Any such star contributes $n/k^2$ to expected spread size of the output of $\alg'$. We can thus upper-bound the expected spread size from seeding the output of $\alg'$, as therefore $\alg$, as follows: 
\begin{align}
  (2\beta k)\frac{n}{k^2} + k + 1 + \frac{n}{k^2} < ((2\beta+1) k + 2)\frac{n}{k^2} = \mu \opt - \epsilon n, 
\end{align} using $n>k^2$ in the last inequality. The proof follows as with $\alg$ being optimal, no algorithms can provide an approximation guarantee that exceeds  $\mu\opt - \epsilon n$, using less than $\beta k^2$ influence samples on this example. 

\subsection{Proof of Theorem \ref{thm:edge_query_complexity-additive-loss-is-unavoidable}: Hardness of approximation with edge queries}\label{app:proof:thm:edge_query_complexity-additive-loss-is-unavoidable}

We present our hard example for $k=1$ and $p=1$. Moreover, for simplicity of presentations we assume that $3/\mu$, ${1}/{\mu^2}$, and $\mu^2 n/9$ are integers. Consider the following two graphs.
\begin{itemize}
    \item \textbf{$G$: } This graph consists of ${9}/{\mu^2}$ cliques, each of size $\mu^2 n/9$.
    \item \textbf{$G'$: } This graph is constructed from $G$ via the following random process. We select $\frac{3}{\mu}$ clusters uniformly at random. Then we select one edge from each selected cluster uniformly at random. Let $(v_1,u_1),(v_2,u_2),...(v_{3/\mu},u_{3/\mu})$ be the list of the selected edges. we remove $(v_1,u_1),(v_2,u_2),...(v_{3/\mu},u_{3/\mu})$ and replace them by $(u_1,v_2),$ $(u_2,v_3),$ $\dots,$ $(u_{3/\mu-1},v_{3/\mu}),$ $(u_{3/\mu},v_1)$. Note that this process connects all of the selected clusters while preserving the node degree (see Figure \ref{fig:mu_n}). 
\end{itemize}
    Let $\alg$ be an arbitrary (potentially randomized) algorithm for influence maximization that queries less than $({\mu^3}/{27})\binom{\mu^2 n/9}{2}$ edges. Note that with $k=1$ an optimum seed on $G'$ spreads to $\opt=\mu n/3$ nodes. Next we show that the expected spread size form seeding the output of $\alg$ on $G'$ is less than $\mu^2 n/3$, which means that $\alg$ is not an $\mu$-approximation algorithm. This implies that there is no $\mu$-approximation algorithm that queries less than $({\mu^3}/{27})\binom{\mu^2 n/9}{2} \in O_{\mu}(n^2)$ edges as claimed.
    
    We use the \revision{result} of $\alg$ on $G$ to analyze the run of $\alg$ on $G'$. Note that due to symmetric construction of $G$ we can assume that $\alg$ seeds one of the nodes of $G$ uniformly at random. Observe that the expected spread size of a random seed in $G'$ is 
    \begin{align}  
        \color{green} \left( \normalcolor 1-\frac{3/\mu}{9/\mu^2} \color{green} \right) \normalcolor  \frac{\mu^2 n}{9} +  \frac{3/\mu}{9/\mu^2} \frac 3 {\mu}  \frac{\mu^2 n}{9} \leq \frac{2\mu^2 n}{9}\cdot
    \end{align}
    Moreover, note that the run of $\alg$ on $G$ and $G'$ is the same unless $\alg$ queries one of the positions (i.e., edges) that we change in $G$ to construct $G'$. In what follows, we upper-bound the probability that $\alg$ queries one of the changed positions in $G$ by ${\mu}/ 3$. This implies that $\alg$ cannot be a $\mu$-approximation because its expected spread size is strictly less than $\mu\opt$:
    \begin{align}
        \color{green} \left( \normalcolor 1-\frac {\mu} {3} \color{green} \right) \normalcolor  \frac{2\mu^2 n}{9} + \frac {\mu} {3}   \frac 3 {\mu} \frac{\mu^2 n}{9} < \frac{\mu^2 n}{3} = \mu\opt.
    \end{align}
    
To finish the proof, we need to bound the probability that $\alg$ queries one of the changed positions in $G$. Let $A_i$ be the (possibly random) number of edges that $\alg$ queries from the $i$-th clique. Recall that by assumption, $\alg$ queries less than $({\mu^3}/{27})\binom{\mu^2 n/9}{2}$ edges. Hence, with probability one (with respect to randomness of $\alg$), we have $A_i < ({\mu^3}/{27})\binom{\mu^2 n/9}{2}$. Therefore, the total probability (with respect to randomness of $\alg$ and $G'$) that $\alg$ queries a changed position in the $i$-th clique is upper-bounded by 
    \begin{align}
        \frac{(\mu^3/27)\binom{\mu^2 n/9}{2}}{\binom{\mu^2 n/9 }{2}} = \frac{\mu^3}{27} \cdot 
    \end{align} By a union bound over all $9/\mu^2$ cliques we can upper-bound the probability that $\alg$ queries a changed position by $(9/\mu^2)(\mu^3/27) =\mu/3$, as claimed. This completes the proof of the theorem.

\subsection{Proof of Lemma \ref{lem:approx_tech}: Combining additive and multiplicative approximation loss}\label{app:proof:lem:approx_tech} 

Starting with an approximate solution satisfying ${\Gamma}_{\delta}({\Lambda}'_{\delta}) \geq \alpha {\opt}_{\delta} - \beta n$ on the one hand, we have
\begin{align}\label{ineq:approx10}
{\Gamma}({\Lambda}'_{\delta}) + \epsilon n \geq {\Gamma}_{\delta}({\Lambda}'_{\delta}),
\end{align} since $|{\Gamma}_{\delta}({\Lambda}'_{\delta}) - {\Gamma}({\Lambda}'_{\delta})| \leq \epsilon n$. 

On the other hand, consider ${\Lambda}_{\delta}$ and ${\Lambda}$, which are the optimum seed sets for ${\Gamma}_{\delta}$ and ${\Gamma}$, respectively. By assumption we have ${\Gamma}_{\delta}({\Lambda}'_{\delta})\geq \alpha{\Gamma}_{\delta}({\Lambda}_{\delta}) - \beta n$, and since, by optimality of ${\Lambda}_{\delta}$ for ${\Gamma}_{\delta}$, ${\Gamma}_{\delta}({\Lambda}_{\delta}) \geq {\Gamma}_{\delta}({\Lambda})$, we get
\begin{align}\label{ineq:approx20}
{\Gamma}_{\delta}({\Lambda}'_{\delta})&\geq \alpha{\Gamma}_{\delta}({\Lambda}) - \beta n \geq \alpha{\Gamma}({\Lambda}) - (\beta+ \alpha\epsilon) n
\end{align} where, in the last inequality, we have again invoked the $|{\Gamma}_{\delta}({\Lambda}) - {\Gamma}({\Lambda})| \leq \epsilon n$ property. The proof is complete upon combining \eqref{ineq:approx10} and \eqref{ineq:approx20} to get that ${\Gamma}({\Lambda}'_{\delta}) \geq \alpha{\Gamma}({\Lambda}) - (\beta+ (\alpha+1)\epsilon) n$.

\subsection{Proof of Lemma \ref{lem:sample_loss_bound}: Sub-sampling the node set}\label{app:proof:lem:sample_loss_bound}
Fix a seed set $S$ and let \begin{align}
    \frac{(2+\epsilon)(\delta'+\log2 )}{2 n \epsilon^2 } \leq \rho \leq 1.
\end{align} We use a Hoeffding-Bernstein bound to claim that with probability at least $1-e^{-\delta'}$ we have 
\begin{align}
\left|\Gamma_{\rho}(\mathcal{S}) - \Gamma(\mathcal{S})\right| \leq  \epsilon n. 
\label{eq:hoeffding}  
\end{align} Let $X_v$ be the random variable that is zero if $v$ is not in ${\mathcal{V}}_{\rho}$ and $\phi(v,\mathcal{S})$ otherwise. Consider their summation and note that 
\begin{align}\sum_{v\in\mathcal{V}}X_v  = \sum_{v \in {\mathcal{V}}_{\rho}} \phi(v,\mathcal{S}) = \rho \Gamma_{\rho}\left(\mathcal{S}\right).
\end{align}

Hoeffding-Bernstein inequality \cite[Lemma 2.14.19]{VaartWellner} provides that
\begin{align}
\mathbb{P}\left[\left|\frac{1}{n\rho}\sum_{v \in {\mathcal{V}}_{\rho}} \phi(v,\mathcal{S}) - \frac{1}{n}\Gamma(\mathcal{S})\right| \geq \epsilon\right] &=
\mathbb{P}\left[\left|\Gamma_{\rho}(\mathcal{S}) - \Gamma(\mathcal{S})\right| \geq \epsilon n \right]\\& \leq 2\exp\left(-\frac{2n\rho{\epsilon}^2}{2\sigma^2_n + \epsilon\Delta_n}\right), \label{eq:hoefdding_bernstein_bound}
\end{align}
where $\Delta_n = \max_{v \in \mathcal{V}}\phi(v,\mathcal{S}) - \min_{v \in \mathcal{V}}\phi(v,\mathcal{S}) \leq 1$ and \begin{align}
\sigma^2_n &= \frac{1}{n}\sum_{v\in\mathcal{V}}\left(\phi(v,\mathcal{S}) - \frac{1}{n}\Gamma(\mathcal{S})\right)^2 \\&= \frac{1}{n}\sum_{v\in\mathcal{V}}\phi(v,\mathcal{S})^2 - \left(\frac{1}{n}\Gamma(\mathcal{S})\right)^2 \\ & \leq \frac{1}{n}\sum_{v\in\mathcal{V}}\phi(v,\mathcal{S}) - \left(\frac{1}{n}\Gamma(\mathcal{S})\right)^2 \\&= \ell - \ell^2  \leq \ell \leq 1.  
\end{align} In the last equality, we used the notation $\ell := ({1}/{n})\sum_{v\in\mathcal{V}}\phi(v,\mathcal{S}) = ({1}/{n})\Gamma(\mathcal{S})$. The bound in \eqref{eq:hoefdding_bernstein_bound} subsequently simplifies
\begin{align}
\mathbb{P}\left[\left|\Gamma_{\rho}(\mathcal{S}) - \Gamma(\mathcal{S})\right| \geq \epsilon n \right]\leq 2\exp\left(-\frac{2n\rho{\epsilon}^2}{2 + \epsilon}\right).
\end{align} Using  $n\rho \geq {(2+\epsilon)(\delta'+\log 2)}/{2{\epsilon}^2}$,  we get that for all $\delta' > 0$
\begin{align}
\mathbb{P}\left[\left|\Gamma_{\rho}(\mathcal{S}) - \Gamma(\mathcal{S})\right| \geq \epsilon n \right] & \leq 2\exp(-(\delta' + \log 2)) = e^{-\delta'}.
\label{ineq:before_union_bound}
\end{align}

To complete the proof we use a union bound to claim that \eqref{eq:hoeffding} holds for all choices of the seed set $\mathcal{S}$ simultaneously. To claim a union bound over all $\binom{n}{k}$ choices of the seed sets $\mathcal{S}$, it suffices to choose $\delta' = k\delta\log n$ in \eqref{ineq:before_union_bound}. 

\subsection{Proof of Lemma \ref{lem:probing_chernoff_bound}: Probing the extended neighborhoods}\label{app:proof:lem:probing_chernoff_bound}

We begin by considering a fixed $\mathcal{S}\subset\mathcal{V}$. Recall ${\Gamma}^{(T)}_{\rho}(\mathcal{S}) = (1/\rho) \sum_{v\in\mathcal{V}_{\rho}}{\phi}^{(T)}(v,\mathcal{S})$ and ${\phi}^{(T)}(v,\mathcal{S}) = ({1}/{T})\sum_{i=1}^{T}Y^{(i)}(v,\mathcal{S})$ for $v\in\mathcal{V}_{\rho}$. When $v\in\mathcal{V}_{\rho}$ is fixed, the Bernoulli variables $Y^{(1)}(v,\mathcal{S}), \ldots, Y^{(T)}(v,\mathcal{S})$ are independent and identically distributed with mean ${\phi}(v,\mathcal{S})$. By Chernoff bound to ${\phi}^{(T)}(v,\mathcal{S})$,  we get that:
\begin{align}
\mathbb{P}\left[\left|{\phi}^{(T)}(v,\mathcal{S}) - {\phi}(v,\mathcal{S}) \right| > \epsilon\right]  \leq 2\exp(-{\epsilon^2 T}/{3}), \mbox{ for any fixed }  v\in\mathcal{V}_{\rho}.
\end{align} Using  $T = T_{\epsilon,\delta}^{n,k}$, by union bound over the choice of $\binom{n}{k}$ seed sets $\mathcal{S}\subset\mathcal{V}$, $\mbox{card}(\mathcal{S})=k$,  and  $n$ nodes $v \in \mathcal{V}$, we obtain that:
\begin{align}
\mathbb{P}\left[\left|{\phi}^{(T)}(v,\mathcal{S}) - {\phi}(v,\mathcal{S}) \right| > \epsilon,  \mbox{ for all  $\mathcal{S}$ and  $v$}\right]  \leq 2\exp(-\delta-\log 2)=e^{-\delta}.
\end{align} The proof is complete upon considering the summation over $v\in\mathcal{V}_{\rho}$:
\begin{align}
&\mathbb{P}\left[\left|\Gamma^{(T)}_{\rho}(\mathcal{S}) - \Gamma_{\rho}(\mathcal{S})\right| \leq \epsilon n, \mbox{ for all  $\mathcal{S}$}\right] = \\ 
&\mathbb{P}\left[\left|\sum_{v\in\mathcal{V}_{\rho}}{\phi}^{(T)}(v,\mathcal{S}) - \sum_{v\in\mathcal{V}_{\rho}}{\phi}(v,\mathcal{S}) \right| \leq \epsilon n \rho ,  \mbox{ for all  $\mathcal{S}$} \right] \geq  \\   
& \mathbb{P}\left[\left|{\phi}^{(T)}(v,\mathcal{S}) - {\phi}(v,\mathcal{S}) \right| \leq \epsilon,  \mbox{ for all  $\mathcal{S}$ and  $v$}\right]  \geq  1 - e^{-\delta}.
\end{align}

\subsection{Proof of Lemma \ref{lem:Limiting_neighborhood}: Limiting the size of the probed neighborhoods}\label{app:proof:lem:Limiting_neighborhood} 

Following the notation in Definition \ref{def:kIM}, let us use ${\Lambda}^{(T)}_{\rho}$ and ${{\opt}^{(T)}_{\rho}}$ to denote the maximizer of ${\Gamma}^{(T)}_{\rho} $ and its maximal value subject to the size constraint: $\mbox{card}({\Lambda}^{(T)}_{\rho}) = k$. Similarly, let us denote the optimal solution to $k$-IM on ${\Gamma}^{(T)}_{\rho,\tau}$ and its value by ${\Lambda}_{\rho,\tau}^{(T)}$ and ${\opt}^{(T)}_{\rho,\tau}$, respectively. Moreover, following Definition \ref{def:alphaAPPROX}, let us use $\Lambda_{\rho}^{\alpha,(T)}$ and $\Lambda_{\rho,\tau}^{\alpha,(T)}$ to denote the $\alpha$-approximate solutions to $k$-IM on ${\Gamma}^{(T)}_{\rho} $ and ${\Gamma}^{(T)}_{\rho,\tau}$, respectively. Our goal is to show that for $\tau = \tau^{n,k}_{\epsilon}$ any $\Lambda^{\alpha,(T)}_{\rho,\tau}$ is also $\Lambda^{\alpha(1-\epsilon),(T)}_{\rho}$.

It is useful to think of  $\mathcal{G}^{(1)}_{\rho,\tau}, \ldots, \mathcal{G}^{(T)}_{\rho,\tau}$ as subgraphs of  $\mathcal{G}^{(1)}_{\rho}, \ldots, \mathcal{G}^{(T)}_{\rho}$. An immediate consequence of this observation is that for any set of nodes $\mathcal{S}$, we have $\Gamma^{(T)}_{\rho}\left(\mathcal{S}\right)  \geq \Gamma^{(T)}_{\rho,\tau}\left(\mathcal{S}\right)$. We call the imaginary process whereby $\mathcal{G}^{(i)}_{\rho,\tau}$ is obtained after removing some nodes and edges from $\mathcal{G}^{(i)}_{\rho}$ a $\tau$-cutting, and subsequently, we refer to $\mathcal{G}^{(i)}_{\rho,\tau}$  and $\mathcal{G}^{(i)}_{\rho}$ as the cut and uncut copies, respectively. Finally, it is also useful to define ${\phi}^{(T)}_{\tau}(v,\mathcal{S})$ in the exact same way as \eqref{eq:phi_T} but using the $\tau$-cut copies $\mathcal{G}^{(1)}_{\rho,\tau}, \ldots, \mathcal{G}^{(T)}_{\rho,\tau}$. 

The proof follows \cite[Lemma 2.4]{Bateni:2017:AOS:3087556.3087585} closely. In particular, we first note that it suffices to show the existence of a set $\mathcal{L}$, $\mbox{card}(\mathcal{L}) = k$ satisfying $\Gamma^{(T)}_{\rho,\tau}(\mathcal{L}) \geq (1-\epsilon){{\opt}^{(T)}_{\rho}}$. Because if there exists such a set $\mathcal{L}$, then for any $\alpha$-approximate solution $\Lambda^{\alpha,(T) }_{\rho,\tau}$ we can write (recall $\Gamma^{(T)}_{\rho}\left(\mathcal{S}\right)  \geq \Gamma^{(T)}_{\rho,\tau}\left(\mathcal{S}\right)$ for all ${\mathcal{S}}$):
\begin{align}
    \Gamma^{(T)}_{\rho}\left(\Lambda^{\alpha,(T)}_{\rho,\tau}\right) & \geq \Gamma^{(T)}_{\rho,\tau}\left(\Lambda^{\alpha,(T)}_{\rho,\tau}\right)  \geq  \alpha\Gamma^{(T)}_{\rho,\tau}\left({\Lambda^{(T)}_{\rho,\tau}}\right)  \geq  \alpha\Gamma^{(T)}_{\rho,\tau}\left(\mathcal{L}\right)  \geq  (1-\epsilon)\alpha {{\opt}^{(T)}_{\rho}},
\end{align} implying that $\Lambda^{\alpha,(T)}_{\rho,\tau}$ is also $\Lambda^{\alpha(1-\epsilon),(T)}_{\rho}$.  To show the existence of such a set $\mathcal{L}$ we use a probabilistic argument by constructing a random set $\mathbf{L}$, satisfying $\mathbb{E}\left\{\Gamma^{(T)}_{\rho,\tau}\left(\mathbf{L}\right)\right\} \geq (1-\epsilon){\opt}^{(T)}_{\rho}$. The set $\mathbf{L}$ is constructed is as follows: Starting from $\Lambda^{(T)}_{\rho}$, remove $\epsilon k$ of its nodes randomly, and replace them with $\epsilon k$ nodes chosen uniformly at random from $\mathcal{V}$. To see why $\mathbb{E}\left\{\Gamma^{(T)}_{\rho,\tau}\left(\mathbf{L}\right)\right\} \geq (1-\epsilon){{\opt}^{(T)}_{\rho}}$, consider
\begin{align}
{{\opt}^{(T)}_{\rho}}  &=  \frac{1}{\rho}\sum_{v \in \mathcal{V}_{\rho} }{\phi}^{(T)}\left(v,{\Lambda^{(T)}_{\rho}}\right), \mbox{ and  } 
\mathbb{E}\left\{\Gamma^{(T)}_{\rho,\tau}\left(\mathbf{L}\right)\right\} = \sum_{v \in \mathcal{V}_{\rho} }\mathbb{E}\left\{{\phi}^{(T)}_{\tau}\left(v,\mathbf{L}\right) \right\}.   
\end{align}  The inequality, $\mathbb{E}\left\{\Gamma^{(T)}_{\rho,\tau}\left(\mathbf{L}\right)\right\} \geq (1-\epsilon){{\opt}^{(T)}_{\rho}}$, would follow if for any node $v \in \mathcal{V}$ we have,
\begin{align}\mathbb{E}\left\{{\phi}^{(T)}_{\tau}\left(v,\mathbf{L}\right) \right\} \geq (1-\epsilon){\phi}^{(T)}(v,{\Lambda^{(T)}_{\rho}}) + \epsilon \geq {\phi}^{(T)}(v,{\Lambda^{(T)}_{\rho}}). 
\end{align}
It only remains to verify the truth of the former inequality, $\mathbb{E}\left\{{\phi}^{(T)}_{\tau}\left(v,\mathbf{L}\right) \right\} \geq (1-\epsilon){\phi}^{(T)}(v,{\Lambda^{(T)}_{\rho}}) + \epsilon$. First note that  $\mathbb{E}\left\{{\phi}^{(T)}_{\tau}\left(v,\mathbf{L}\right)\right\}$ represents the probability of node $v$ being connected to one of the nodes in the random set $\mathbf{L}$ averaged over the $T$ subsampled graphs ($\mathcal{G}^{(1)}_{\rho,\tau}, \ldots, \mathcal{G}^{(T)}_{\rho,\tau}$). Consider each of the $T$ copies in our uncut sketch, $\mathcal{G}^{(1)}_{\rho}, \ldots, \mathcal{G}^{(T)}_{\rho}$, and the connections between node $v$ and the optimal set $\Lambda_{\rho}^{(T)}$ in these uncut copies. If these connections remain unchanged in the $\tau$-cut copies $\mathcal{G}^{(1)}_{\rho,\tau}, \ldots, \mathcal{G}^{(T)}_{\rho,\tau}$, then with probability at least $(1-\epsilon)$ they remain unchanged after $\epsilon k$ nodes in ${\Lambda^{(T)}_{\rho}}$ are randomly replaced. If, however, any of these connections are affected by the $\tau$-cutting, then this is an indication that $v$ belongs to a connected component of size $\tau_{\epsilon}^{n,k}$. This connected component is large enough to contain one of the $\epsilon k$ random nodes of $\mathbf{L}$ with probability at least $\epsilon$. Indeed, the probability that none of the $\tau_{\epsilon}^{n,k} = \lceil{n \log{(1/\epsilon)}}/{(\epsilon k)}\rceil$ nodes is chosen is upper-bounded by $\epsilon$:
 \begin{align}
 \left(1 - \frac{\tau_{\epsilon}^{n,k}}{n}\right)^{\epsilon k} \leq \left(1 - \frac{\log (1/\epsilon)}{\epsilon k}\right)^{\epsilon k} \leq e^{-\log(1/\epsilon)}=\epsilon. 
 \end{align}

\subsection{Proof of Lemma \ref{lem:bounding_approximation_loss}: Total approximation loss from subsampling and limited probing} \label{app:proof:lem:bounding_approximation_loss}

Following the notation in the proof of Lemma \ref{lem:Limiting_neighborhood} (Appendix \ref{app:proof:lem:Limiting_neighborhood}), consider any $\Lambda^{\alpha,(T)}_{\rho,\tau}$. Lemma \ref{lem:Limiting_neighborhood} implies that $\Lambda^{\alpha,(T)}_{\rho,\tau}$ is also $\Lambda^{(1-\epsilon)\alpha,(T)}_{\rho}$ because the loss in approximation factor from limited probing is at most $(1-\epsilon)$ when $\tau = \tau^{n,k}_{\epsilon}$. Next note that Lemma \ref{lem:probing_chernoff_bound}, together with Lemma \ref{lem:approx_tech}, implies that for $T= T_{\epsilon,\delta}^{n,k}$ with probability at least $1 - e^{-\delta}$, the value of $\Lambda^{(1-\epsilon)\alpha,(T)}_{\rho}$ for $\Gamma_{\rho}$ can be lower-bounded as follows: $\Gamma_{\rho}(\Lambda^{(1-\epsilon)\alpha,(T)}_{\rho}) \geq (1-\epsilon)\alpha {\opt}_{\rho} - ((1-\epsilon)\alpha+1)\epsilon n$. Finally, another application of Lemma \ref{lem:approx_tech} with Lemma \ref{lem:sample_loss_bound} yields that with at least $1 - e^{-\delta}$ probability, $\Gamma(\Lambda^{(1-\epsilon)\alpha,(T)}_{\rho}) \geq (1-\epsilon)\alpha {\opt} - 2((1-\epsilon)\alpha+1)\epsilon n$. The proof is complete upon combining the preceding statements to get that, with total probability at least $1 - 2e^{-\delta}$, $\Gamma(\Lambda^{\alpha,(T)}_{\rho,\tau}) \geq (1-\epsilon)\alpha {\opt} - 2((1-\epsilon)\alpha+1)\epsilon n$.   

\subsection{Proof of Theorem \ref{thm:total_cost_bound}: Upper-bounding the total number of edges queries}\label{app:proof:thm:total_cost_bound}
 To begin, we bound the total number of edges used in our sketch, i.e., the $T$ subsampled graphs ($\mathcal{G}^{(1)}_{\rho,\tau}, \ldots, \mathcal{G}^{(T)}_{\rho,\tau}$). Let us also denote the set of all edges that appear in our sketch by $\mathcal{E}_{T}$. Fix a choice of $\tau$ nodes in one of the subsampled graphs. Let $\mathbf{X}$ be the number of edges between these $\tau$ nodes. Note that $\mathbf{X}$ is a random variable and its distribution is fixed by PROBE$(\rho,T,\tau,p)$. Using the Chernoff upper tail \revision{bound} and the fact that $\mathbb{E}\left[\mathbf{X}\right] \leq p \binom{\tau}{2}$, we can upper-bound $\mathbf{X}$  as follows: 
\begin{align}
    \mathbb{P}\left[\mathbf{X} \geq p \tau(\tau-1)/2 + \delta'  \sqrt{p \tau(\tau-1)/2} \right]  &\leq \mathbb{P}\left[\mathbf{X} \geq  \mathbb{E}\left[\mathbf{X}\right] + \delta'  \sqrt{\mathbb{E}\left[\mathbf{X}\right]} \right]  \\  & \leq  e^{-{\delta'}^2/4}\cdot
    \label{ineq:chernoff_bound_edge_limits}
\end{align} Recall from \eqref{eq:Enkep} that ${E}^{n,k}_{\epsilon,p}  =  p{\tau_{\epsilon}^{n,k}(\tau_{\epsilon}^{n,k} - 1)}/2 = p \binom{\tau}{2}$ with $\tau = \tau_{\epsilon}^{n,k}$. Setting $\tau = \tau_{\epsilon}^{n,k}$ and  
\begin{align}
\delta' = 2\sqrt{(\tau_{\epsilon}^{n,k}\log{n}+\log{T})\delta}  \geq 2\sqrt{\delta\log \color{green}\left(\normalcolor  T\binom{n}{\tau_{\epsilon}^{n,k}} \color{green} \right) \normalcolor}\ccomma
\end{align} in \eqref{ineq:chernoff_bound_edge_limits} is enough to ensure that, by union bound, with probability at least $1 - e^{-\delta}$ for any subset of size $\tau_{\epsilon}^{n,k}$ in all of the $T$ subsampled graphs, we have $\mathbf{X}  \leq \overline{X}$, where
\begin{align}
    \overline{X} :=  {E}^{n,k}_{\epsilon,p} + \sqrt{\delta(\tau_{\epsilon}^{n,k}\log{n}+\log{T}){E}^{n,k}_{\epsilon,p}}\cdot 
    \label{eq:single_component_edge_upper_bound}
\end{align} Next note that starting from any of the $n\rho$ nodes in $\mathcal{V}_{\rho}$ we never hit more than $\tau$ nodes following the limited probing procedure --- see the ``while loop'' condition in step $8$ of the PROBE$(\rho,T,\tau,p)$ algorithm. Moreover, the only way for the connected component of one of the $n\rho$ initial nodes in  $\mathcal{G}^{(i)}_{\rho,\tau}$ to get larger than $\tau$ is if an edge is added which combines two connected components each containing at least one node in $\mathcal{V}_{\rho}$. Upon inclusion of any such edge in $\mathcal{G}^{(i)}_{\rho,\tau}$, no further edges will be added to the corresponding connected component because of the ``while loop'' condition in step $8$ of PROBE. Hence, there could be at most $n{\rho}$ such edges in each of the subsampled graphs, and in total there are at most  $n{\rho}T$ such edges in our sketch. Upon removing all such edges, any connected component in the remainder of our sketch will have size at most $\tau$ and their total number is always less than $n{\rho}T$. Given \eqref{eq:single_component_edge_upper_bound}, we can bound the total number of edges in our sketch by $n{\rho}T + n{\rho}T{\overline{X}}$. More precisely, with probability at least $1 - e^{-\delta}$, we have: 
\begin{align}
    \mbox{card}(\mathcal{E}_T) \leq n \rho T (1 + \overline{X}) = C^{n,k}_{\epsilon,\delta}\ccomma
    \label{eq:visible_edge_query_bound}
\end{align} where $\rho = \rho^{n,k}_{\epsilon,\delta}$, $T = T_{\epsilon,\delta}^{n,k}$, $\tau = \tau_{\epsilon}^{n,k}$, and $C^{n,k}_{\epsilon,\delta}$ is defined in \eqref{eq:Enkep}.

Note that during the construction of $\mathcal{G}^{(i)}_{\rho,\tau}$, if the revealed neighbor ($\nu_{\iota}$) in step $10$ of the PROBE algorithm is previously probed, then the following ``if statement'' in step $11$ prevents the queried edge from being added to $\mathcal{G}^{(i)}_{\rho,\tau}$. Such edges are queried but not added to $\mathcal{G}^{(i)}_{\rho,\tau}$ because they have already got their chance of appearing in $\mathcal{G}^{(i)}_{\rho,\tau}$ once (during the probing of $\nu_{\iota}$). We can bound the number of such edges in each copy as follows. Let $A^{(i)}_e$ be the indicator variable for the event that both nodes incident to edge $e$ are probed; let $B^{(i)}_e$ be the indicator that edge $e$ is queried on its second chance, i.e., when the second of the two nodes incident to $e$ is probed. Finally, let $C^{(i)}_{e}$ be the indicator that edge $e$ is queried when the second of its two incident nodes is probed,  conditioned on both of its incident nodes being probed (i.e., $B^{(i)}_e$ conditioned on $A^{(i)}_e = 1$).  The edges $e$ for which $B^{(i)}_e = 1$, are those which are queried but do not appear in $\mathcal{G}^{(i)}_{\rho,\tau}$. In \eqref{eq:visible_edge_query_bound} we bound the total number of edges belonging to $\mathcal{E}_T$, i.e., the edges that are queried and appear in one or more of the $T$ subsampled graphs. Our next goal is to provide a complementary bound on $\sum_{i}\sum_{e}B^{(i)}_e$, thus controlling the total number of edge queries.

We begin by noting that $B^{(i)}_e = \sum_e A^{(i)}_e C^{(i)}_e$. The indicator variables $C^{(i)}_e, e \in \mathcal{E}$, are i.i.d. Bernoulli variables with success probability $p$. Using the Chernoff upper tail bound, conditioned on the realizations of $A^{(i)}_e$ for all $e \in \mathcal{E}$, we have:  
\begin{align}
    &\mathbb{P}\left[\sum_e A^{(i)}_e C^{(i)}_e \geq p \sum_e A^{(i)}_e +  2 n\sqrt{\delta + \log T} \right] \\ & \leq   \exp\left(-\frac{4 n^2 \left(\delta + \log T\right)}{2\left(p\sum_e A^{(i)}_e + n\sqrt{\delta + \log T}\right)}\right) \\ & \leq \exp\left(- \delta - \log T\right) =  \frac{1}{T}e^{- \delta},
    \label{ineq:chernoff_bound}
\end{align} where in the last inequality we have used $p\sum_e A^{(i)}_e \leq n^2$ and $\sqrt{\delta + \log T} \leq n$. Union bound over $i = 1 , \ldots , T$ provides that with probability at least $1 - e^{-\delta}$,  for all $i$:
\begin{align}
\sum_e B^{(i)}_e = \sum_e A^{(i)}_e C^{(i)}_e &\leq p \sum_e A^{(i)}_e +  2 n\sqrt{\delta + \log T}.    
\label{ineq:lowerbound_0}
\end{align}

To proceed, for any edge $e$, let $D^{(i)}_e$ be the indicator of the event that edge $e$ gets at least one chance to appear in $\mathcal{G}^{(i)}_{\rho,\tau}$, i.e., at least one of the nodes incident to $e$ are probed. Note that, by definition, $A^{(i)}_e \leq D^{(i)}_e$ for all $i$ and $e$; hence, replacing in \eqref{ineq:lowerbound_0} yields:  
\begin{align}
\sum_e B^{(i)}_e \leq p \sum_e D^{(i)}_e +  2 n\sqrt{\delta + \log T},    
\label{ineq:lowerbound}
\end{align} with probability at least $1 - e^{-\delta}$ for all $i$. In the next step, let $E^{(i)}_e$ be the indicator of the event that edge $e$ is reported on its first chance --- i.e., the first time that one of its incident nodes is probed.  Note that $E^{(i)}_e = 1$ , $e \in \mathcal{E}$, are those edges which are queried and appear in $\mathcal{G}^{(i)}_{\rho,\tau}$. Hence, from \eqref{eq:visible_edge_query_bound} we have: 
\begin{align}
\sum_{i}\sum_{e}E^{(i)}_e =\mbox{card}(\mathcal{E}_T) \leq C^{n,k}_{\epsilon,\delta},
\label{ineq:edge_bound_revisited}
\end{align}  with probability at least $1 - e^{-\delta}$. Finally, let $F^{(i)}_e$ be the indicator of the event that edge $e$ is reported on its first chance, conditioned on at least one of its incident nodes being probed (i.e., $E^{(i)}_e$ conditioned on $D^{(i)}_e = 1$). By definition, $F^{(i)}_e$ are i.i.d. Bernoulli variables with success probability $p$, and $E^{(i)}_e = D^{(i)}_e F^{(i)}_e$.  Similarly to \eqref{ineq:chernoff_bound}, using a Chernoff lower tail bound we can guarantee that, with high probability, $\sum_{e}E^{(i)}_e = \sum_{e}D^{(i)}_e F^{(i)}_e$ is not much smaller than  $p \sum_{e}D^{(i)}_e$. Subsequently, we can upper-bound $\sum_{e}B_{e}^{(i)}$ in \eqref{ineq:lowerbound} in terms of $\sum_{e}E^{(i)}_e$. These details are spelled out next.

Application of the Chernoff lower tail bound to $\sum_{e}E^{(i)}_e = \sum_{e}D^{(i)}_e F^{(i)}_e$, yields:    
\begin{align}
    &\mathbb{P}\left[\sum_{e}E^{(i)}_e = \sum_e D^{(i)}_e F^{(i)}_e \leq p \sum_e D^{(i)}_e -   n\sqrt{2\left(\delta + \log T \right)} \right] \\ & \leq   \exp\left(-\frac{ n^2 \left(\delta + \log T \right)}{p\sum_e D^{(i)}_e}\right)  \leq \exp\left(- \delta - \log T \right) =  \frac{1}{T}e^{- \delta},
    \label{ineq:chernoff_lower_tail_bound}
\end{align} where in the second inequality we use $p\sum_e D^{(i)}_e \leq n^2$. Union bound over $i = 1 , \ldots , T$ provides that with probability at least $1 - e^{-\delta}$, for all $i$: 
\begin{align}
p \sum_e D^{(i)}_e \leq \sum_e E^{(i)}_e  +  n\sqrt{2\left(\delta + \log T\right)}.    
\label{ineq:anotherlowerbound}
\end{align} Combing \eqref{ineq:lowerbound} and \eqref{ineq:anotherlowerbound} and taking the summation over $i =1, \ldots, T$ gives that with probability at least $1-2e^{-\delta}$: 
\begin{align}
\sum_i\sum_e B^{(i)}_e   & \leq \sum_i\sum_{e}E^{(i)}_e +  \left(2+\sqrt{2}\right)T n\sqrt{\delta + \log T}.    
\label{ineq:yetanotherlowerbound}
\end{align} To complete the proof, we combine \eqref{ineq:edge_bound_revisited} and \eqref{ineq:yetanotherlowerbound} to get the claimed upper bound on the total number of edge queries:
\begin{align}
    q = \sum_i\sum_e B^{(i)}_e + \mbox{card}(\mathcal{E}_T) \leq 2C^{n,k}_{\epsilon,\delta}  +  \left(2+\sqrt{2}\right)T_{\epsilon,\delta}^{n,k} n\sqrt{\delta + \log T_{\epsilon,\delta}^{n,k}},
\end{align} with probability at least $1 - 3e^{-\delta}$.

\subsection{Proof of Theorem \ref{thm:edge_query_complexity}: Lower-bounding the required number of edge queries}\label{app:proof:thm:edge_query_complexity}

For this proof, we expand on our construction in Appendix \ref{app:proof:thm:edge_query_complexity-additive-loss-is-unavoidable}. We set $k=1$ and prove the hardness result for fixed $0< \mu < 1$ and $0\leq\epsilon<\mu^2/18$, while assuming that $3/\mu$, ${1}/{\mu^2}$, and $n\mu^2/9$ are all integers for simplicity, without any loss in generality. We present our hard example for 
\begin{align}
p > \frac{9}{\mu^2} \frac{\log n + c}{n} \text{ and } n > (18e^{\bar{c}}/\mu^2)^4, \label{eq:edgeprobabilitylowerbouand}
\end{align} where $\bar{c}  = c +  \log(\gamma\mu/3)/c$, $c$ is a constant satisfying $c> \max\{5+2\log{2}, 1-\log(\gamma/6), -\log(\gamma\mu/3)\}$, and $0< \gamma < \mu/6$ is a constant that is fixed arbitrarily. Note that $p$ is allowed to vary with $n$ subject to \eqref{eq:edgeprobabilitylowerbouand}. As in Appendix \ref{app:proof:thm:edge_query_complexity-additive-loss-is-unavoidable}, we rely on a modification of a collection of  ${9}/{\mu^2}$ cliques by connecting $3/\mu$ of them at random around a circle. One key difference in our construction here is that rather than connecting each clique to the next one by rewiring a single link (as in Figure \ref{fig:mu_n}), we do so using a collection of  $-p^{-1}\log(\gamma\mu/6)$ edges, where $0< \gamma <\mu/6$ is a fixed constant. Let graph $G$ be the collection of ${9}/{\mu^2}$ isolated cliques, each of size $\mu^2 n/9$. We construct  $G'$, our hard input graph, from $G$ via the following random process: 
\begin{enumerate}
    \item Select ${3}/{\mu}$ cliques at random and label them by $i = 1, 2, 3, \ldots, 3/\mu$.
    \item Select one edge from each of the $3/\mu$ selected cliques uniformly at random. Let $(v_1,u_1),(v_2,u_2),...(v_{3/\mu},u_{3/\mu})$ be the list of the first ${3/\mu}$ selected edges.
    \item Remove $(v_1,u_1),(v_2,u_2),...(v_{3/\mu},u_{3/\mu})$ and replace them by $(u_1,v_2),$ $(u_2,v_3),$ $\dots,$ $(u_{3/\mu-1},v_{3/\mu}),$ $(u_{3/\mu},v_1)$. Note that this process connects all of the selected clusters while preserving the degree distribution (see Figure \ref{fig:mu_n}).
    \item Repeat this process until $-p^{-1}\log(\gamma\mu/6)$ edges are chosen from each clique, $i$, and rewired to connect to the proceeding clique, $i+1$: $1\to2\to 3\to\ldots\to i \to i+1\to\ldots\to3/\mu\to1$.
\end{enumerate}
We refer to these $3/\mu$ cliques in graph $G'$ as {\it rewired cliques}; see Figure \ref{fig:mu_n}. Let $\alg$ be an arbitrary (potentially randomized) algorithm for influence maximization that queries less than $ pC^{\epsilon}_{\mu,\gamma}n^2$ edges, where 
    \begin{align*}
        C^{\epsilon}_{\mu,\gamma} = \frac{-(\mu^6 - 9\mu^4\epsilon)}{14580\log(\gamma\mu/6)} .
    \end{align*} We are interested in the run of $\alg$ on $G'$. Note that with $k=1$ the optimum on $G'$ is to seed one of the $n\mu/3$ nodes in the rewired cliques. The gist of our proof here is the same as the one in Appendix \ref{app:proof:thm:edge_query_complexity-additive-loss-is-unavoidable}: we show that the optimum seed on $G'$ induces an expected spread size of at least $(1-\gamma)n\mu/3$; on the other hand, seeding the output of any algorithm that queries less than $ pC^{\epsilon}_{\mu,\gamma}n^2$ edges of $G'$, induces an expected spread size that is at most $(1-\gamma)\mu^2 n/3 < \mu\opt$.
    
    To begin, note that the active edges on each clique constitute an Erd\H{o}s-R{\'e}nyi random graph with edge probability $p$ on $n\mu^2/9$ nodes, except for the $3/\mu$ rewired cliques. The edge probability, $p$, satisfying \eqref{eq:edgeprobabilitylowerbouand} is large enough to induce a connected random graph on each of the $9/\mu^2 - 3/\mu$ isolated cliques (with high probability as $n \to \infty$). Similarly, since only, $-p^{-1}\log(\gamma\mu/6) \in O(n/\log n)$, of the $\binom{n\mu^2/9}{2}$ edges in each of the $3/\mu$ cliques are randomly chosen and rewired, the induced random graphs on each of the $3/\mu$ rewired cliques are also going to be connected with high probability as $n \to \infty$. Hence, any time that a node in any of the cliques is activated, all of the nodes in that clique are going to be activated. Formally, the probability that a pair of nodes in any of the clique are connected by an active edge can be lower-bounded as follows:
    \begin{align} \frac{9(\log n + c)}{n\mu^2} - \frac{-p^{-1}\log(\gamma\mu/6)}{\binom{n\mu^2/9}{2}} & = \frac{9(\log n + c)}{n\mu^2} + \frac{2(
    9/\mu^2)\log(\gamma\mu/3)}{(\log n + c)(n\mu^2/9 - 1)}  & n> e^c\\  
    & > \frac{9(\log n + c)}{n\mu^2} + \frac{(9/c\mu^2)\log(\gamma\mu/3)}{(n\mu^2/9 - 1)} &  n > n\mu^2/9 - 1 \\  
    & > \frac{9(\log n + c)}{n\mu^2} + \frac{(9/c)\log(\gamma\mu/3)}{n\mu^2} = \frac{ \log n + \bar{c}}{\bar{n}} =: \bar{p}, & 
    \end{align} where $\bar{c}  = c +  \log(\gamma\mu/3)/c$ is a constant term that is corrected for the effect of the $-p^{-1}
    \log(\gamma\mu/6)$ edges that are removed at random from each of the $3/\mu$ rewired cliques and $\bar{n} = n\mu^2/9$ is the size of each clique. Note that by assumption we have $c> -\log(\gamma\mu/3)$ which implies that $c - 1 < \bar{c} < c$. These edges are used to construct the graph $G'$ by connecting the rewired cliques together (Figure \ref{fig:mu_n}). Our next lemma allows us to upper-bound the probability that the active edges on a fixed clique do not constitute a connected graph.
    
    \begin{lemma}\label{lem:RandomGraphConnectivityLemma}
    Let $C$ be a random graph on $\bar{n}$ nodes with edge probability $\bar{p} = (\log n + \bar{c})/\bar{n}$, where $n = 9\bar{n}/\mu^2$, $0<\mu<1$, and $\bar{c} > 2(2+\log{2})$ are fixed constants. Let $\bar{\mathcal{C}}$ be the event that $C$ is not connected. If $n > (18 e^{\bar{c}}/\mu^2)^4 $, then $\mathbb{P}(\bar{\mathcal{C}})< {\mu^2e^{-\bar{c}}}/3$.
    \end{lemma}
    \proof{Proof.} We use a common technique in random graph theory to upper-bound the probability that a random graph on $\bar{n}$ nodes with edge probability $\bar{p}$ is not connected. Following \cite[Theorem 7.3]{bollobas2001random}, $\mathbb{P}(\bar{\mathcal{C}})$ is upper-bounded by the sum of the expected values of the number of the connected components of sizes $j = 1, 2, \ldots, \bar{n}/2$:  
    \begin{align}
        \mathbb{P}(\bar{\mathcal{C}}) < \sum_{j = 1}^{\bar{n}/2} \mathbb{E}(X_j), \label{eq:connectivitybouand}
    \end{align} where $X_j$ is a random variable that counts the number of connected components of size $j$ in a random graph on $\bar{n}$ nodes with edge probability $\bar{p}$. We next bound the different terms in \eqref{eq:connectivitybouand} individually. Starting with $j = 1$, we have: 
    \begin{align}
        \mathbb{E}(X_1) = \bar{n}(1- \bar{p})^{\bar{n} - 1} < \bar{n}\exp(-\bar{p}\bar{n}) = \frac{\mu^2e^{-\bar{c}}}{9} \cdot \label{ineq:term1}
    \end{align} For $j = 2$, we have: 
    \begin{align}
        \mathbb{E}(X_2) &= \binom{\bar{n}}{2}\bar{p}(1- \bar{p})^{2(\bar{n} - 2)} < \bar{p}\bar{n}^2 \exp(-2\bar{p}\bar{n}) \\ 
        & =  \frac{e^{-2\bar{c}}\mu^2}{9}  \frac{ \log n + \bar{c}}{n} & \bar{c} < \log{n} \\ 
        &  < \frac{e^{-2\bar{c}}\mu^2}{9} \frac{ 2\log n}{n} & {2\log{n}}/{n} \leq {2}/{e} < 1 \\ 
        & < \frac{e^{-2\bar{c}}\mu^2}{9} \cdot &   \label{ineq:term2}
    \end{align} For $3 \leq  j \leq \bar{n}/2$, we can bound $\mathbb{E}(X_j)$ by the expected number of trees on $j$ nodes as follows: 
    \begin{align}
        \mathbb{E}(X_j) & \leq & & \binom{\bar{n}}{j} j^{j-2} \bar{p}^{j-1}(1- \bar{p})^{j(\bar{n} - j)} <  \left( {\bar{n} e}/{j}\right)^j j^{j-2} \bar{p}^{j-1}(1- \bar{p})^{j(\bar{n} - j)} &  \\  
        & < & & j^{-2}\exp(j + j\log \bar{n} + (j-1)\log\bar{p} - \bar{p}j(\bar{n} - j)) & j^{-2} < 1 \\ 
        & < & & \exp(j(1 -\bar{c}) -j(1-\mu^2/9)\log{n} + (j-1)\log\bar{p} + j^2\bar{p}) &  j^2\bar{p} < {j(\log n + \bar{c})}/{2} \mbox{ \color{green} for all } {j \leq \bar{n}/2} \\
        & < & & \exp(j(1 -\bar{c}/2 ) - (j/2-\mu^2/9)\log{n} + (j-1)\log((\log n + \bar{c})))   &  \mu^2/9 < 1 \text{  and  } \bar{c} < \log{n} \\
        & < & & \exp(j(1 -\bar{c}/2 ) - (j-1)\log{n}/2 + (j-1)\log(2\log n))) & -\log{2} < 0  \\ 
        & < & & e^{j(1 + \log{2} -\bar{c}/2)} \left({\log n}/{\sqrt{n}}\right)^{j-1} &  2(2+\log{2}) < c-1 < \bar{c} \text{ and } 3 \leq j \\ 
        & < & & e^{-j} {(\log n)^2}/{n}. & 
        \label{ineq:otherterms}
    \end{align} Replacing \eqref{ineq:term1}, \eqref{ineq:term2}, and \eqref{ineq:otherterms} in \eqref{eq:connectivitybouand} and using the fact that $\sum_{j = 1}^{\bar{n}/2}e^{-j}< \frac{e}{e-1} < 2$ we get:
    \begin{align}
        \mathbb{P}(\bar{\mathcal{C}}) & < 2\frac{\mu^2e^{-\bar{c}}}{9} + 2\frac{(\log n)^2}{n} &    \\ 
        &=  2\frac{\mu^2e^{-\bar{c}}}{9} + \frac{(\log n)^2}{n^{3/4}} \frac{2}{n^{1/4}} & \frac{(\log n)^2}{n^{3/4}} < 1 \\ 
        & < 2\frac{\mu^2e^{-\bar{c}}}{9} + \frac{2}{n^{1/4}} &  (18 e^{\bar{c}}/\mu^2)^4  < n \\
        & < \frac{\mu^2e^{-\bar{c}}}{3} \ccomma &   \label{ineq:barPcontrol}
    \end{align} finishing the proof. 
    \endproof
    
    Fix any of the $9/\mu^2$ cliques and let $\bar{P}$ be the probability that it is not connected. Note that this probability is decreasing in the probability of the edges being active and because we lower-bounded the latter by $\bar{p}$, it suffices to upper-bound $\bar{P}$ assuming that the probability of having an active edge between each pair of nodes is $\bar{p}$. Equipped with Lemma \ref{lem:RandomGraphConnectivityLemma}, we know that $\bar{P} < \mu^2e^{-\bar{c}}/3$ if $n > (18 e^{\bar{c}}/\mu^2)^4 $. Next, consider the event that the active edges induce a disconnected graph on at least one of the $9/\mu^2$ cliques and denote this event by $\mathcal{E}_1$. By a union bound over the $9/\mu^2$ cliques, we obtain that $\mathbb{P}(\mathcal{E}_1) \leq (9/\mu^2)\bar{P} < 3e^{-\bar{c}}$, which upon choosing $\bar{c} > -\log(\gamma/6) $ can be upper-bounded by $\gamma/2$:
    \begin{align}
        \mathbb{P}(\mathcal{E}_1) < \frac{\gamma}{2}, \text{ for }  n > (18 e^{\bar{c}}/\mu^2)^4  \text{ and } \bar{c} > -\log(\gamma/6) \cdot \label{ineq:connectovityupperbouand} 
    \end{align}

    We next consider the event that for at least one connected pair of rewired cliques, none of the $-p^{-1}\log(\gamma\mu/6)$ rewired edges between them are active. Let us denote this event by $\mathcal{E}_2$. we upper-bound $\mathbb{P}(\mathcal{E}_2)$ by $\gamma/2$. Consider the $-p^{-1}
    \log(\gamma\mu/3)$ edges connecting the $i$-th rewired clique to the $(i+1$)-th rewired clique and let $A_i$ be the random variable that is the number of active edges among these $-p^{-1}
    \log(\gamma\mu/3)$ edges. Random variable $A_i$ has a Bernoulli distribution with parameters  $-p^{-1}
    \log(\gamma\mu/3)$ and $p$. Subsequently, we can bound the  probability that $A_i = 0$ as follows:
    \begin{align}
            \mathbb{P}(A_i = 0) = (1-p)^{-p^{-1}
    \log(\gamma\mu/6)} < e^{\log(\gamma\mu/6)} = \frac{\gamma\mu}{6} \cdot
    \end{align} By a union bound over the $3/\mu$ rewired cliques, we obtain that, with probability at most $\gamma/2$, there is a pair of rewired cliques that are connected together but none of the $-p^{-1}
    \log(\gamma\mu/6)$ edges that connect them  are active: 
    \begin{align}
        \mathbb{P}(\mathcal{E}_2) < \gamma/2 \cdot \label{ineq:e2upperbound}
    \end{align} 
    Having thus upper-bounded both $\mathbb{P}(\mathcal{E}_1)$ and $\mathbb{P}(\mathcal{E}_2)$ by $\gamma/2$ in \eqref{ineq:connectovityupperbouand} and \eqref{ineq:e2upperbound}, we conclude that $\mathbb{P}(\mathcal{E}_1\cup\mathcal{E}_2) < \mathbb{P}(\mathcal{E}_1) + \mathbb{P}(\mathcal{E}_2) < \gamma$. Subsequently, we can give a $1-\gamma$ lower bound on the probability of the complement event, $\mathcal{E}_1^c\cap\mathcal{E}_2^c$, that the active edges induce a connected graph on each of the cliques and at least one of the $-p^{-1}\log(\gamma\mu/6)$ rewired edges between each pair of connected cliques is active. Hence, with probability at least $1-\gamma$, after seeding one of the $n\mu/3$ nodes on the rewired cliques, the activation is going to spread from one rewired clique to the next. In other words, seeding one of rewired clique nodes activates all of the $n\mu/3$ nodes on the rewired cliques with probability at least $1-\gamma$, and the optimal expected spread size on $G'$ is at least $(1-\gamma)n\mu/3$.
    
    So far we have shown that the optimum on $G'$ is at least $(1-\gamma)n\mu/3$. Next we show that the expected spread size from seeding the output of $\alg$ on $G'$ is less than $(1-\gamma)\mu^2 n/3$, which means that $\alg$ cannot be a $\mu$-approximation algorithm. This implies that there is no $\mu$-approximation algorithm that queries less than $p C^{\epsilon}_{\mu,\gamma}n^2 \in O_{\mu}(pn^2)$ edges as claimed.
    
    We use the run of $\alg$ on $G$ to analyze the run of $\alg$ on $G'$. Note that due to symmetric construction of $G$, $\alg$ seeds one of the nodes of $G$ at random. Observe that the expected spread size of a random seed in $G'$ is upper-bounded as follows: 
    \begin{align}  
        \color{green} \left( \normalcolor 1-\frac{3/\mu}{9/\mu^2} \color{green} \right) \normalcolor \frac{n\mu^2}{9} + \color{green} \left( \normalcolor \frac{3/\mu}{9/\mu^2} \color{green} \right) \normalcolor  \frac{n\mu}{3} & =  \color{green} \left( \normalcolor 2-\frac{\mu}{3} \color{green} \right) \normalcolor  \frac{\mu^2 n}{9} < 2(1-\gamma)\frac{\mu^2 n}{9}\ccomma \label{eq:randomseedupperbound}
    \end{align} where the last inequality follows since $\gamma <\mu/6$. Moreover, note that the runs of $\alg$ on $G$ and $G'$ are the same unless $\alg$ queries one of the positions (i.e., edges) that we rewire in $G$ to construct $G'$. Next we upper-bound the probability that $\alg$ queries one of the rewired edges by ${\mu}/3 - 3\epsilon/\mu$. Using the upper bound in \eqref{eq:randomseedupperbound}, this implies that the expected spread size from seeding the output of $\alg$ on $G'$ is at most ${(1-\gamma)\mu^2 n}/{3} - \epsilon n$, as claimed: 
    \begin{align*}
        \color{green} \left( \normalcolor 1-\frac {\mu} {3} + \frac{3\epsilon}{\mu} \color{green} \right) \normalcolor \frac{2(1-\gamma)\mu^2 n}{9} + \color{green} \left( \normalcolor \frac{\mu}{3} - \frac{3\epsilon}{\mu} \color{green} \right) \normalcolor  \frac{n\mu}{3} & =  \left(3 -2\gamma-\frac{2\mu} {3}(1-\gamma) + \frac{6\epsilon}{\mu}(1-\gamma)\right) \frac{\mu^2 n}{9} - \epsilon n \\ 
        & < (3 -2\gamma-\frac {\mu} {3}(1-\gamma))\frac{\mu^2 n}{9} - \epsilon n\\ 
        & < \frac{(1-\gamma)\mu^2 n}{3} - \epsilon n\ccomma
    \end{align*} where the penultimate is because ${6\epsilon}/\mu < \mu/3$ and the last inequality is because ${\mu}(1-\gamma)/ {3} > \gamma$. To see why, recall the assumptions  $\epsilon < \mu^2/18$ and $\gamma < \mu/6$. In particular, with the latter we have $\gamma < \mu(1 - 1/6)/3 <  \mu(1 - \mu/6)/3 < \mu(1-\gamma)/3$.
    
    To finish the proof, it only remains to verify that $\mu/3 - 3\epsilon/\mu$ is an upper bound on the probability of $\alg$ querying one of the rewired edges of $G'$. To this end, let $\mathcal{A}_i$ be the (possibly random) set of all edges belonging to the $i$-th clique in $G$ that are queried by $\alg$. Conditioned on $\mathcal{A}_i$, consider any of the edges belong to $\mathcal{A}_i$. The probability that this specific edge is rewired when constructing $G'$ from $G$ is at most:
    \begin{align*}
        \frac{-p^{-1}\log(\gamma\mu/6)}{\binom{\mu^2 n/9 }{2}}\cdot 
    \end{align*} Note that this probability is conditioned on the realization of $\mathcal{A}_i$ and is with respect to the random process by which $G'$ is constructed from $G$. Therefore, conditioned on $\mathcal{A}_i$, the probability that a rewired edge belonging to the $i$-th clique is queried by $\alg$ is at most: 
    \begin{align*}
        \mbox{card}(\mathcal{A}_i)\frac{-p^{-1}\log(\gamma\mu/6)}{\binom{\mu^2 n/9 }{2}} \frac{3/\mu}{9/\mu^2}  & \leq \left(\frac{-p n^2 (\mu^6 - 9\mu^4\epsilon)}{14580\log(\gamma\mu/6)}\right) \left( \frac{-\log(\gamma\mu/6)}{p\binom{\mu^2 n/9 }{2}}\right)  \frac{\mu}{3} \\ 
        & <  \frac{n (\mu^2-9\epsilon)}{90(n - 9)}  \frac{\mu}{3} < \frac{\mu^3}{27} - \epsilon\frac{\mu}{3}\cdot  
    \end{align*} where, in the fist inequality, we have used the assumption that $\alg$ queries less than $p{C}^{\epsilon}_{\mu,\gamma}n^2$ edges; therefore, $\mbox{card}(\mathcal{A}_i) \leq p{C}^{\epsilon}_{\mu,\gamma}n^2$ with probability one. In the last inequality, we use the fact that $n/(90(n - 9)) < 1/9$ for $n>10$, and in particular, for $n > (18 e^{\bar{c}}/\mu^2)^4  > 18^4$ in \eqref{eq:edgeprobabilitylowerbouand}. By averaging over all realization of $\mathcal{A}_i$ and a union bound over all of the ${9}/{\mu^2}$ cliques, we can bound the probability that $\alg$ queries any rewired edge by $({9}/{\mu^2})  ({\mu^3}/{27} - \epsilon\mu/3) = {\mu}/{3} - 3\epsilon/\mu$, completing the proof.

\subsection{Proof of Theorem \ref{thm:main}: Approximation guarantees with bounded edge queries} \label{app:proof:thm:main}

Given $\epsilon' > 0$, fix $\epsilon = \epsilon'/7$, $\delta = 2\log n$ and set $\rho = \rho_{\epsilon,\delta}^{n,k}$, $T = T_{\epsilon,\delta}^{n,k}$ and $\tau = \tau_{\epsilon}^{n,k}$ according to Algorithm~\ref{alg:seeding-queries}. First we consider the approximation guarantee of SEED$(\epsilon,k)$. Running Algorithm PROBE$(\rho,T,\tau,p)$, provides access to the submodular function $\Gamma_{\rho,\epsilon}^{(T)}$ which has the $k$-IM optimal solution  $\Lambda_{\rho,\epsilon}^{(T)}$ with value ${\opt}_{\rho,\epsilon}^{(T)}$. Using SEED$(\epsilon,k)$, we obtain an approximate solution to $k$-IM on $\Gamma_{\rho,\epsilon}^{(T)}$.  Call this solution $\Lambda_{\rho,\epsilon}^{\star(T)}$. The analysis of \cite[Theorem 1]{mirzasoleiman2015lazier} implies that  
\begin{align*}\mathbb{E}\left[\Gamma_{\rho,\epsilon}^{(T)}(\Lambda_{\rho,\epsilon}^{\star(T)})\right] 
&\geq (1-1/e-\epsilon)\Gamma_{\rho,\epsilon}^{(T)}(\Lambda_{\rho,\epsilon}^{(T)}) \\ 
& = (1-1/e-\epsilon){\opt}_{\rho,\epsilon}^{(T)}\ccomma
\end{align*} where the expectation is with respect to the randomness of the SEED algorithm. We can combine the claims of Lemma \ref{lem:bounding_approximation_loss} and Theorem \ref{thm:total_cost_bound} with the assumption $n \geq (30/\epsilon')^2 > \sqrt{35/\epsilon'} =  \sqrt{5/\epsilon}$ to guarantee that with probability at least $1 - 5e^{-\delta} = 1 - 5/n^2  \geq  1- \epsilon$  we have 
\begin{align*}
\Gamma(\Lambda_{\rho,\epsilon}^{(T)}) = {\opt}_{\rho,\epsilon}^{(T)} \geq (1-\epsilon){\opt} - 2(2-\epsilon)\epsilon n.    
\end{align*} Thus the expected number of nodes that are covered by the output of SEED algorithm, $\Lambda_{\rho,\epsilon}^{\star(T)}$, can be lower-bounded as follows:
\begin{align*}
\mathbb{E}\left[\Gamma(\Lambda_{\rho,\epsilon}^{\star(T)})\right] & \geq (1-1/e-\epsilon)\mathbb{E}\left[{\opt}_{\rho,\epsilon}^{(T)}\right]  
\\& \geq (1-1/e-\epsilon)(1-\epsilon)((1-\epsilon){\opt} - 2(2-\epsilon){\epsilon}n) 
\\ & \geq (1-1/e-\epsilon)(1-2\epsilon) {\opt} - 4\epsilon n 
\\&  \geq (1-1/e)(1-2\epsilon){\opt} -(\epsilon)(1-2\epsilon){\opt} - 4{\epsilon}n
\\ & \geq (1-1/e) {\opt} - 7{\epsilon}n = (1 - {1}/{e}) {\opt} - {\epsilon'}n,
\end{align*} where the first expectation is with respect to the randomness of both SEED and PROBE, and the second one is with respect to the randomness of PROBE. 

Next we analyze the running times of PROBE$(\rho,T,\tau,p)$ and SEED$(\epsilon,k)$ algorithms. Let us denote the expected number of edge queries by $q$. Using Theorem \ref{thm:total_cost_bound}, we can upper bound $q$ as follows: 
\begin{align}
    q &\leq (1 - 3 e^{-\delta}) Q^{n,k}_{\epsilon,\delta} + 3e^{-\delta}(2 T n^2) \leq  Q^{n,k}_{\epsilon,\delta} + 6 T\\
    &\in {O}_{\epsilon'}(p n^2\log^4 n + \sqrt{k p} n^{1.5}\log^{5.5} n + k n\log^{3.5}{n}), 
    \label{eq:edgeboundtimeconcequences}
\end{align} where $Q^{n,k}_{\epsilon,\delta}$ is given by \eqref{eq:edgequeryupperbound} with $\epsilon = \epsilon'/7$ and $\delta = 2 \log n$; and the $2 T n^2$ upper bound is by probing every node (so that each edge is queried twice). Note that the $n \geq \sqrt{\delta + \log T}$ condition in Theorem \ref{thm:total_cost_bound} is always satisfied for $n \geq (30/\epsilon')^2$ with our choice of $\delta$ and $T$. Because with $\delta = 2\log n$ we have $\log T = \log({3(\delta+\log 2) (k+1)\log n }/{\epsilon^2}) < \log({18 n \log^2 n }/{\epsilon^2}) < 4\log n$ for all $n \geq (30/\epsilon')^2 > 18/\epsilon^2$; therefore, $\sqrt{\delta + \log T} < \sqrt{6\log n} < n$, the later being true for all $n \geq 3$.

Every iteration of the ``while loop'' in step $8$ of the PROBE algorithm leads to an edge query, and therefore, the expected run time of PROBE$(\rho,T,\tau,p)$ is $O(q)$. Note that the expected number of edges in the $T$ subsampled graphs is upper bounded by $q$. Hence, using any typical graph traversal algorithm (such as DFS or BFS), we can identify the connected components of all the $T$ subsampled graphs in a time that is $O(q)$. To compute $\Delta(v|\Lambda^{\star})$, we go over the connected components of $\nu$ in each of the $T$ subsampled graphs and add up their values. These values are pre-computed for each connected component, and hence this operation takes $O(T)$ time. Recall that the value of a connected component is initially set equal to its number of initial nodes (belonging to $\mathcal{V}_{\rho}$). After adding a node $v$ to $\Lambda^{\star}$, we reset the value of the connected components containing $v$ in each of the $T$ subsampled graphs to zero (see steps $10$ and $11$ of the SEED algorithm). This ensures that if we later pick another node from these components we do not double count those initial nodes that are already covered by $v$. This process can be done in $O(T) = O_{\epsilon'}({k\log^2 n})$ time, where $T = T_{\epsilon,\delta}^{n,k}$ in \eqref{eq:edgequeriesparametersetup} with $\epsilon = \epsilon'/7$ and $\delta = 2\log n$. Finally, note that as step $9$ of Algorithm~\ref{alg:seed} is repeated $k$ times, the submodular maximization algorithm that we are using makes no more than $k\cdot\mbox{card}(\mathcal{R}) = n\log(1/\epsilon) = n\log(7/\epsilon') =  O_{\epsilon'}(n)$ queries to the function $\Delta(\cdot|\cdot)$. Combining the preceding bounds with \eqref{eq:edgeboundtimeconcequences} puts the total running time of SEED and PROBE algorithms in $O_{\epsilon'}(p n^2\log^4 n + \sqrt{k p} n^{1.5}\log^{5.5}n + k n\log^{3.5}{n})$ as claimed.

\subsection{Proof of Theorem \ref{thm:hardness_discrepency}: Hardness of approximation with query discrepancy}\label{app:proof:thm:hardness_discrepency} Consider the probe and seed cascade probabilities, and denote their minimum and maximum by $\check{p} := \min\{p,p'\}$ and $\hat{p}:=\max\{p,p'\}$; by assumption, we have $\hat{p}/\check{p} = 1+\delta$. We show when the query and seed cascade probability are different (given by $p$ and $p'$), then for $\epsilon$ satisfying
\begin{align}
\epsilon < \frac{8\mu (\delta-\delta^2)c_{\delta}}{41(2-\mu) + 16(\delta-\delta^2)c_{\delta}} < \frac{\mu}{2} \ccomma \text{  where } c_{\delta} = 1+(\delta - \delta^2)/2, \label{eq:epsiloncondition}
\end{align} there can be no $k$-IM approximation algorithm that guarantees an expected spread size of at least $\mu\opt - \epsilon n$ on any input graph with optimum expected spread size $\opt$. Let $\alg$ be any such $k$-IM approximation algorithm. Define $n_{\epsilon,\mu} := (1-2\epsilon/\mu)n$ and note that under \eqref{eq:epsiloncondition} we have $2\epsilon/\mu < 1$. The input graph for our hard example is comprised of two subgraphs: $\mathcal{G}_1$ and $\mathcal{G}_2$. The former is a clique of size $2\epsilon{n}/\mu$ and the latter is a random graph with edge probability $(1-\delta/2)/(n_{\epsilon,\mu}\check{p})$ on the remaining $n_{\epsilon,\mu}$ nodes. We prove \revision{our} hardness results for $k=1$ under the following specifications:
\begin{align}
    & n > \frac{256}{(\epsilon- 3\mu{e}^{-c} - \delta^2)^2} \ccomma \text{ and } \label{eq:LargeNetConstraint} \\
    & \check{p} = \min\{p,p'\} > \frac{\log n + c}{2\epsilon n/\mu} \ccomma \text{ where }  \label{eq:LargeCascadeProbConstraint} \\
    & c = \max\left\{2(2+\log 2),\log\left(\frac{3\mu}{3 - \delta^2}\right)\right\} \cdot 
\end{align}

First note that the expected spread size when seeding one of the $\mathcal{G}_1$ clique nodes is at most $2\epsilon{n}/\mu$, irrespective of whether the cascade probability is $p$ or $p'$. We use Lemma \ref{lem:RandomGraphConnectivityLemma}, with $\bar{n} = 2\epsilon{n}/\mu$ and $\bar{p} = \check{p}$ satisfying \eqref{eq:LargeCascadeProbConstraint}, to provide a companion lower bound on the expected spread size when seeding $\mathcal{G}_1$. Accordingly, if $n > (\mu{e}^{c}/\epsilon)^4$, then the probability that active edges on $\mathcal{G}_1$ do not form a connected graph is at most $3\mu{e}^{-c}/(2\epsilon)$. Having established this lower bound for $\check{p} = \min\{p,p'\}$, it holds true for either of the two cascade probabilities ($p$ and $p'$) because increasing the cascade probability can only increase the expected spread size.  

So far we have shown that the expected spread size from seeding one of the $\mathcal{G}_1$ nodes is at most $2\epsilon{n}/\mu$ and at least 
\begin{align}
    \left(1- \frac{3\mu{e}^{-c}}{2\epsilon}\right)\frac{2\epsilon n}{\mu} = \frac{2\epsilon n}{\mu} - 3{n}{e}^{-c},
\end{align} both bounds being true irrespective of whether the cascade probability is $p$ or $p'$. We next show that an opposite situation holds in case of $\mathcal{G}_2$. The expected spread size from seeding one of the nodes in $\mathcal{G}_2$ depends critically on the cascade probability and can be much larger or much smaller than the range that we have established for $\mathcal{G}_1$. We use techniques from the analysis of the giant connected component in random graphs \cite[Theorem 5.4]{janson2011random} to lower-bound the expected spread size from seeding $\mathcal{G}_2$ when the cascade probability is $\hat{p}$ and to upper-bound it when the cascade probability is $\check{p}$. Comparing the two bounds with the range that we have established for the expected spread size on $\mathcal{G}_1$ reveals that $\alg$ should necessarily seed $\mathcal{G}_2$ when cascade probability is $\hat{p}$ and $\mathcal{G}_1$ when the cascade probability is $\check{p}$. Therefore, there can be no approximation algorithm that provides a $\mu\opt -\epsilon n$ guarantee while the query and seed cascade probabilities are different. We show this for the small enough $\epsilon$ in \eqref{eq:epsiloncondition}, based on the discrepancy between the query and seed cascade probabilities ($\hat{p}/\check{p} = 1+\delta$).

We begin by lower-bounding the expected spread size from seeding a random node in $\mathcal{G}_2$ when the cascade probability is $\hat{p}$. Note that if the cascade probability is $\hat{p}$, then the active edges after the independent cascade on $\mathcal{G}_2$ constitute a random graph with edge probability:
\begin{align}
    \hat{p}\frac{1-\delta/2}{n_{\epsilon,\mu}\check{p}} = \frac{(1+\delta)(1-\delta/2)}{n_{\epsilon,\mu}} = \frac{c_{\delta}}{n_{\epsilon,\mu}} > \frac{1}{n_{\epsilon,\mu}} \cdot
\end{align} In this case we show that the expected spread size from seeding a random node in $\mathcal{G}_2$ is at least $n_{\epsilon,\delta}/2$, where
\begin{align}
    n_{\epsilon,\delta} := \frac{{16}n_{\epsilon,\mu}(\delta-\delta^2)c_{\delta}}{41} \cdot
\end{align} We do so by upper-bounding the probability that a random node on $\mathcal{G}_2$ belongs to a component of size less than $n_{\epsilon,\delta}$. Let us denote this event by $\mathcal{E}$. To upper-bound $\mathbb{P}(\mathcal{E})$, we consider a random node $\nu$ and search the component containing $\nu$ according to the following procedure: 

\begin{Procedure}\label{proc:gcc}
Starting from $\nu$, initialize the set of discovered nodes to include only $\nu$ and set $X_0 = 1$. At step one ($i=1$), declare all neighbors of $\nu$ to be discovered and set $X_{1}$ equal to the number of newly discovered nodes. Also declare $\nu$ to be explored. At any step $i>1$, choose an unexplored node at random from the set of discovered but unexplored nodes; declare it to be explored; add all of its undiscovered neighbors to the discovered set; and set $X_{i}$ equal to the number of newly discovered nodes at step $i$. 
\end{Procedure}

\begin{fact}\label{fact:gcc}
Consider $\tau \geq 1$ iterations of Procedure \ref{proc:gcc}. The total number of discovered but unexplored nodes after $\tau$ iterations is given by: $\sum_{i=0}^{\tau}X_i - \tau$. Therefore, at any step $\tau$, the size of the connected component containing $\nu$ is greater than $\tau$ if, and only if, $\sum_{i=0}^{\tau}X_i > \tau$; the size of the connected component containing $\nu$ is equal to $\tau$ if, and only if, $\sum_{i=0}^{\tau}X_i = \tau$; and the size of the connected component containing $\nu$ is less than $\tau$ if, and only if, $\sum_{i=0}^{\tau}X_i < \tau$. 
\end{fact}

Fact \ref{fact:gcc} implies that $\mathbb{P}(\mathcal{E}) = \mathbb{P}\left(\textstyle\sum_{i =0}^{n_{\epsilon,\delta}}X_{i} < n_{\epsilon,\delta}\right)$. To upper-bound $\mathbb{P}(\mathcal{E})$, let $X^{-}_i$ be i.i.d. random variables with their common distribution set to:
\begin{align}
    \mbox{Binomial}\left(n_{\epsilon,\mu}-n_{\epsilon,\delta},\frac{c_{\delta}}{n_{\epsilon,\mu}}\right) \cdot
\end{align}
Note that for all $i \leq n_{\epsilon,\delta}$, $X_{i}$ stochastically dominates $X^{-}_{i}$; hence,
\begin{align}
    \mathbb{P}(\mathcal{E}) 
    & = \mathbb{P}\left(\textstyle\sum_{i =0}^{n_{\epsilon,\delta}}X_{i} < n_{\epsilon,\delta}\right) \leq \mathbb{P}\left(\textstyle\sum_{i =0}^{n_{\epsilon,\delta}}X^{-}_i \leq n_{\epsilon,\delta} \right) \cdot \label{eq:giantcomponentprobabilityupperbound} 
\end{align} To ease the notation, define $X^{-} = \textstyle\sum_{i =0}^{n_{\epsilon,\delta}}X^{-}_i$ and note that $X^{-}$ also has a binomial distribution with parameters given by:
\begin{align}
    \mbox{Binomial}\left(n_{\epsilon,\delta}(n_{\epsilon,\mu} - n_{\epsilon,\delta}),\frac{c_{\delta}}{n_{\epsilon,\mu}}\right) \cdot
\end{align} We next point out that, 
\begin{align}
\mathbb{E}[X^{-}] &= n_{\epsilon,\delta}(n_{\epsilon,\mu} - n_{\epsilon,\delta})\frac{c_{\delta}}{n_{\epsilon,\mu}} \\
                  & = n_{\epsilon,\delta}\left(1 - \frac{16}{41}(\delta-\delta^2)c_{\delta}\right)c_{\delta} \\ 
                  & = n_{\epsilon,\delta}\left(1 + \frac{\delta-\delta^2}{2}(1 - \frac{32}{41}c^2_{\delta})\right) \\
                  & > n_{\epsilon,\delta} \ccomma \label{eq:meanupperbound}
\end{align} where the last inequality follows because $0 <$ $\delta - \delta^2$ $\leq 1/4$ which implies $c_{\delta} = 1+(\delta - \delta^2)/2 \leq 9/8$ and ${32c^2_{\delta}}/{41} \leq 81/82 < 1$, for all $0 < \delta < 1$. Combining \eqref{eq:giantcomponentprobabilityupperbound} and \eqref{eq:meanupperbound}, we get:    
\begin{align}
    \mathbb{P}(\mathcal{E}) \leq \mathbb{P}\left(X^{-} \leq n_{\epsilon,\delta} \right) \leq \mathbb{P}\left(X^{-} < \mathbb{E}[X^{-}] \right) < \frac{1}{2}\cdot
\end{align} We have established that the probability of the event $\mathcal{E}$ that the size of the connected component of a random node in $\mathcal{G}_2$ is less than $n_{\epsilon,\delta}$ is at most $1/2$. This implies that the expected spread size from seeding a random node in $\mathcal{G}_2$ is at least  $n_{\epsilon,\delta}/2$. Under \eqref{eq:epsiloncondition}, we have $2\epsilon{n}/\mu < \mu n_{\epsilon,\delta}/2 - \epsilon n $; hence, whenever the cascade probability is $\hat{p}$, $\alg$ should necessarily seed one of the nodes in $\mathcal{G}_2$.
 
 Next, we upper-bound the expected spread size from seeding a node in $\mathcal{G}_2$ when the cascade probability is $\check{p}$. For a fixed node $\nu$ in $\mathcal{G}_2$, let $\bar{\mathcal{E}}$ be the event that the size of the connected component containing $\nu$ is greater than 
 \begin{align}
     \bar{n}_{\epsilon,\delta} := \frac{16\log n_{\epsilon,\mu}}{\delta^2} \cdot
 \end{align} Let $X_i$ be defined in the same way as in Procedure \ref{proc:gcc}, then Fact \ref{fact:gcc} continues to hold and we have: \begin{align}
     \mathbb{P}(\bar{\mathcal{E}}) = \mathbb{P}\left(\textstyle\sum_{i=0}^{\bar{n}_{\epsilon,\delta}} X_i > \bar{n}_{\epsilon,\delta}\right) \cdot
 \end{align} To upper-bound $\mathbb{P}(\bar{\mathcal{E}})$, let $X^{+}_i$ be i.i.d. random variables with common $\mbox{Binomial}(n_{\epsilon,\mu},(1-\delta/2)/n_{\epsilon,\mu})$ distribution and note that for all $i$, $X^{+}_{i}$ stochastically dominates $X_{i}$; hence,
\begin{align}
    \mathbb{P}(\bar{\mathcal{E}}) & = \mathbb{P}\left(\textstyle\sum_{i =0}^{\bar{n}_{\epsilon,\delta}}X_{i} > \bar{n}_{\epsilon,\delta} \right) \\
    & \leq \mathbb{P}\left(\textstyle\sum_{i =0}^{\bar{n}_{\epsilon,\delta}}X^{+}_i > \bar{n}_{\epsilon,\delta} \right)\\ 
    & = \mathbb{P}\left(\textstyle\sum_{i =1}^{\bar{n}_{\epsilon,\delta}}X^{+}_i \geq \bar{n}_{\epsilon,\delta}(1-\delta/2) + \bar{n}_{\epsilon,\delta}\delta/2 \right) \\
    & \leq \exp\left(\frac{-\bar{n}^2_{\epsilon,\delta}\delta^2/4}{2\bar{n}_{\epsilon,\delta}(1-\delta/2) + \bar{n}_{\epsilon,\delta}\delta/3}\right) \\
    & \leq \exp\left(\frac{-\bar{n}_{\epsilon,\delta}\delta^2}{8}\right) = \frac{1}{n_{\epsilon,\mu}^{2}}\ccomma \label{eq:giantcomponentprobabilitylowerbound} 
\end{align} where the penultimate inequality follows by applying \revision{a} Chernoff bound to $\sum_{i =0}^{\bar{n}_{\epsilon,\delta}}X^{+}_i$ which is a binomial random variable with mean $\bar{n}_{\epsilon,\delta}(1-\delta/2)$. By union bound, the probability of having at least one node that belongs to a connected component of size greater than $\bar{n}_{\epsilon,\delta}$ is at most $n_{\epsilon,\mu}\mathbb{P}(\bar{\mathcal{E}}) = 1/n_{\epsilon,\mu}$; or equivalently, with probability at least $1-1/n_{\epsilon,\mu}$, all of the nodes in $\mathcal{G}_2$ belong to connected components of size less than or equal to $\bar{n}_{\epsilon,\delta}$. Hence, when the cascade probability is $\check{p}$, the expected spread size from seeding any of the nodes in $\mathcal{G}_2$ is at most
\begin{align}
    \left(1-\frac{1}{n_{\epsilon,\mu}}\right)\bar{n}_{\epsilon,\delta} + \frac{1}{n_{\epsilon,\mu}}n_{\epsilon,\mu} < \bar{n}_{\epsilon,\delta} + 1.
\end{align} Therefore, whenever cascade probability is $\check{p}$ and 
\begin{align}
    \bar{n}_{\epsilon,\delta} + 1 < \mu (2\epsilon{n}/\mu)(1- 3\mu{e}^{-c}/(2\epsilon)) - \epsilon{n} = (\epsilon- 3\mu{e}^{-c}){n},
\end{align} $\alg$ would necessarily seed one of the nodes in $\mathcal{G}_1$ and achieve $(2\epsilon{n}/\mu)(1- 3\mu{e}^{-c}/(2\epsilon))$ expected spread size. The conditions in \eqref{eq:LargeNetConstraint} are sufficient to ensure $\bar{n}_{\epsilon,\delta} + 1 < (\epsilon- 3\mu{e}^{-c}){n}$. This completes the proof showing that any approximation algorithm $\alg$ seed\revision{s} different nodes for different cascades probabilities on our hard example. Therefore, the approximation guarantees cannot be achieved when the query and seed probability differ under the specified conditions.

\revision{
\subsection{A pruning algorithm to correct for query discrepancy}\label{app:alg:discrepency}
\begin{algorithm}[htb]
    \SetAlgoLined
    \DontPrintSemicolon
    \vspace{5pt}
    \KwIn{$\mathcal{G'}^{(1)}_{\rho,\tau}, \ldots, \mathcal{G'}^{(T)}_{\rho,\tau}$ and $\mathcal{V}_{\rho}$ generated by PROBE$(\rho,T,\tau,p')$, and target cascade probability $p$}
    \KwOut{$\mathcal{G}^{(1)}_{\rho,\tau}, \ldots, \mathcal{G}^{({T})}_{\rho,\tau}$ simulating the output of PROBE$(\rho,T,\tau,p)$}
    \KwRequire $p' > p$. \;
    Set $h = p/p'$.\;
    \For{$i$ \KwFrom $1$ \KwTo $T$}{
        Initialize $\mathcal{G}^{(i)}_{\rho,\tau} \leftarrow \mathcal{G'}^{(i)}_{\rho,\tau}\cdot$ \;
        \For{every edge $e$ in $\mathcal{G}^{(i)}_{\rho,\tau}$}{
            Let $H$ be an independent $\mbox{Bernoulli}$ draw with $\mathbb{P}\left[H=1\right]=h$. \;
            \If{$H = 0$}{
                    Remove $e$ from $\mathcal{G}^{(i)}_{\rho,\tau}$.\;
                    }
        }
        \For{every node $\nu$ in $\mathcal{G}^{(i)}_{\rho,\tau}$}{
            \If{none of the nodes in $\mathcal{V}_{\rho}$ are reachable from $\nu$}{
                    Remove $\nu$ from $\mathcal{G}^{(i)}_{\rho,\tau}$.\;
                    }
        }
        }
    \Return{$\mathcal{G}^{(1)}_{\rho,\tau}, \ldots, \mathcal{G}^{({T})}_{\rho,\tau}\cdot$}
\caption{PRUNE$(p,p')$}
\label{alg:prune}
\end{algorithm}
}

\section{Extensions to other influence models}\label{app:ext-inf-models}

We presented our results for undirected graphs with a homogeneous cascade probability ($p$). In subsection \ref{app:ext}, we present the extension of our results to directed influence graphs. In subsection \ref{app:ext:linth}, we explore other influence models beyond the independent cascade and provide a pathway to perform edge queries in the general class of triggering influence models.   

\subsection{Directed influence graphs}\label{app:ext}

To perform edge queries on directed graphs, in steps $6$ and $7$ of the PROBE algorithm we let $\mathcal{N}_{\nu}$ denote the set of \emph{incoming} neighbors (i.e., nodes that influence $\nu$) such that the edge query in step $10$ reveals the $\iota$-th incoming neighbor of $\nu$. Two directed edges that are between the same pair of nodes in opposite directions ($u \to \nu$ and $\nu \to u$) are distinguished. Therefore, as long as we do not probe a node more than once, each directed edge will get at most one chance of appearing in $\mathcal{G}^{(i)}_{\rho,\tau}$. Subsequently, the ``if statement'' in step $11$ of the PROBE algorithm should be removed when performing edge queries on directed graphs: any queried edge in step $10$ is always added to $\mathcal{G}^{(i)}_{\rho,\tau}$ in the following steps. As a further consequence, our edge query upper bound in the directed case consists entirely of the edges that appear in the sketch (denoted by $\mathcal{E}_{T}$ in Appendix \ref{app:proof:thm:total_cost_bound}) and is given by \eqref{eq:visible_edge_query_bound}: with probability at least $1 - e^{\delta}$ no more than $C^{n,k}_{\epsilon,\delta} \in {O}_{\epsilon,\delta}(p n^2\log^2n + \sqrt{k p} n^{1.5}\log^{2.5}n + {k^2}\log^{2}n)$ edges are queried --- a slight improvement over our edge query upper bound in the \revision{un}directed case.

In the ``while loop'' condition in step $8$ of the PROBE algorithm, instead of considering the size of the connected component of node $\nu$, we count the number of nodes that are reachable via directed paths from $\nu$ (i.e., the size of its realized \emph{cone of influence}). For example, in Figure \ref{fig:directed_edge_queries}(\subref{fig:reversecascadered}), the reachable set for all of the initial nodes is empty, therefore we proceed to probe the incoming neighbors until there are no new nodes to probe. In fact, of all the initial nodes in all three cascades in Figure  \ref{fig:directed_edge_queries}, only the leftmost initial node in Figure \ref{fig:directed_edge_queries}(\subref{fig:reversecascadeorange}) has a non-empty reachable set that is a singleton.

\begin{figure}[htb]
\centering
\begin{subfigure}[b]{0.32\textwidth}
\includegraphics[width=\textwidth]{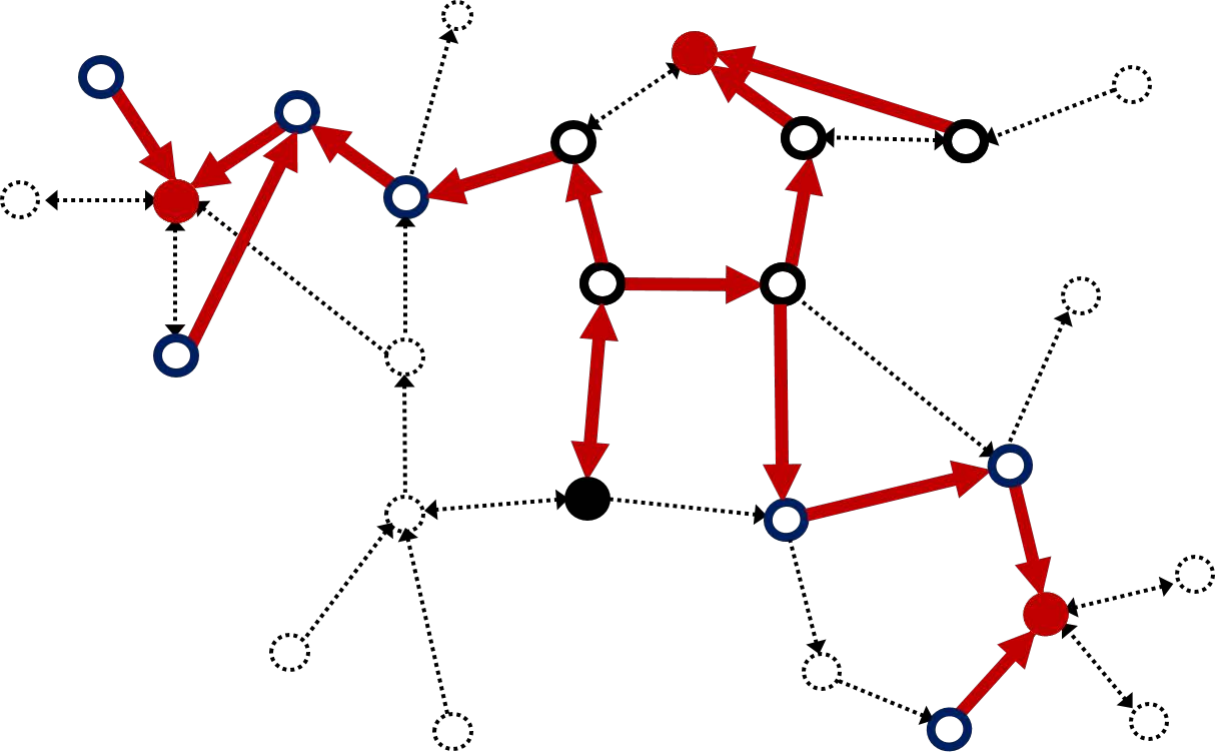}
\caption{ }
\label{fig:reversecascadered}
\end{subfigure}~\begin{subfigure}[b]{0.32\textwidth}
\includegraphics[width=\textwidth]{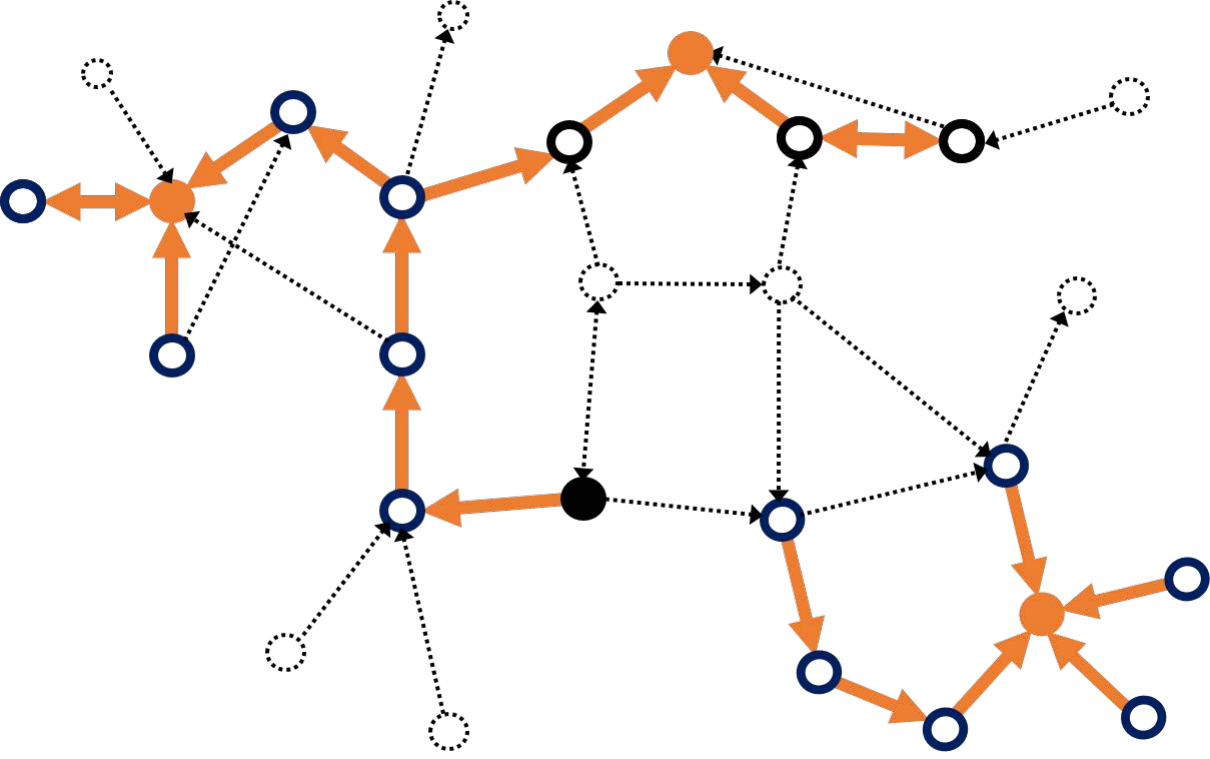}
\caption{ }
\label{fig:reversecascadeorange}
\end{subfigure}~\begin{subfigure}[b]{0.32\textwidth}
\includegraphics[width=\textwidth]{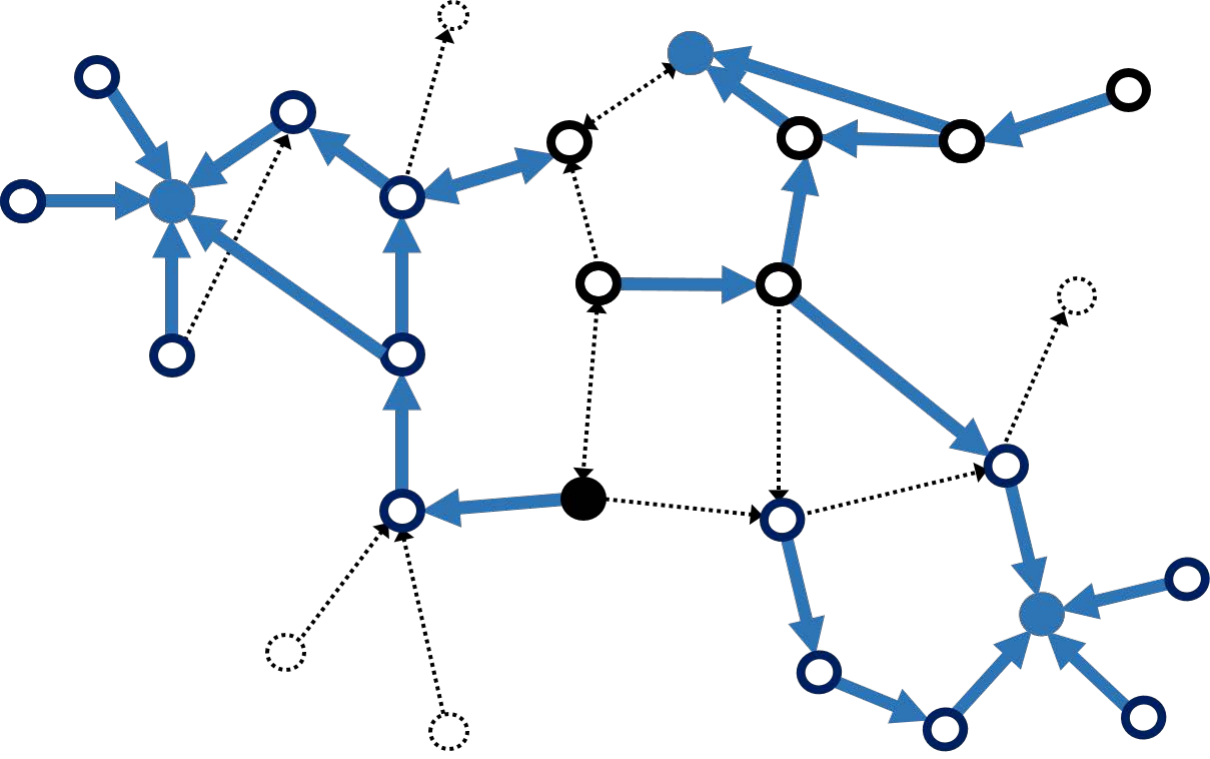}
\caption{ }
\label{fig:reversecascadeblue}
\end{subfigure}
\caption{Three reverse cascades are depicted in (\subref{fig:reversecascadered}) red, (\subref{fig:reversecascadeorange}) orange, and (\subref{fig:reversecascadeblue}) blue. All cascades start from the same random initial nodes which are marked in the same color as the cascades. As the cascades diffuse in reverse, each node reveals its incoming edges at random.  To score the nodes, we count the number of reachable initial nodes in each cascade and add them up. For example, the node that is marked in black scores three in the red cascade, two in the orange cascade, and one in the blue cascade. In total, it scores as high as or higher than other nodes across the three cascades. The unsampled nodes and edges are dotted.}
\label{fig:directed_edge_queries}
\end{figure}

We need to make similar changes to the way candidates are scored in the SEED algorithm. First, in step $3$ of SEED we set the value of each initial node (belonging to $\mathcal{V}_{\rho}$) to be one. When evaluating the marginal increments ($\Delta(v|\Lambda^{\star})$) in step $7$ of SEED, we add the values of the initial nodes that are reachable via directed paths from $v$ in each of the $T$ subsampled graphs ($\mathcal{G}^{(1)}_{\rho,\tau},\ldots, \mathcal{G}^{(T)}_{\rho,\tau}$), rather than summing the value of its connected components (see Figure \ref{fig:directed_edge_queries}). Finally, in step $11$ of SEED if an initial node is reachable from the chosen seed, then we nullify its value for scoring the subsequent candidates.

\begin{figure}[ht]
\centering
\includegraphics[scale=0.5]{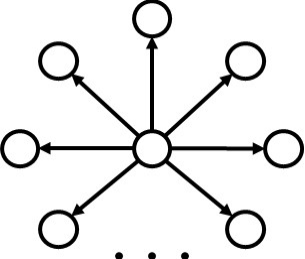}
\caption{With $o(n)$ influence samples, one is unlikely to discover the center of the directed star network. In such a case it is impossible to guarantee a $\mu\opt - \epsilon n$ expected spread size with fewer than $\Omega(n)$ queries.}
\label{fig:star}
\end{figure}

In contrast, finding an approximately optimal seed set by influence sampling in a directed graph is very hard. For example, consider a star graph with $n$ leafs where all of the edges are directed away from the center toward the leafs (see Figure \ref{fig:star}). Assume that the cascade probability on each edge is $1$. In this case, if we seed a leaf we only observe an isolated node and hence we need $\Omega(n)$ influence samples to find the center of the star and seed it. In a situation where running reverse spreads is a plausible way of acquiring network information (e.g., by reverse influence sampling as \citet{borgs2014maximizing} do), our algorithm and proofs continue to hold exactly the same. In particular, we can estimate the marginal increase of a candidate node on the current seed set by counting the number of times that it has appeared in the output of the reverse influence samples without any of the currently chosen seeds (i.e., the number of times that the random initial nodes are reachable from the candidate node but not from any of the currently chosen seeds). Following Algorithm~\ref{alg:sample} and Theorem \ref{thm:main_spreading_queries}, we can collect ${O}_{\epsilon}(k^2\log{n})$ reverse influence samples and choose the best $k$ seeds by approximating the greedy steps.

\subsection{Triggering models} \label{app:ext:linth}

In general, the influence graph may be directed and cascade probabilities may differ in each direction and across the edges. When performing edge queries, the probed nodes should reveal each of their incoming neighbors (influencers), with the cascade probability associated with that edge. Here we explain how a triggering set technique that is proposed by \cite{kempe2015maximizing} helps us devise edge queries in a large class of influence models, including the independent cascade and \emph{linear threshold} models. Recall that the influence function $\Gamma$ maps a seed set to a positive real number that is the (expected) number of adopters under a (randomized) model of diffusion. \cite{kempe2003maximizing,kempe2005influential,kempe2015maximizing} --- through a conjecture that is positively resolved by \cite{doi:10.1137/080714452} --- identify a broad class of threshold models for which the influence function is non-negative, monotone, and submodular. Influence maximization in such models can be solved to within a $(1-1/e)$ approximation guarantee, following a natural greedy node selection algorithm.

In a general threshold model, each node $v$ has an activation function $f_v$ and a threshold $\theta_v \in [0,1]$. The activation function maps subsets of neighbors of $v$ to a real number between zero and one. Node $v$ becomes an adopter at time $t$ if $f_{v}(\mathcal{A}_{v,t-1}) \geq \theta_v$, where $\mathcal{A}_{v,t-1} \subset \mathcal{N}_v$ is the set of all active (adopter) neighbors of node $v$. Approximate influence maximization with deterministic thresholds is known to be very hard --- see, e.g., \citep[Section 3.2]{kempe2015maximizing} and \citep{Schoenebeck:2019:BWA:3323875.3313904}. To avoid the intractable settings, \cite{kempe2003maximizing,kempe2005influential,kempe2015maximizing} consider a randomized model where thresholds are i.i.d. uniform $[0,1]$ variables. \cite{doi:10.1137/080714452} show that if the ``local'' activation functions $f_v$ are submodular, then the ``global'' influence function $\Gamma$ is also submodular and influence maximization can be achieved with strong approximation guarantees \citep[Theorem 1.6 and Corollary 1.7]{doi:10.1137/080714452}.  In the special case of the \emph{linear} threshold model, each node $v$ is influenced by its incoming neighbours $u\in\mathcal{N}_v$ according to their edge weights $b_{uv}$. Node $v$ becomes an adopter at time $t$ if the total weight of her adopting neighbors exceeds her threshold, i.e., if $f_{v}(\mathcal{A}_{v,t-1}) = \sum_{u\in\mathcal{A}_{v,t-1}}b_{uv} \geq \theta_v$. 

At the heart of the proofs of \cite{kempe2003maximizing,kempe2015maximizing} lie a triggering set technique.  Accordingly, each node $v$ chooses a random subset of its incoming neighbors, which we call its triggering set and denote it by $\mathcal{T}_v \subset\mathcal{N}_v$. Node $v$ becomes an adopter at time $t$ if any of the nodes in $\mathcal{T}_v$ is an adopter at time $t-1$. The distribution according to which the triggering sets are drawn is determined by the diffusion model. However, not all diffusion processes can be reduced to a triggering set model. For those that do, their influence function is guaranteed to be submodular \cite[Lemma 4.4]{kempe2015maximizing}. 

For example, in the independent cascade model, the triggering set $\mathcal{T}_v$ includes each of the neighbors $u \in \mathcal{N}_v$ with cascade probability associated with the $u\to v$ edge ($p_{uv}$), independently at random. In the case of the linear threshold model, \cite{kempe2015maximizing} devise the following construction, assuming $\sum_{u\in\mathcal{N}_{v}}b_{uv} \leq 1$: The triggering set $\mathcal{T}_v$ is comprised of a single node or no nodes at all. For $u \in \mathcal{N}_v$, the probability that $\mathcal{T}_v = \{u\}$ is equal to $b_{uv}$, and  $\mathcal{T}_v = \varnothing$ with probability  $1 - \sum_{u\in\mathcal{N}_{v}}b_{uv}$.

\begin{figure}[t]
\centering
\begin{subfigure}[b]{0.32\textwidth}
\includegraphics[width=\textwidth]{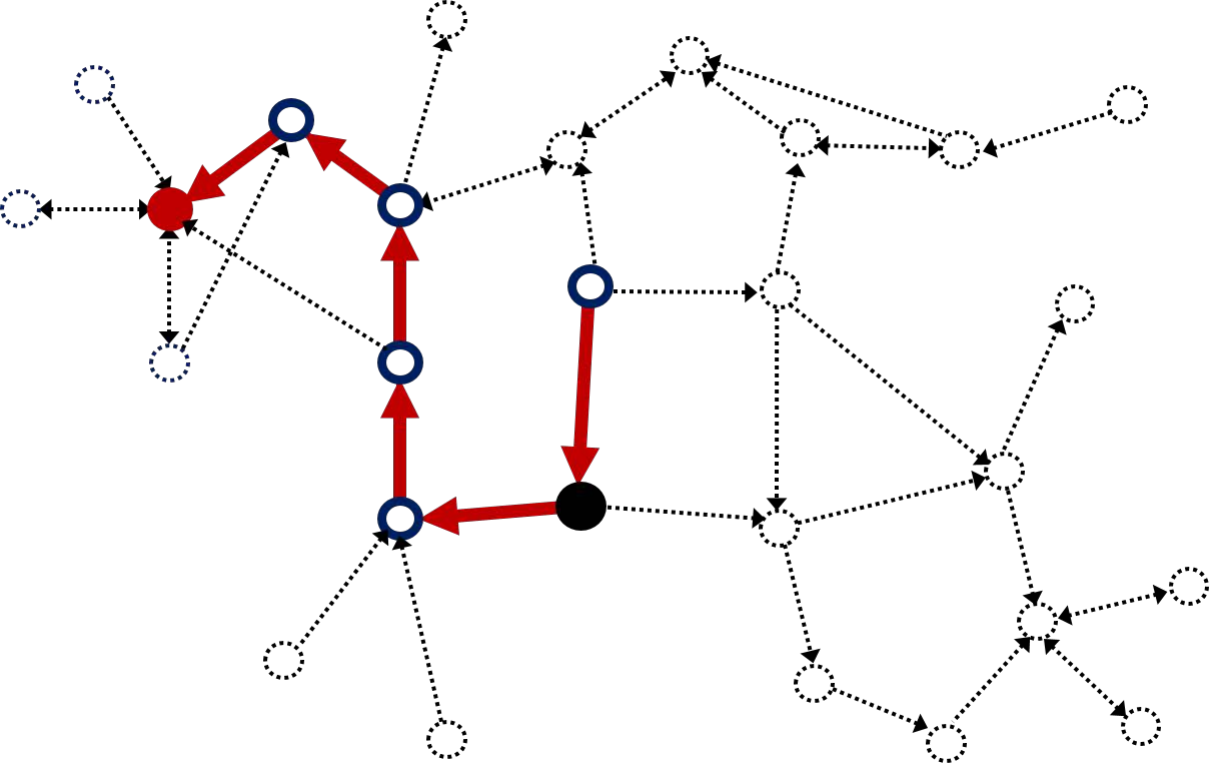}
\caption{ }
\label{fig:linear_thresholdred}
\end{subfigure}~\begin{subfigure}[b]{0.32\textwidth}
\includegraphics[width=\textwidth]{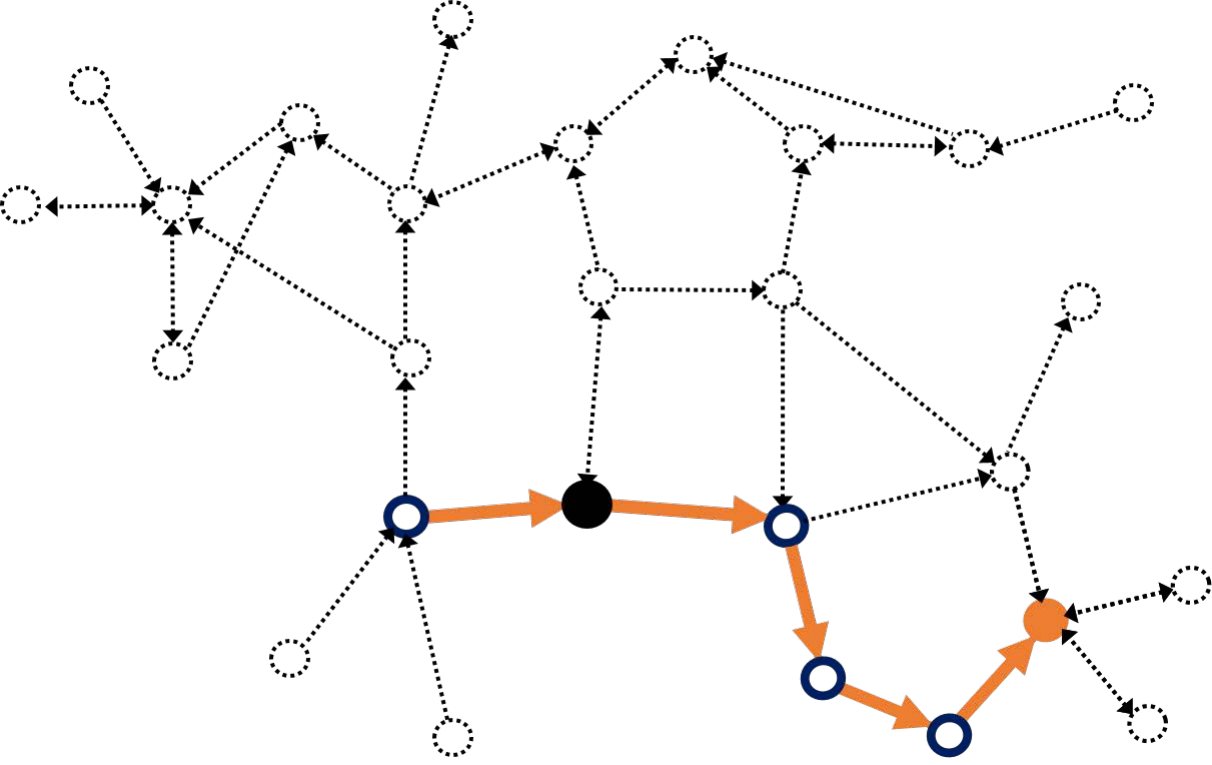}
\caption{ }
\label{fig:linear_thresholdorange}
\end{subfigure}~\begin{subfigure}[b]{0.32\textwidth}
\includegraphics[width=\textwidth]{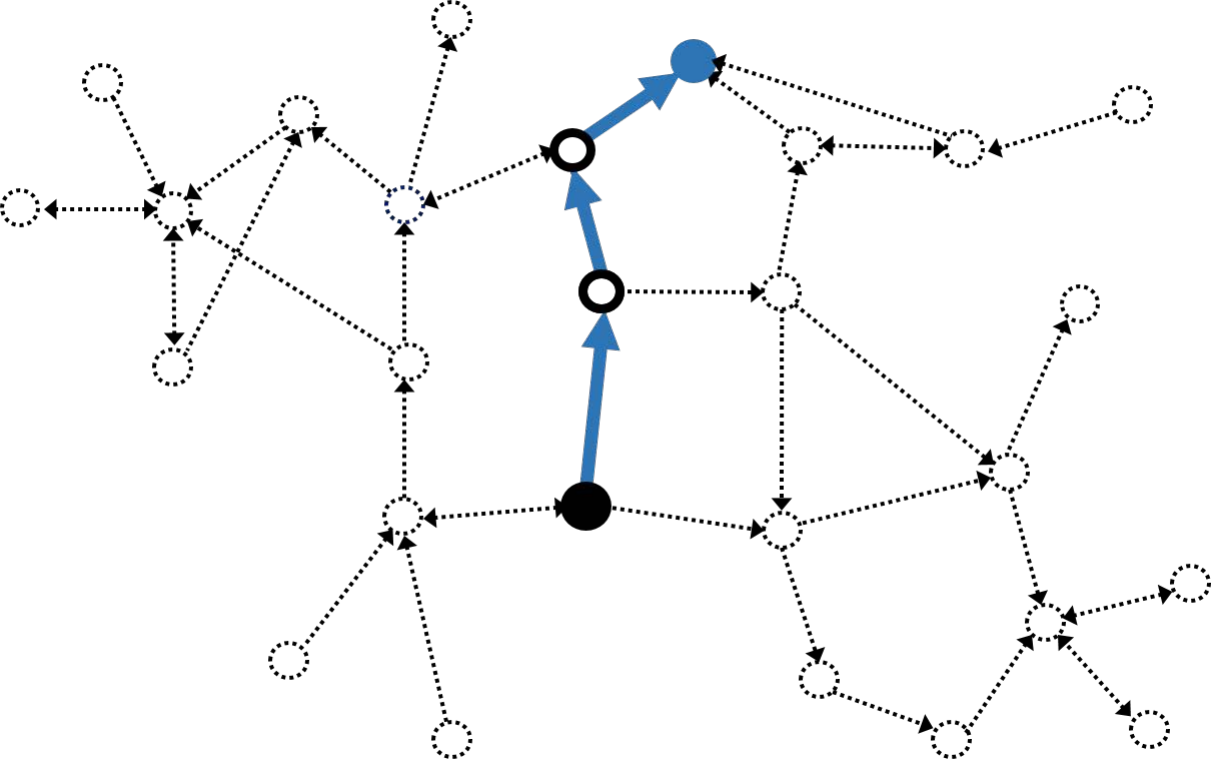}
\caption{ }
\label{fig:linear_thresholdblue}
\end{subfigure}
\caption{Some diffusion processes can be reduced to a triggering set model; whereby, after randomly drawing a triggering set from among her neighbors, a node becomes active if any of the nodes in her triggering set are active. In such models, we can devise edge queries by having a probed node reveal a random realization of her triggering set, and then proceeding to probe the revealed nodes. In the case of the linear threshold model, the triggering sets are either empty or a singleton chosen randomly according to the edge weights. The reverse cascades in (\subref{fig:linear_thresholdred}), (\subref{fig:linear_thresholdorange}) and (\subref{fig:linear_thresholdblue}) show three queries where the nodes sequentially reveal their triggering sets, starting from a random initial node. The black node appears in the output of all three ``reversed cascades'', and therefore scores the highest as a seed candidate.}
\label{fig:linear_threshold_queries}
\end{figure}

For diffusion processes that can be cast as a triggering set model, we can implement queries by having the probed nodes reveal their triggering sets. Starting from random initial nodes, each triggering set corresponds to a batch of directed edges that are incoming to the probed node. We can use the number of reachable initial nodes to implement an approximate greedy heuristic as described in Appendix \ref{app:ext} --- see Figure \ref{fig:linear_threshold_queries}. One needs to analyze the specific diffusion process and the triggering distributions to provide approximation guarantees using a bounded number of queries. 

In the particular case of the linear threshold model, the proof of Theorem \ref{thm:main_spreading_queries}, and its adaptation to directed graphs in Appendix \ref{app:ext}, continue to hold with little change. Using $k$ batches of $\rho$ reversed cascades, we can provide an approximate $k$-IM solution for the linear threshold model over directed graphs. At each stage we choose $\rho$ initial nodes at random and implement $\rho$ cascades in reverse. In each reversed cascade, we  start from the initial node, reveal her triggering set, and then proceeding to probe the node that is revealed in the triggering set, etc.  After each batch of $\rho$ reversed cascades, we choose the node that appears the most number of times, discarding those cascades that include any of the already chosen seeds. Following Theorem \ref{thm:main_spreading_queries}, we can provide a $(1-1/ e) {\opt}-\epsilon n$ approximation guarantee, by running $k\rho = k\color{green} \lceil \normalcolor {81 k \log (\frac {6nk}{\epsilon})}/{\epsilon^3} \color{green} \rceil \normalcolor$ reversed cascades.

To bound the number of edge queries, recall that each triggering set in the linear threshold model consists of at most a single node. Hence, each reversed cascade corresponds to a path of length at most $n$ --- see Figure~\ref{fig:linear_threshold_queries}. Therefore, we can bound the total number of queried edges by $nk\rho = n k \color{green} \lceil \normalcolor {81 k \log (\frac {6nk}{\epsilon})}/{\epsilon^3}  \color{green} \rceil \normalcolor$. In Theorem \ref{theo:main_lin_threshold_queries}, we state our results formally for the case of $k$-IM with the linear threshold model over directed graphs.

\end{APPENDICES}

%\printendnotes

\bibliographystyle{informs2014} 
\bibliography{ref}

\subsubsection*{Dean Eckles}is an associate professor at the MIT Sloan School of Management. His research interests include social influence, networks, and causal inference.

\subsubsection*{Hossein Esfandiari}is a senior research scientist at Google. His works are in theoretical computer science, including approximation algorithms and sublinear-time algorithms. 

\subsubsection*{Elchanan Mossel}works in probability, combinatorics, and inference. He is on the senior faculty of the Mathematics Department, with a jointly core faculty appointment at the Statistics and Data Science Center of Massachusetts Institute of Technology’s Institute for Data, Systems, and Society.

\subsubsection*{M. Amin Rahimian}is an assistant professor of Industrial Engineering at University of Pittsburgh. He works at the intersection of networks, data, and decision sciences. He borrows tools from applied probability, statistics, algorithms, as well as decision and game theory to address problems of distributed inference and decentralized interventions in large-scale sociotechnical systems.

\end{document}